ORIGINAL ARTICLE

# Comparison of Deep Space Navigation Using Optical Imaging, Pulsar Time-of-Arrival Tracking, and/or Radiometric Tracking

Todd Ely[1] · Shyam Bhaskaran[1] · Nicholas Bradley[1] · T. Joseph W. Lazio[1] · Tomas Martin-Mur[1]

## Abstract

Recent advances with space navigation technologies developed by NASA in space-based atomic clocks and pulsar X-ray navigation, combined with past successes in autonomous navigation using optical imaging, brings to the forefront the need to compare space navigation using optical, radiometric, and pulsar-based measurements using a common set of assumptions and techniques. This review article examines these navigation data types in two different ways. First, a simplified deep space orbit determination problem is posed that captures key features of the dynamics and geometry, and then each data type is characterized for its ability to solve for the orbit. The data types are compared and contrasted using a semi-analytical approach with geometric dilution of precision techniques. The results provide useful parametric insights into the strengths of each data type. In the second part of the paper, a high-fidelity, Monte Carlo simulation of a Mars cruise, approach, and entry navigation problem is studied. The results found complement the semi-analytic results in the first part, and illustrate specific issues such as each data type's quantitative impact on solution accuracy and their ability to support autonomous delivery to a planet.

✉ Todd Ely
Todd.A.Ely@jpl.nasa.gov

[1] Jet Propulsion Laboratory, California Institute of Technology, Pasadena, CA, USA

## Introduction

Recent successes with NASA's Station Explorer for X-ray Timing and Navigation Technology (SEXTANT) experiment on the International Space Station using a pulsar X-ray detector to collect arriving photons from target pulsars and time tagging them to formulate time-of-arrival (TOA) measurements for use in orbit determination has raised interest in the suitability of pulsar TOA measurements for autonomous deep space navigation [1–5]. Another recent technology advancement is the development and space demonstration of NASA's Deep Space Atomic Clock (DSAC) launched June 2019 to prove DSAC's stability for forming precise one-way radiometric tracking data that also could be used for autonomous deep space navigation [6, 7]. These new developments follow the success in the late 1990s of NASA's Deep Space 1 technology demonstration mission that proved the viability of autonomous navigation using JPL's AutoNav software processing optical imaging of solar system bodies from its onboard camera system. This autonomous navigation technique was later used for critical mission operations of Stardust and Deep Impact [8]. Another emerging, nascent technology is StarNAV that uses relativistic perturbations in the observed wavelengths of stars and their associated locations to determine a spacecraft's absolute velocity, directly, and position, indirectly [9]. StarNAV is a promising development that warrants further investigation; however, the current study focuses on radio, optical, and pulsar-based navigation technologies that have had an initial space demonstration; future work will compare StarNav to the preceding measurement types. These technological advances create an opportunity to extend autonomous deep space navigation capabilities towards new levels of performance, robustness, and reliability that would enable its use and reduce the need for traditional ground-based deep space navigation. With this in mind, we will consider how these three data types - pulsar TOA tracking, one-way radiometric tracking, and optical imaging - compare and what their relative strengths are for use in autonomous deep space navigation.

To examine the autonomous navigation problem, it is natural to first consider traditional ground-based deep space navigation. Since the advent of space exploration in the 1960's, deep space navigation has relied primarily on processing radiometric tracking to produce trajectory solutions and maneuvers to maintain flight path control in the presence of trajectory disturbances and solution errors. The radiometric data for most NASA missions, typically two-way range and Doppler, and, since the early 2000's, double differenced one-way range – called DDOR for short – is obtained via tracking by NASA's Deep Space Network (DSN). An extensive history of deep space navigation has been documented by Wood [10–14] in a series of papers that describes developments in tracking methods and accuracy of the DSN; the use of optical navigation; and deep space navigation techniques and methods, including the early experiences with autonomous navigation. Fundamental, to the ground-based navigation is the use of two-way range and Doppler, 'line-of-sight' measurements between an Earth ground-station and the spacecraft, for orbit determination. These measurements have extreme precision (at X-Band, typically, 1–3 m for 1-σ range error and less than 0.1 mm/s for 1-σ range rate error) with ground-based atomic frequency standards forming the basis of this precision [15]. Two-way range and Doppler has provided sufficient navigation capabilities for many decades (examples include Voyager, Galileo, Mars Pathfinder, etc.), but in the early 2000's the loss of the Mars Climate Orbiter due to inconsistent navigation solutions prompted NASA to add DDOR as another radiometric measurement type to augment two-way Doppler and two-way range for future missions [16]. DDOR utilizes the baseline geometry between two receiving DSN stations to provide a precise 'plane-of-sky' measurement (with a 1-σ error of 2–3 nanoradians) that is complementary to Doppler and range. The combination of all three data types provides for robust ground-based navigation that is standard for most NASA missions today. When considering onboard, autonomous navigation, a similar approach – combining disparate and complimentary navigation data – should be considered for obtaining accurate solutions that are naturally fault tolerant; this is explicitly examined in Part 2.

Another measurement that is used to complement radiometric data is optical navigation (OpNav for short) using onboard imaging of target objects. This provides the bearing between the spacecraft and the observed target and, like DDOR, is complementary to Doppler and range. For ground-based navigation, the images are telemetered to the ground and processed along with the radiometric data to yield navigation solutions, and has been well documented by Wood [10–14]. OpNav provides direct information relative to the imaged targets of interest, which become fiducial objects when their ephemerides are well-known and can then be used for absolute navigation. A recent work by Broschart [17] examines the ability of OpNav to determine trajectories by imaging asteroids and concludes that OpNav is a good candidate for onboard absolute autonomous navigation. That is, their known positions plus associated uncertainties are used as reference locations in an image, which provides sufficient information for orbit determination (OD) to converge (where the process has also been initialized with a priori spacecraft state information, the typical scenario encountered in space navigation).[1] Furthermore, OpNav naturally transitions into target relative navigation as a spacecraft nears a destination object, which is essential for accurate planetary entry or close flybys. Naturally, the target relative navigation problem must consider the target's extended body when the spacecraft is close to the target. In this case, terrain matching and/or limb tracking (see Christian [18] for a recent autonomous navigation example) can be useful. However, the current study's focus is on absolute navigation using OpNav of point sources (additional references on using point source objects for OD are provided by Riedel [19], Vasile [20], and Enright [21]).

Since OpNav is already an onboard measurement, it is a natural measurement type for use in an autonomous navigation system (as has already been demonstrated on a limited scale by Deep Space 1), but to obtain the robustness and accuracy similar to the state-of-the-art ground-based navigation other onboard navigation data types would be needed [17]. Collecting radiometric data onboard is most effective with one-way transmissions (versus two-way coherent Doppler and range), but to get measurements with accuracy similar to two-way data requires an extremely stable onboard frequency reference/clock. DSAC provides the frequency stability necessary for making this possible [22]. Similarly, the ability to take pulsar TOA measurements onboard represents another technological advance that could be used for autonomous navigation [1, 2]. The comparison of these data types is discussed herein and consists of two parts:

1. Part 1: A simplified two-dimensional problem is examined that captures the fundamental geometric and dynamic characteristics of deep space cruise navigation, and facilitates a *qualitative* comparison of the navigation characteristics of the three different data types. The measurement models and the dynamic model have been reduced to their leading order, fundamental representations so that the most significant effects can be examined and compared using a semi-analytical approach.
2. Part 2: A high-fidelity Mars cruise, approach, and entry navigation problem is developed that is similar to NASA's Mars InSight mission, and uses combinations of the three data types being investigated. This problem has been formulated using the same models and characteristics developed by InSight's navigation team during the development stages of the mission [23–25]. High fidelity models for the optical and one-way radiometric measurements will be utilized as well as a representative pulsar TOA model that are sufficient for making *quantitative* comparisons of the data types for navigating a spacecraft to Mars. The results will also augment the qualitative findings from Part 1, and provide insight into navigating in the outer solar system.

[1] The so called 'lost-in-space' problem where there is no a priori knowledge of the spacecraft state has not been addressed in the in this research.

## Part 1: 2-d Qualitative Analysis of Simplified Cruise Phase Deep Space Navigation

In the qualitative analysis, the dynamics are restricted to Keplerian two-body motion about the Sun, and the spacecraft is in a circular orbit at either a representative inner solar system radius or outer solar system radius with radii for Mars' and Neptune orbits selected, respectively. Furthermore, the geometry is reduced from three-dimensions (3-d) to just 2-d by restricting the spacecraft orbit and all the celestial body orbits or positions (if inertially fixed such as a pulsar) to lie in the plane of the ecliptic. That is, no out of plane motions are considered. While these models are extremely simplified, they do capture the key lowest-order geometry and dynamics (i.e., inverse square effects, relative geometries, orbital rates, etc.) of the deep space cruise navigation problem, and facilitate comparing the relative merits of the measurement types, also, using simplified models.

There are many complicating factors that affect the accuracy, precision, and performance of the three measurement types being examined. Examples include: availability of a pulsar catalog with stable sources, stability of available clocks, and sensitivity of camera systems to detecting distant asteroids. In the qualitative analysis, it is assumed that these issues have been ‘solved’ and are not significant factors that need to be considered. This reduces each measurement type to its fundamental geometric characteristics with only simple, unbiased measurement noise affecting measurement precision. The combination of simple 2-d dynamic models with simplified observation models that capture the essential geometries yields a problem that can be analyzed analytically or semi-analytically and can be used to characterize the fundamental information content that each measurement type has to offer; facilitating comparisons between them.

The deep space navigation problem fundamentally begins with an observation problem to determine a spacecraft’s orbit (or trajectory) and then considers flight path control to guide a spacecraft on the estimated (determined) trajectory back to a desired trajectory. Our analysis is restricted to the orbit determination problem, and, since an orbit is defined by both position and velocity, it is necessary to estimate both simultaneously; position-only determination is not sufficient. Hence, our 2-d geometry requires a 4-d estimation state that consists of the two components of the initial position vector and the two components of the initial velocity vector. Also typical of deep space navigation, observations are required to be collected over time in order to estimate the orbit accurately. Indeed, all three data types being considered are sensitive to only one or two dimensions at any given time, thus collecting measurements at different times (with the commensurate geometry change induced by the dynamics in the problem) is necessary to have a solvable 4-d problem. To improve the accuracy of these estimates requires an observation interval over a period of time and, to compare the data types on an equal basis, a common observation interval. A representative interval is a full sidereal day. This corresponds to continuous tracking by the Deep Space Network (DSN) with up to three tracking passes (one at each DSN complex) of about 8 h each, or 144 different pulsar TOA measurements separated by 10 min each, or 144 different asteroid images, also, separated by 10 min each. In the case of ranging with the DSN, one-day represents the minimum amount of time to complete a full set of tracks from the three different complexes before returning to the first complex. In that sense, it represents a ‘canonical’ period that will guide the comparison with the other observation methods. In addition to analyzing relative solution uncertainties achievable with one-day of data, we will investigate multi-day arcs of data to characterize the effect that a changing geometry has on the solution knowledge vs the increased knowledge gained by simply adding more data (which, for normally distributed and uncorrelated random variables, should improve knowledge proportionally to the square root of the number of data points collected).

## Geometry, Dynamics, and Assumptions

The geometry of the simplified 2-d problem is shown in Fig. 1. The spacecraft nominal trajectory is a heliocentric, circular orbit with a radial distance $r$ from the solar system barycenter (SSB) and associated central angle $\theta$ with respect to the inertial, $x-axis$ which can be related to orbital elements as follows

$$\{r, \theta\} = \left\{a, n\left(t - t_0\right) + \theta_0\right\} \tag{1}$$

where the orbit's semi-major axis is $a$, its mean motion (or constant angular velocity) is $n = \sqrt{\mu/a^3}$, and $\theta_0$ is the initial angle. Note that since the orbit is nominally circular, all the anomalies (True, Eccentric, and Mean) are equivalent to $\theta$; hence, no further distinction will be made between the anomalies. Without loss of generality, we will set $t_0=0$ so that time represents an elapsed time since epoch, $\theta_0=0$ so that the axes $\{x, y\}$ of the inertial frame are aligned with the spacecraft's initial position.

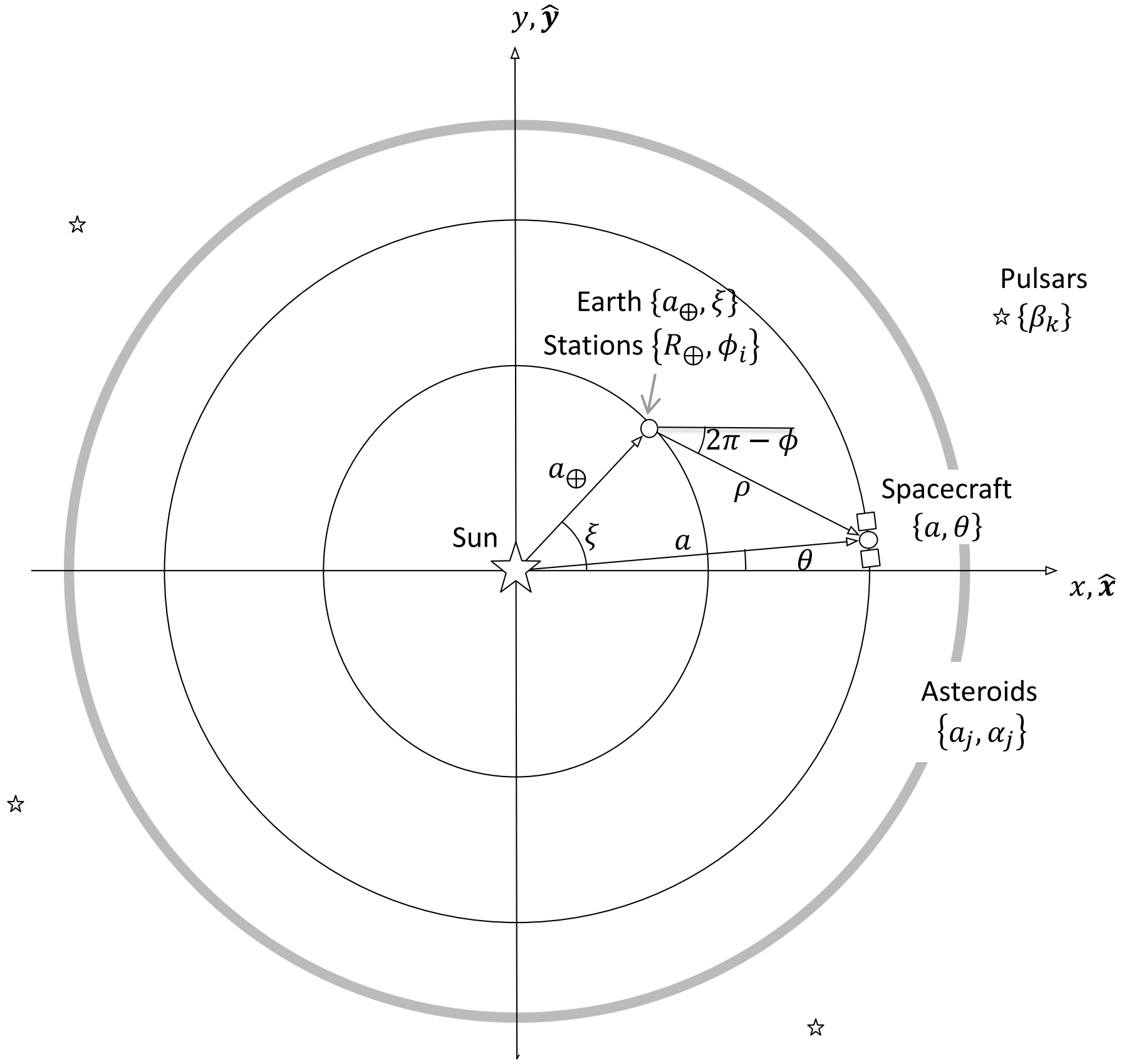

**Fig. 1** Simplified two-dimensional solar system and trajectory geometry

The pulsar ascension angle $\beta_k$ and asteroid ascension angle $a_j$ are delta angles with respect to the $x$-axis.

Definitions for the position and pointing vectors include:

1. The spacecraft's heliocentric position vector is given by $\mathbf{r} = a\cos nt\,\hat{\mathbf{x}} + a\ \sin\ nt\,\hat{\mathbf{y}}$ where, as noted previously, the semi-major axis $a$ is nominally constant and $\theta = nt$ because the orbit is circular. Its associated inertial velocity is $\mathbf{v} = -an\sin nt\,\hat{\mathbf{x}} + an\cos nt\,\hat{\mathbf{y}}$ . Note that the nominal initial conditions are $\mathbf{r}_0 = a\hat{\mathbf{x}}$ and $\mathbf{v}_0 = an\,\hat{\mathbf{y}}$.
2. The Earth's heliocentric position vector is given by $\mathbf{r}_\oplus + a_\oplus cos\xi\hat{\mathbf{x}} + a_\oplus \sin\xi\hat{\mathbf{y}}$ where $a_\oplus$ is a constant, $\xi = n_\oplus t + \xi_0$, and $\{n_\oplus, \xi_0\}$ are also constants.
3. The heliocentric position vector of the $i$-th Earth tracking (and/or beacon) station $S_i$ is $\mathbf{r}_{s_i} = R_\oplus \cos\phi_i\hat{\mathbf{x}} + R_\oplus \sin\phi_i\hat{\mathbf{y}} + \mathbf{r}_\oplus$ where $R_\oplus$ is the Earth's radius (for this simplified problem, we assume the stations are located at a zero latitude), $\phi_i = \omega_\oplus t + \phi_{i,0}$, and $\{\omega_\oplus, \phi_{i,0}\}$ are also constants.
4. The heliocentric position vector of the $j$-th asteroid $(A_j)$ is $\mathbf{r}_{s_i} = R_\oplus \cos\phi_i\hat{\mathbf{x}} + R_\oplus \sin\phi_i\hat{\mathbf{y}} + \mathbf{r}_\oplus$ where $a_{A_j}$ is a constant. Since any given asteroid $A_j$ is being observed at a discrete time, its intrinsic motion is not relevant; therefore, we can treat $\alpha_j$ as a constant.
5. The inertial pointing unit vector of the $k$-th pulsar $(P_k)$ is $\hat{\mathbf{n}}_{P_k} = \cos\beta_k\hat{\mathbf{x}} + \sin\beta_k\hat{\mathbf{y}}$ where the angle $\beta_k$ is a constant since no proper motion is being considered.

For the analytic study we will consider two important cruise navigation cases, one at Mars' distances and the other at Neptune distances from the SSB. These are representative cases for inner and outer solar system navigation. Below are lists of representative values for the quantities being investigated and their associated order assumptions suitable for an asymptotic analysis.

1. Zeroth-order $O(1)$ quantities:

   a. $a$ is ~1.5 AU (Mars' distance) or a ~ 30 AU (Neptune's distance)
   b. $a_\oplus$ is ~1 AU
   c. For navigating out to distances as far as Jupiter, the main-belt asteroids can be utilized [17]. The mid-point distance of these asteroids is $a_A$ is~2.7 AU (with a range from 2.2 AU to 3.2 AU). Navigating in the outer solar system using images of minor planets such as the centaurs, scattered disk objects (SDOs), and/or Kuiper belt objects (KBOs), while more challenging, could be possible with improvements in camera sensitivity, image processing, and improved catalogs. For the Neptune case, we assume the existence of a catalog with a set of known objects in the range of 5 AU $< a_{A_j} <$ 50AU.
   d. All the absolute angles and associated initial values $\{\theta, \theta_0, \xi, \xi_0, \phi_i, \phi_{i,0}, \alpha_j, \alpha_{j,0}, \beta_k\}$ are assumed to lie in the range $\{0, 2\pi\}$.
   e. The only angular rate that is zeroth-order is the Earth's rotation rate at $\omega_\oplus$ is~$2\pi$ rad/day (for simplicity the distinction between sidereal and solar days is ignored).

f. The derived angular shift for the Earth tracking station $\Delta\phi = \omega_{\oplus}T$ over the fundamental observation period being examined $T \equiv 1$ day is a full $2\pi$ radians.

2. First-order $O(\varepsilon)$ quantities:

a. $n$ is ~$2\pi$ rad/6 87 days = 0.009 rad/day (Mars' distance)
b. $n$ is ~$2\pi$ rad/60182 days = 0.001 rad/day (Neptune's distance)
c. $n_{\oplus}$ is ~$2\pi$ rad/365 days = 0.017 rad/day
d. The derived angular shifts for the spacecraft $\Delta\theta = nT$ and the Earth $\Delta\xi = n_{\oplus}T$ over the one-day period is 0.009 rad and 0.017 rad, respectively.
e. $R_{\oplus}$ is ~6378 km, therefore $R_{\oplus}/a$ is ~$2.8 \times 10^{-5}$.

Later, when we are doing an asymptotic analysis to eliminate higher order terms, we will use the ordering parameter $\varepsilon$ explicitly in the equations for those terms that are $O(\varepsilon)$ or higher and when the analysis is complete set $\varepsilon$ to one.

## State Model

The quantities of interest are the spacecraft's heliocentric position vector $\mathbf{r}(t)$ and velocity vector $\mathbf{v}(t)$. It proves more convenient for our analysis to convert the velocity vector into a position displacement so that all state vector elements have a dimension of length and yield an information matrix for a given observation scenario with a consistent set of units. This is easily done via a change of scale by multiplying the velocity components with the canonical one-day observation period $T$ of interest. The result is a linear displacement $\Delta\mathbf{r}(t)$ defined as

$$\Delta\mathbf{r}(t) \equiv T\mathbf{v}(\mathrm{t}) \tag{2}$$

The four-dimensional state vector $\mathbf{x}(t)$ is the following combination of the position vector and linear displacement vector

$$\mathbf{x}(t) \equiv \begin{bmatrix} \mathbf{r}(t) \\ \Delta\mathbf{r}(t) \end{bmatrix} = \mathbf{S}\begin{bmatrix} \mathbf{r}(t) \\ \mathbf{v}(t) \end{bmatrix} \tag{3}$$

with

$$\mathbf{S} = \begin{bmatrix} 1 & 0 & 0 & 0 \\ 0 & 1 & 0 & 0 \\ 0 & 0 & T & 0 \\ 0 & 0 & 0 & T \end{bmatrix} \text{ and } \mathbf{S}^{-1} = \begin{bmatrix} 1 & 0 & 0 & 0 \\ 0 & 1 & 0 & 0 \\ 0 & 0 & 1/T & 0 \\ 0 & 0 & 0 & 1/T \end{bmatrix} \tag{4}$$

The goal is to determine bounds on the information content and estimation uncertainty of the state using observations collected under several different scenarios (i.e., ranging, passive optical imaging, or pulsar timing). The dynamic and observation models are nonlinear; however, we utilize the standard linearization step and seek uncertainties and the associated information content associated with estimates of the linear state deviations around a priori nominal values for the full state. That is, we estimate information for

$$\delta\mathbf{x}(t) \equiv \begin{bmatrix} \delta\mathbf{r}(t) \\ \delta\Delta\mathbf{r}(t) \end{bmatrix} = \begin{bmatrix} \mathbf{r}(t) - \mathbf{r}_n(t) \\ T\left(\mathbf{v}(t) - \mathbf{v}_n(t)\right) \end{bmatrix} \tag{5}$$

where the subscript $n$ represents the known nominal values for the position and velocity.

### State Transition Matrix

Recall that the state transition matrix (STM) from time $t_0$ to time $t$ relates the state deviations at these times as follows

$$\delta\mathbf{x}(t) = \mathbf{\Phi}(t, t_0)\ \delta\mathbf{x}(t_0) \tag{6}$$

where, for notational simplicity, we will drop the time argument and use the following definitions for the state deviation components $\delta\mathbf{x} = \delta\mathbf{x}(t)$, $\delta\mathbf{x}_0 = \delta\mathbf{x}(t_0)$, $\delta\mathbf{r} = \delta\mathbf{r}(t)$, $\delta\mathbf{r}_0 = \delta\mathbf{r}(t_0)$, $\delta\Delta\mathbf{r} = \delta\Delta\mathbf{r}(t)$, and $\delta\Delta\mathbf{r}_0 = \delta\Delta\mathbf{r}(t_0) = T\delta\mathbf{v}(t_0)$. The STM $\mathbf{\Phi}(t, t_0)$ associated with the spacecraft's circular orbit can be formulated analytically for a Keplerian orbit in three dimensions using the method outlined in Chapter 9 of Battin [26]. Since the orbit is circular and in two dimensions, the STM and its constituent $2\times2$ submatrices are readily found and take the form

$$\mathbf{\Phi}\left(t, t_0\right) \equiv \begin{bmatrix} \mathbf{\Phi}_{\mathbf{rr}}\left(t, t_0\right) & \mathbf{\Phi}_{\mathbf{r}\Delta\mathbf{r}}\left(t, t_0\right) \\ \mathbf{\Phi}_{\Delta\mathbf{rr}}\left(t, t_0\right) & \mathbf{\Phi}_{\Delta\mathbf{r}\Delta\mathbf{r}}\left(t, t_0\right) \end{bmatrix} = \mathbf{S}\begin{bmatrix} \mathbf{\Phi}_{\mathbf{rr}}\left(t, t_0\right) & \mathbf{\Phi}_{\mathbf{rv}}\left(t, t_0\right) \\ \mathbf{\Phi}_{\mathbf{vr}}\left(t, t_0\right) & \mathbf{\Phi}_{\mathbf{vv}}\left(t, t_0\right) \end{bmatrix}\mathbf{S}^{-1} \tag{7}$$

with

$$\mathbf{\Phi}_{\mathbf{rr}}\left(t, t_0\right) = \begin{bmatrix} \frac{1}{2}(-3 + 4\cos nt + \cos\ 2nt + 6nt\sin nt) & (1 - \cos\ nt\ )\sin\ nt \\ \frac{1}{2}(4\sin nt + \sin 2nt - 6nt\cos nt) & 1 - (1 - \cos\ nt)\cos nt \end{bmatrix} \tag{8}$$

$$\mathbf{\Phi}_{\mathbf{r}\Delta\mathbf{r}}\left(t, t_0\right) = \frac{1}{nT}\begin{bmatrix} (2 - \cos nt)\sin nt & -3 + 2\cos nt + \cos 2nt + 3nt\sin nt \\ (1 - \cos nt)^2 & 2\sin nt + \sin\ 2nt - 3nt\cos nt \end{bmatrix} \tag{9}$$

$$\mathbf{\Phi}_{\Delta\mathbf{rr}}\left(t, t_0\right) = \frac{nT}{4}\begin{bmatrix} \sin\ nt + 2\ \sin\ 2nt - 3\sin 3nt + 6nt + 6nt\cos 2nt & \cos\ nt - 4\cos 2nt + 3\cos 3nt + 6nt\sin 2nt \\ 6 - 7\ \cos\ nt - 2\cos 2nt + 3\cos 3nt + 6nt\sin 2nt & -2\ \sin\ nt(1 + 4\cos nt - 3\cos 2nt - 6nt\sin nt) \end{bmatrix} \tag{10}$$

$$\mathbf{\Phi}_{\Delta\mathbf{r}\Delta\mathbf{r}}\left(t, t_0\right) = \begin{bmatrix} 2\ \cos\ nt - \cos 2nt & \sin\ nt - 2\sin 2nt + 3nt\cos nt \\ 2(1 - \cos\ nt)\ \sin\ nt & -\cos nt + 2\cos 2nt + 3nt\sin nt \end{bmatrix}. \tag{11}$$

We will take advantage of the fact that a sequence of measurements collected over short periods of time (such as a day or several days) support use of the following approximations for the STM of a heliocentric orbit. The STM $\boldsymbol{\Phi}\,(t,t_0)$ can be expanded asymptotically in the form

$$\boldsymbol{\Phi}\left(t,t_0\right)=\boldsymbol{\Phi}^{\underline{0}}\left(t,t_0\right)+\varepsilon\boldsymbol{\Phi}^{\underline{1}}\left(t,t_0\right)+\varepsilon^2\boldsymbol{\Phi}^{\underline{2}}\left(t,t_0\right)+\varepsilon^3\boldsymbol{\Phi}^{[\underline{3}]}\left(t,t_0\right) \tag{12}$$

where the order of the function is annotated with an underlined superscript. That is, $f^{\underline{n}}(\bullet)$ represents the n-th order function in an asymptotic expansion of $f(\bullet)$. This notation will assist in distinguishing between the function's order and an exponent. The use of the square bracketed superscript $[\underline{3}]$ in Eq. (12) identifies the lowest order of all the neglected terms and defines the 'order of the approximation.' In this case, the asymptotic approximation includes terms for the zeroth, first, and second order and is accurate to the third-order (i.e., a third-order approximation). Specific expressions for the $\boldsymbol{\Phi}^{\underline{n}}\left(t,t_0\right)$ include

$$\boldsymbol{\Phi}^{\underline{0}}\left(t,t_0\right)\equiv\begin{bmatrix}1&0&\frac{t}{T}&0\\0&1&0&\frac{t}{T}\\0&0&1&0\\0&0&0&1\end{bmatrix},\boldsymbol{\Phi}^{\underline{1}}\left(t,t_0\right)=\mathbf{0}_{4\times4},\boldsymbol{\Phi}^{\underline{2}}\left(t,t_0\right)\equiv\begin{bmatrix}n^2t^2&0&\frac{n^2t^3}{3T}&0\\0&-\frac{1}{2}n^2t^2&0&-\frac{-n^2t^3}{6T}\\2n^2tT&0&n^2t^2&0\\0&-n^2tT&0&-\frac{1}{2}n^2t^2\end{bmatrix}. \tag{13}$$

Note that the zeroth-order term $O(0)$ is actually good to second-order $(\varepsilon^2)$ . That is, there is no first order term present $O(\varepsilon)$. For convenience, the following truncated order state transition matrices are defined

$$\boldsymbol{\Phi}^{(\underline{1})}\left(t,t_0\right)\equiv\boldsymbol{\Phi}^{\underline{0}}\left(t,t_0\right)+\varepsilon\boldsymbol{\Phi}^{\underline{1}}\left(t,t_0\right) \tag{14}$$

where the parenthesis notation $(\underline{1})$ identifies the expansion as a k-jet, in this case a 1-jet, that includes all terms up to and including k, and

$$\boldsymbol{\Phi}^{(\underline{2})}\left(t,t_0\right)\equiv\boldsymbol{\Phi}^{\underline{1}}\left(t,t_0\right)+\varepsilon^2\boldsymbol{\Phi}^{\underline{2}}\left(t,t_0\right) \tag{15}$$

It is informative to compare the error introduced by using the above approximations on the heliocentric problem of interest using the following relative error functions

$$e_{\Phi}^{[\underline{2}]}\equiv\frac{\left\|\boldsymbol{\Phi}^{(\underline{1})}\left(t,t_0\right)-\boldsymbol{\Phi}\left(t,t_0\right)\right\|_F}{\left\|\boldsymbol{\Phi}\left(t,t_0\right)\right\|_F}=\frac{\left\|\boldsymbol{\Phi}^{[\underline{2}]}\left(t,t_0\right)\right\|_F}{\left\|\boldsymbol{\Phi}\left(t,t_0\right)\right\|_F} \tag{16}$$

and

$$e_{\Phi}^{[\underline{3}]}\equiv\frac{\left\|\boldsymbol{\Phi}^{[\underline{2}]}\left(t,t_0\right)-\boldsymbol{\Phi}\left(t,t_0\right)\right\|_F}{\left\|\boldsymbol{\Phi}\left(t,t_0\right)\right\|_F}=\frac{\left\|\boldsymbol{\Phi}^{[\underline{3}]}\left(t,t_0\right)\right\|_F}{\left\|\boldsymbol{\Phi}\left(t,t_0\right)\right\|_F} \tag{17}$$

The notation $\|\bullet\|_F$ represents the Frobenius matrix norm that gives a measure of the 'size' the matrix being operated on. For square matrices, the Frobenius norm is equal to the root sum square (RSS) of the matrix eigenvalues [27]. Recall that at Mars' distances, $n \cong 0.009$ rad/day, the associated error functions $e_\Phi^{[2]}$ and $e_\Phi^{[3]}$ plotted over one-day and 14-day periods obtain the values shown in Fig. 2.

Over the one-day interval, the second-order error $e_\Phi^{[2]}$ at one day is $<0.01\%$ and the third-order error $e_\Phi^{[3]}$ is $<8.0 \times 10^{-5}\%$. Over a more extended 14-day interval, the values obtained are $e_\Phi^{[2]} < 0.5\%$ and $e_\Phi^{[3]} < 0.05\%$. The error growth over these time periods yields results that are reasonable as compared to the order of the approximation and provide a measure of their bounds (i.e., $< 1\%$ error for a second-order analysis, $< 0.1\%$ error for a third-order analysis).

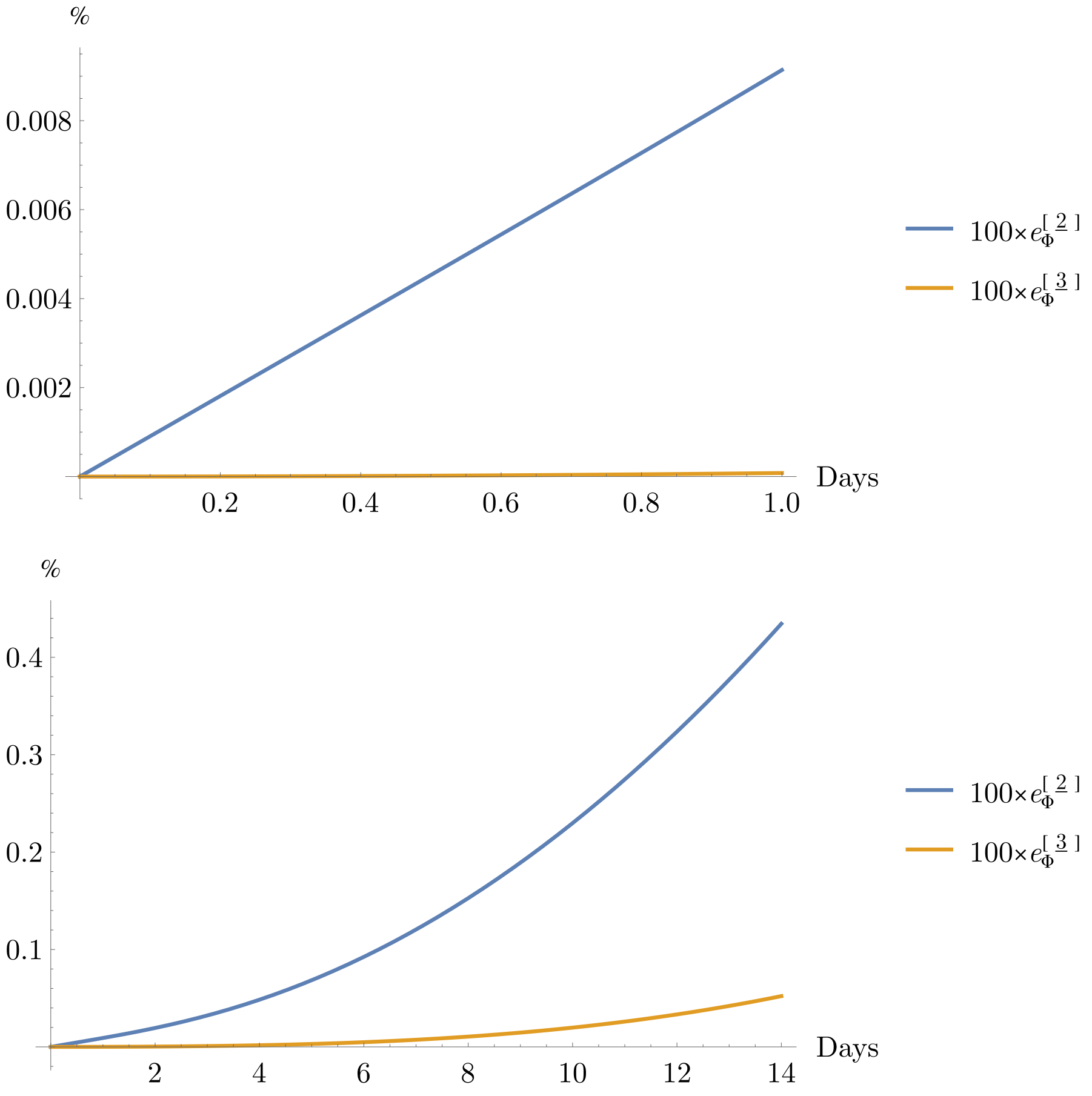

**Fig. 2** Error Functions for the State Transition Matrix Approximations (top plot one-day, bottom plot 14-days)

## Observable Models Suitable for an Analytic Analysis

### Simple OpNav Using Camera Sample Measurements

We now define models for the observables beginning with camera imaging of known bodies (i.e., asteroids) in a simplified 2-d geometry, as shown in Fig. 3. The displacement vector between the spacecraft (technically the origin of the camera focal plane) and the asteroid of interest (assumed to be a point source) is defined as

$$\delta \mathbf{r}_A \equiv \mathbf{r}_A - \mathbf{r} = \delta \mathbf{r}_z \hat{\mathbf{z}}_C + \delta \mathbf{r}_x \hat{\mathbf{x}}_C \tag{18}$$

and

$$\delta r_A = \sqrt{(\mathbf{r}_A - \mathbf{r}) \bullet (\mathbf{r}_A - \mathbf{r})} \tag{19}$$

where, for simplicity, the index $j$ has been dropped for the time being. The camera frame $(\hat{\mathbf{z}}_C, \hat{\mathbf{x}}_C)$ is related to the inertial frame $\hat{\mathbf{x}}, \hat{\mathbf{y}}$ as follows

$$\begin{bmatrix} \hat{\mathbf{z}}_C \\ \hat{\mathbf{x}}_C \end{bmatrix} = \begin{bmatrix} \cos\gamma & \sin\gamma \\ -\sin\gamma & \cos\gamma \end{bmatrix} \begin{bmatrix} \hat{\mathbf{x}} \\ \hat{\mathbf{y}} \end{bmatrix} \tag{20}$$

and inverse

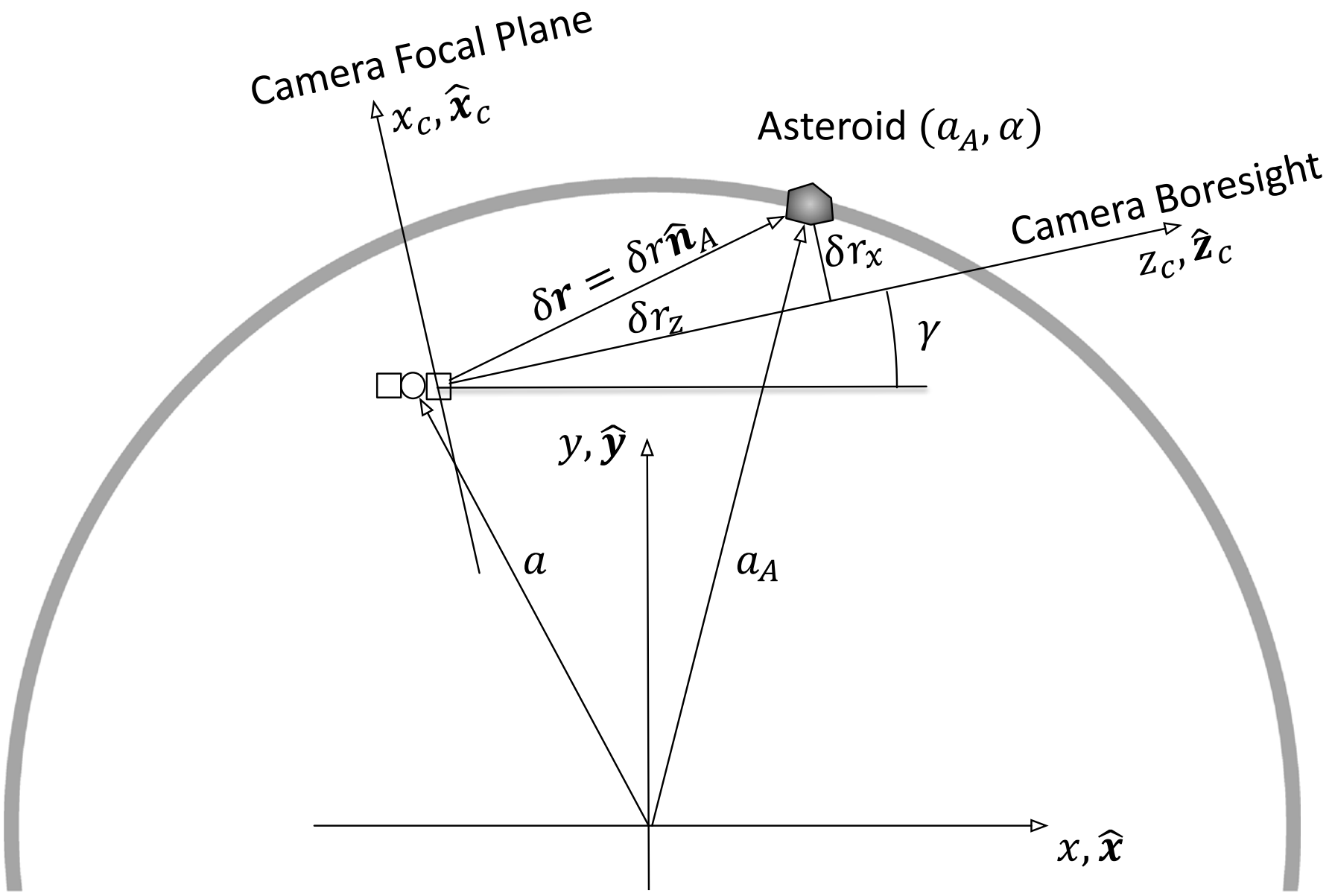

**Fig. 3** Camera frame and image geometry for onboard optical navigation

$$\begin{bmatrix} \hat{\mathbf{x}} \\ \hat{\mathbf{y}} \end{bmatrix} = \begin{bmatrix} \cos\gamma & -\sin\gamma \\ \sin\gamma & \cos\gamma \end{bmatrix} \begin{bmatrix} \hat{\mathbf{z}}_C \\ \hat{\mathbf{x}}_C \end{bmatrix} \tag{21}$$

with the pointing angle $\gamma$ defining the axis of the camera boresight with respect to the inertial x-axis. The camera will image the asteroid and, in our two-dimensional example, will appear at a particular pixel location $s$ (called its sample) in a defined pixel array that is aligned with the focal plane. The sample $s$ can be related to the components of $\delta\mathbf{r}_A$ as follows

$$s = \frac{1}{\Theta}\frac{\delta r_x}{\delta r_z} \tag{22}$$

where $\Theta$ is the resolution of a single camera pixel in radians. For our 2-d problem, we simplify the observation geometry further and constrain the camera boresight to align along the unit vector $\hat{\mathbf{n}}_A$ between the spacecraft and asteroid. That is, the unit vectors $\hat{\mathbf{z}}_C$ and $\hat{\mathbf{n}}_A$ are parallel, such that the following identity holds true

$$\begin{gathered} \hat{\mathbf{z}}_C = \hat{\mathbf{n}}_A = \frac{\delta\mathbf{r}_A}{\delta r_A} = \frac{\mathbf{r}_A - \mathbf{r}}{\sqrt{(\mathbf{r}_A - r)\bullet(\mathbf{r}_A - r)}} \rightarrow \\ \cos\ \gamma\hat{\mathbf{x}} + \sin\ \gamma\hat{\mathbf{y}} = \frac{a_A\cos\alpha - a\cos\theta}{\delta r_A}\hat{\mathbf{x}} + \frac{a_A\sin\alpha - a\sin\theta}{\delta r_A}\hat{\mathbf{y}}. \end{gathered} \tag{23}$$

For simplicity, all quantities are evaluated instantaneously (with no light time delays present). The identity $\hat{\mathbf{z}}_C = \hat{\mathbf{n}}_A$ will be used later to derive the partials of $\gamma$ with respect to the initial state deviation $\mathbf{x}_0$. Note that nominally $\delta r_x = 0$; therefore, in the absence of any noise or errors the sample is also at the origin with $s = 0$. The measurement of interest has now been reduced to the pointing angle to the asteroid $A_j$ at time $t_j$ with the following representative model

$$\tilde{\gamma}_j = \gamma_j + v_{\gamma_j} \tag{24}$$

where the index $j$ has been made explicit, $v_{\gamma_j}$ is the camera observation error to asteroid $A_j$ expressed in radians. For the analytical analysis, this error includes effects of camera measurement noise, pointing error, and, as will be discussed, the asteroid's ephemeris error. We also assume that time is known perfectly. The later quantitative analysis will treat pointing and ephemeris errors as separate filter parameters and introduce errors from an imperfect spacecraft clock for all of the observables.

## Simple Pulsar Time of Arrival Measurements

The pulsar time of arrival measurement is based on detecting wave fronts of pulses from a known pulsar to determine an average pulse over a specified integration interval and comparing that average pulse to its template. The simplified 2-d geometry for the measurement is illustrated in Fig. 4. Once a positive correlation has been determined to a specified resolution, the local spacecraft clock is read and compared to the time that specified wave front should arrive at the solar system barycenter. As with optical measurements, the local time is assumed perfect for now.

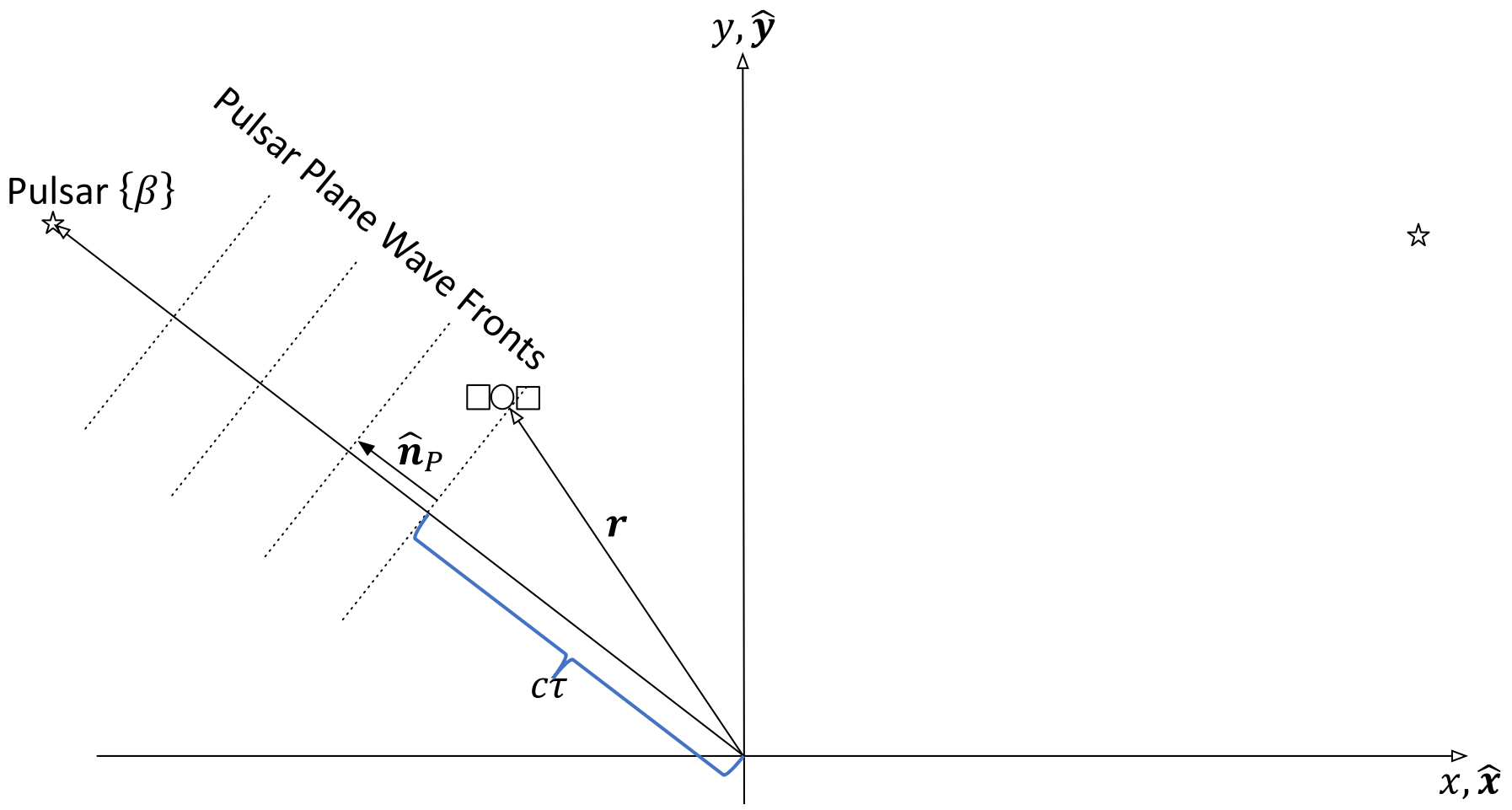

**Fig. 4** Pulsar wave front geometry

As shown in Lorimer and Kramer [28], the simplest form of the time of arrival measurement $\tau$ can be related to the astrodynamics quantities of interest using

$$\tau = \frac{1}{c}\mathbf{r} \bullet \hat{\mathbf{n}}_p. \tag{25}$$

Recall, that $\hat{\mathbf{n}}_p = [\cos\beta, \sin\beta]^T$. To facilitate comparisons to range, the measurement from pulsar $P_k$ at time $t_k$ is recast as a distance (via multiplying by the speed of light $c$) to obtain dimensions of length with the following result

$$\left(c\tilde{\tau}\right)_k = \mathbf{r}^{\mathrm{T}}\hat{\mathbf{n}}_{P_k} + v_{\tau_k} \tag{26}$$

where the dot product has been replaced with the equivalent vector inner product to facilitate obtaining the partials, the index $k$ has been made explicit, and $v_{\tau_k}$ is the observation error (in length) to pulsar $P_k$. The error is a function of the detector design, integration interval, and source signal. Scale factor and bias effects are ignored for now.

### Simple Range Measurements

Finally, we examine the slant range measurement between a transmitting Earth station and the receiving spacecraft. The slant range $\rho_i$ (range for short) between $i$-th Earth tracking station $S_i$ and the spacecraft is defined as

$$\rho_i(t) \equiv \left\| \mathbf{r}(t) - \mathbf{r}_{S_i}\left(t - \frac{\rho_i(t)}{c}\right) \right\| \tag{27}$$

where, to simplify the discussion, the effects of light time delays will be ignored allowing the quantities in the equation to be evaluated at the same time (hence, becoming *instantaneous* slant range). Assuming instantaneous values, the measured range $\tilde{\rho}_i$ at time $t$ that includes additive measurement noise takes the form

$$\tilde{\rho}_i = \sqrt{\left(\mathbf{r} - \mathbf{r}_{S_i}\right) \bullet \left(\mathbf{r} - \mathbf{r}_{S_i}\right)} + v_{\rho_i} \tag{28}$$

where $v_{\rho_i}$ is the range observation error (more on the form and characteristics of this later) and all geometric quantities are evaluated at the common time $t$.

## Optical Image Information Content

Our objective is to examine the information content of a sequence of measurements for estimating $\delta\mathbf{x}_0$ that are collected over the canonical time interval T (one-day) and over a more extended time interval $pT$ ($p$-days), where these intervals are short relative to the orbital periods involved in the problem. This assumption on $pT$ facilitates an asymptotic analysis of the measurement scenario to obtain analytic bounds on the scenario's information content and associated position uncertainties. We seek the partials of the pointing angle $\gamma_j$ to the asteroid $j$ at time $t_j$ with respect to the initial state vector $\mathbf{x}_0$ at time $t_0$. The partials vector can be obtained using the following chain rule expression applied to Eq. (24)

$$\mathbf{h}_\gamma\left(t_j\right) = \frac{\partial \gamma_j}{\partial \hat{\mathbf{n}}^A}\frac{\partial \hat{\mathbf{n}}^A}{\partial \mathbf{r}_j}\frac{\partial \mathbf{r}_j}{\partial \mathbf{x}_0} = \frac{\partial \gamma_j}{\partial \hat{\mathbf{n}}^A}\frac{\partial \hat{\mathbf{n}}^A}{\partial \mathbf{r}_j}\left[\mathbf{\Phi}_{\mathbf{rr}}\left(t_j, t_0\right) \vdots \mathbf{\Phi}_{\mathbf{r\Delta r}}\left(t_j, t_0\right)\right] \tag{29}$$

From the identity $\hat{\mathbf{z}}_C = \hat{\mathbf{n}}_A$, we have that $\gamma_j = \tan^{-1}\left(\left(\hat{\mathrm{n}}_y\right)_{A_j} / \left(\hat{\mathrm{n}}_x\right)_{A_j}\right)$, which can be used to determine the following partials vector

$$\frac{\partial \gamma_j}{\partial \hat{\mathbf{n}}_{A_j}} = \left[\sin\gamma_j, -\cos\gamma_j\right] \tag{30}$$

From the definition for $\hat{\mathbf{n}}_{A_j}$ in Eq. (23) and illustrated in Fig. 3, the partials with respect to the position vector $\mathbf{r}_j \equiv \mathbf{r}(t_j)$ at time $t_j$ can be found as follows

$$\frac{\partial \hat{\mathbf{n}}_{A_j}}{\partial \mathbf{r}_j} = \frac{\partial}{\partial \mathbf{r}_j}\left(\frac{\delta \mathbf{r}_{A_j}}{\left\|\delta \mathbf{r}_{A_j}\right\|}\right) = \frac{1}{\delta \mathrm{r}_{A_j}}\left(\mathbf{I} - \hat{\mathbf{n}}_{A_j} \otimes \hat{\mathbf{n}}_{A_j}\right) = \frac{1}{\delta \mathrm{r}_{A_j}}\begin{bmatrix} \sin^2\gamma_j & -\cos\gamma_j \sin\gamma_j \\ -\cos\gamma_j\sin\gamma_j & \cos^2\gamma_j \end{bmatrix} \tag{31}$$

with $\delta \mathbf{r}_{A_j} \equiv \left\|\delta \mathbf{r}_{A_j}\right\|$ and the outer product $\otimes$ symbol (rather than the vector transpose) has been utilized to make the ensuing equations more readable. Multiplying Eq. (30) with Eq. (31) leads to the following simple result

$$\frac{\partial \gamma_j}{\partial \hat{\mathbf{n}}_{A_j}}\frac{\partial \hat{\mathbf{n}}_{A_j}}{\partial \mathbf{r}_j} = \frac{1}{\delta \mathrm{r}_{A_j}}\left[-\sin\gamma_j, \cos\gamma_j\right] \tag{32}$$

Completing the matrix products in Eq. (29) using Eq. (32) and the appropriate submatrices of $\mathbf{\Phi}^{(1)}(t_j, t_0)$ in Eq. (14) yields the following optical measurement sensitivity gradient

$$\mathbf{h}_{\gamma}\left(t_j\right) = \frac{1}{\delta \mathrm{r}_{A_j}}\left[-\sin\gamma_j,\ \ \cos\gamma_j, -\frac{t_j}{T}\sin\gamma_j, \frac{t_j}{T}\cos\gamma_j\right] + \mathbf{O}\left(\varepsilon^2\right) \tag{33}$$

From Eq. (23), we can derive the following asymptotic expansions that relate the distance between the spacecraft and asteroid $j$ and the pointing angle $\gamma_j$ to the other astrodynamic quantities of interest

$$\begin{aligned}\cos\gamma_j &= \frac{a_{A_j}\cos\alpha_j - a\ \cos\ \theta}{\delta \mathrm{r}_{A_j}} = \frac{a_{A_j}\cos\alpha_j - a}{\delta \mathrm{r}_{A_j}} + O\left(\varepsilon^2\right)\\ \sin\gamma_j &= \frac{a_{A_j}\sin\alpha_j - a\ \sin\ \theta}{\delta \mathrm{r}_{A_j}} = \frac{a_{A_j}\sin\alpha_j - \varepsilon a n t}{\delta \mathrm{r}_{A_j}} + O\left(\varepsilon^2\right)\end{aligned} \tag{34}$$

An asymptotic expansion of $\mathbf{h}_{\gamma_j}\left(t_j\right)$ yields

$$\begin{aligned}\delta \mathrm{r}_{A_j} \equiv \left\|\delta \mathbf{r}_{A_j}\right\| &= \sqrt{\left(\mathbf{r}_{A_j} - \mathbf{r}\right)\bullet\left(\mathbf{r}_{A_j} - \mathbf{r}\right)} = \delta r^0_{A_j} + \varepsilon \delta r^1_{A_j} + O\left(\varepsilon^2\right)\\ &= \delta r^0_{A_j}\left(1 - \varepsilon\frac{a a_{A_j}}{\left(\delta r^0_{A_j}\right)^2} n t_j \sin\alpha_j\right) + O\left(\varepsilon^2\right)\end{aligned} \tag{35}$$

where we have

$$\delta r^0_{A_j} \equiv \sqrt{a^2 + a^2_{A_j} - 2 a a_{A_j}\cos\alpha_j} \tag{36}$$

Substituting Eq. (35) into the measurement sensitivity vector $\sigma_{A_j}$ and expanding to second order yields the following form for the gradient

$$\mathbf{h}_{\gamma}(t_j) = \mathbf{h}^0_{\gamma}(t_j) + \varepsilon \mathbf{h}^1_{\gamma}(t_j) + \mathbf{O}(\varepsilon^2) \tag{37}$$

with

$$\mathbf{h}^0_{\gamma_j}\left(t_j\right) = \frac{1}{\left(\delta r^0_{A_j}\right)^2}\left[-a_{A_j}\sin\alpha_j, \left(a_{A_j}\cos\alpha_j - a\right), -\frac{t_j}{T} a_{A_j}\sin\alpha_j, \frac{t_j}{T}\left(a_{A_j}\cos\alpha_j - a\right)\right] \tag{38}$$

and

$$\mathbf{h}^1_{\gamma_j}\left(t_j\right) = \frac{a}{\left(\delta r^0_{A_j}\right)^4}\begin{bmatrix} n t_j\left(a^2 - 2 a a_{Aj}\cos\alpha_j + a^2_{A_j}\cos 2\alpha_j\right)\\ 2 a_{Aj} n t_j\left(a_{Aj}\cos\alpha_j - a\right)\sin\alpha_j\\ n\frac{t_j^2}{T}\left(a^2 - 2 a a_{Aj}\cos\alpha_j + a^2_{A_j}\cos 2\alpha_j\right)\\ 2 a_{Aj} n\frac{t_j^2}{T}\left(a_{Aj}\cos\alpha_j - a\right)\sin\alpha_j\end{bmatrix}^{\mathrm{T}} \tag{39}$$

Broschart [17] examined kinematic positioning using camera images of the main belt asteroids and, as part of this work, surveyed the distribution and characteristics of these asteroids for use in navigation. Some particular features that are noteworthy include:

1. There are over 50 K known and mapped bright main belt asteroids with magnitudes (M) < 14.9; hence, bright enough to image using navigation grade cameras and use as optical 'beacons' for navigation in the inner solar system,
2. These asteroids are between 2 and 4 AU from the Sun and have a typical position uncertainty $\delta\alpha_j = \left\|\delta \mathbf{r}_{A_j}\right\| / a_{A_j}$ of <100 km, with almost all <200 km. It is noteworthy to examine the sensitivity of the pointing angle measurement to errors in the location of the asteroid, the partial is given by

$$\delta\gamma_j = \frac{\partial\gamma_j}{\partial\alpha_j}\delta\alpha_j = \left[\frac{a_{A_j}^2 - a a_{A_j}\cos\left(\alpha_j - nt\right)}{a_{A_j}^2 - 2 a a_{A_j}\cos\left(\alpha_j - nt\right) + a^2}\right]\frac{\left\|\delta \mathbf{r}_{A_j}\right\|}{a_{A_j}} < \left|\frac{a_{A_j}}{a_{A_j} - a}\right|\frac{\left\|\delta \mathbf{r}_{A_j}\right\|}{a_{A_j}} \tag{40}$$

In the filtering problem, errors of this type (difficult to observe with short arcs of data) would ordinarily be considered; however, considering location errors ($\delta\alpha_j$) would overly complicate the analytic expressions for this analysis. A more straightforward approach is to augment the measurement noise with the location error effects as follows

$$v_j \longmapsto \left|\frac{a_{A_j}}{a_{A_j} - a}\right|\delta\alpha_j + v_j \tag{41}$$

This approach will ultimately yield optimistic results since $\sigma_{\alpha_j} = \sigma_{A_j}/a_{A_j}$ is being treated as a zero-mean error with uncertainty $\sigma_{\alpha_j}$, which is not the case (consider analysis treats $\delta\alpha_j$ as an unknown, unestimated bias with the same uncertainty). However, this approach does provide a lower bound that accounts for the location errors and produces the following result for the Mars case

$$Mars\ distances\ (a = 1.5\ AU) : \left|\frac{a_{\alpha_j}}{a_{\alpha_j} - a}\right|\frac{\sigma_{A_j}}{a_{\alpha_j}} < 0.62\ \mu\text{rad} \tag{42}$$

For navigating in the outer solar system, the main belt asteroids cannot be imaged because they will be within the minimum allowable sun-spacecraft-asteroid angle at these distances and the Sun will appear too bright in the image. Rather, we can consider using the centaurs, scattered disk objects (SDOs), and the Kuiper belt objects (KBOs) as 'beacons.' There are many thousands of these objects with more being continuously discovered and cataloged. The absolute brightness of the known objects between 5 AU out to 50 AU are shown as a function of their orbital semi-major axes in Fig. 5 with the best line-of-fit plotted as the dotted line. The associated apparent brightness (that scales with the distance between the object and observer) for these outer solar system objects will be dim and difficult to image. However, with improvements in camera sensitivities and use of image stacking, coadding, and filtering to improve the signal; it is conceivable that dim objects with apparent magnitudes of up to 26 could be imaged. Indeed, these techniques were recently applied to New Horizons' flyby of a KBO to improve the objects signal in a stack of images [30]. Absolute magnitudes can be converted to apparent magnitudes using the methodology adopted by the International Astronomical Union (Eq. (3) as documented in [17]). For the line of fit shown in Fig. 5, the maximum (dimmest) and minimum (brightest) apparent magnitude $M$ of the objects as a function of spacecraft and object semi-major axes is shown in Fig. 6 for a range of spacecraft semi-major axes $a$ and object semi-major axes $a_A$. Noteworthy, is that for values of $a_A > 43$ the apparent magnitude is $M < 26$; hence, using real-time advanced signal processing coupled with a well surveyed catalog these objects could be used for absolute navigation – the assumption we make for the current investigation.

As with the main belt asteroids for inner solar system navigation, the impact of the outer solar system object ephemeris errors needs to be examined. ESA's Gaia mission, launched in 2013, has been optically observing the solar system and the Milky way to catalog stars and solar system objects. In Gaia Data Release 2 [31] 14,099 solar system objects were cataloged including some outer planet solar system objects and their orbit estimates. Projections were made on the improvement in ephemeris errors for these objects for an extended Gaia mission that showed improvement in orbit knowledge relative to the current results by orders of magnitude. In particular, extrapolating ephemeris uncertainty estimates that could be possible in the future for the KBOs at $a_A$ ~ 40 AU yields $\sigma_A$ ~ 0.00013 AU. Using this estimate for the Neptune case, produces the following upper bound to the ephemeris error's contribution to the angular error

$$\text{Neptune distances } (a = 30 \text{ AU}) : \left| \frac{a_{\alpha_j}}{a_{\alpha_j} - a} \right| \frac{\sigma_{A_j}}{a_{\alpha_j}} \sim 13\ \mu\text{rad} \tag{43}$$

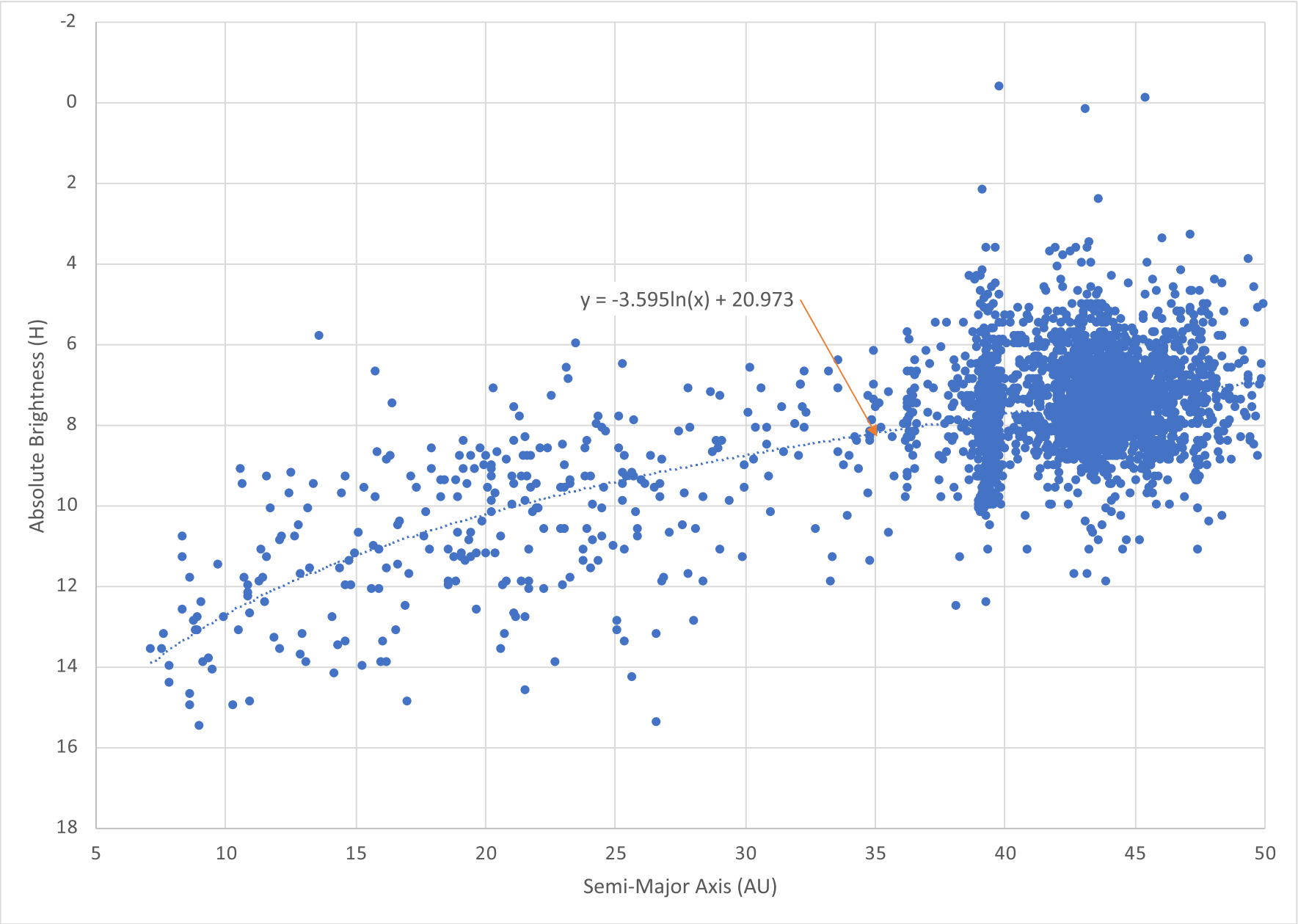

**Fig. 5** Absolute Brightness of the Centaurs, SDOs, & KBOs with Heliocentric Semi-major axes <50 AU. [29]

**Fig. 6** Maximum and minimum apparent magnitudes M of outer solar system objects as a function of spacecraft semi-major axis a and object semi-major axis $a_A$ where the absolute brightness of the objects conforms to the line of fit in Fig. 5

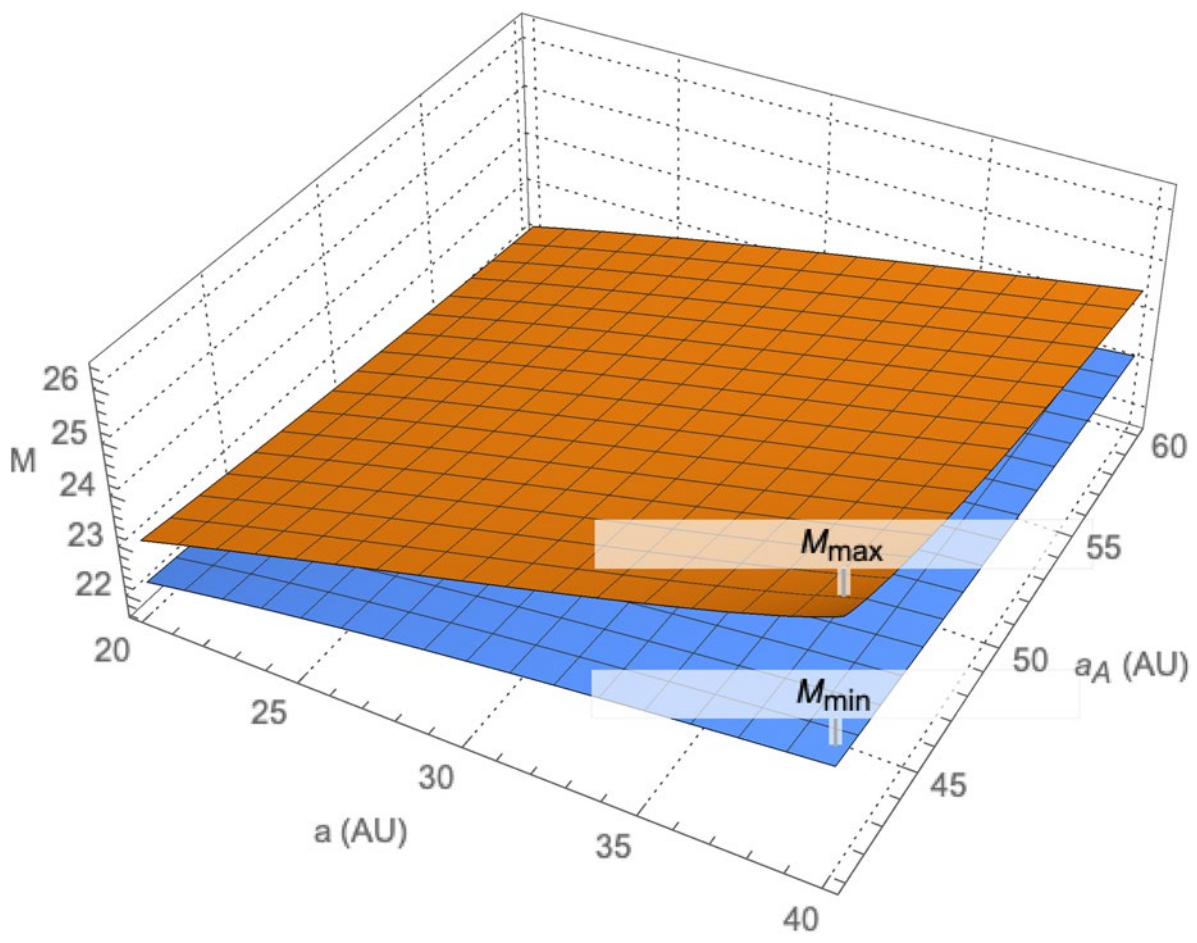

Representative navigation cameras examined by Broschart [17] can be lumped into three categories consisting of low-end, mid-range, and high-end camera features as noted in Table 1. FOV is the camera field-of-view, Θ is field of view of one camera pixel (called the IFOV), $M_{\max}$ is the maximum apparent magnitude of the asteroid that can be detected, $\alpha_{\min}$ is the minimum sun-spacecraft-asteroid angle allowable. Finally, the camera pixel array sample measurement uncertainty $\sigma_s$ selected for each camera is 0.25; a conservative bound factoring errors from center finding, camera calibrations, and camera pointing.

In the present analysis, a high-end camera that employs real time image co-adding is selected for comparison with IFOV of $\Theta = 10$(μrad). The overall sample uncertainty (in pixels) that combines $\sigma_s$ and $\sigma_{\alpha_j}$ yields a rescaled value $\overline{\sigma}_s$, defined as

$$\overline{\sigma}_s \equiv \sqrt{\sigma_s^2 + \frac{1}{\Theta^2}\left|\frac{a_{A_j}}{a_{A_j} - a}\right|^2 \left(\frac{\sigma_{\alpha_j}}{a_{A_j}}\right)^2} \tag{44}$$

**Table 1** Optical Camera Properties

| Name | FOV (deg) | Θ(μrad) | $M_{\max}$ | $\alpha_{min}$(deg) | $\sigma_s$(pixel) |
|---|---|---|---|---|---|
| Low-end | 26.9 | 128 | 9.5 | 30 | 0.25 |
| Mid-level | 7.0 | 60 | 10.5 | 30 | 0.25 |
| High-end | 0.6 | 10.0 | 13.5<br>25 (with real time image co-adding) | 30 | 0.25 |

For a spacecraft at Mars distances with pixel noise of strength $\sigma_s = 0.25$, imaging main-belt asteroids yields $\overline{\sigma}_s \cong 0.26$, hence ephemeris errors have little impact. While at Neptune distances, imaging KBOs produces $\overline{\sigma}_s \cong 1.3$, a relatively significant effect that needs to be taken into account by including the asteroid orbit in the estimation state along with the spacecraft. However, an analytical analysis would quickly become intractable if these states were added (later for the pulsar analysis we do examine a simplified consider analysis), for a lower bound estimate of this effect increasing the measurement noise to $\overline{\sigma}_s \cong 1.3$ is sufficient.

Given that there is, effectively, a dense set of visible known asteroids in the main-belt with well-known ephemerides (and in the future, outer solar system objects) that can be utilized as optical 'beacons', we make a simplifying assumption that there is an available, known asteroid at any selected value for $\alpha$. Furthermore, we also assume the camera is gimballed and is a capable of a full 360° scan over the course of a one-day tracking period $T$. Over $p$-days, the camera will make $p$ full 360° scans.[2] Using these assumptions coupled with a uniform imaging interval $h_\gamma$ yields a hypothetical tracking scenario relationship between consecutive optical measurements over the set of asteroids that spans the full 360° azimuthal range of the camera gimbal as follows

$$\alpha_j = \alpha_{j-1} + \Delta\alpha = \alpha_{j-1} + \frac{2\pi}{n_\gamma} \rightarrow \alpha_j + \frac{2\pi}{n_\gamma} j = \alpha_j + 2\pi\frac{t_j}{T} \quad (\text{full } 360^\circ \text{scan}) \tag{45}$$

where the times between measurements conform to $t_j = jh_\gamma$ and $T = n_\gamma h_\gamma$. This observation schedule repeats on consecutive days, hence over a $p$-day observation schedule a total of $pn_\gamma$ data points would be collected. Since there is no effective integration time for taking an image (shutter speeds are fractions of a second), we are able to approximate the start of the imaging sequence at $t_0$ with the first image taken of an asteroid located at $\alpha_0$. The next asteroid at $\alpha_1$ is imaged at $t_1 = t_0 + h_\gamma$, in accordance with Eq. (45). Our analysis is facilitated by taking a limit ($h_\gamma \ll T$) and examining the continuous case, this allows us to replace the discrete observation scenario in Eq. (45) with the following continuous scenario

[2] Note that a full 360° would include orienting the camera towards the Sun for a fraction of the scan. All the data types, optical, pulsar TOA, and radio tracking, typically have some form of 'keep-out' region that avoids data collection along paths that transit near the Sun in actual operations. For the current qualitative analysis in Part 1 of this work, accounting for the limits imposed by the keep-out zones would overly complicate the analysis while not significantly affecting the results or conclusions. Hence, they will be ignored for all the data types in Part 1; however, in Part 2 for the high-fidelity analysis they are explicitly included and accounted for to ensure accurate quantitative results.

$$\alpha(t) = \alpha_0 + 2\pi\frac{t}{T}, h_\gamma \ll T \tag{46}$$

Alternate observation scenarios may yield a more optimal sequence and could be selected rather than Eq. (46); however, a full scan across the solar system is sufficient to provide a well-observed problem that can be readily compared to the radiometric and pulsar tracking cases. We make one further approximation for the continuous case and replace $a_j$ for individual asteroids being imaged and replace them with the average (and constant) value for the asteroid distance from the Sun, for the main-belt asteroids we use ~2.7 AU and for the KBOs ~40 AU. Normalizing this average distance using the spacecraft heliocentric distance leads to the following definition

$$\left\langle \overline{a}_A \right\rangle \equiv \frac{\langle a_A \rangle}{a} = \frac{1}{a}\left( \frac{1}{n_j}\sum_j a_{A_j} \right) \tag{47}$$

We will investigate the information content of a full scan spread over the multi-day observation period $pT$ and then the construction of the associated estimation uncertainties by inverting the information matrix. To begin, the information content in a *single* optical measurement at time $t_j$ can be ascertained by forming the Fisher information matrix [32] defined for an optical measurement using

$$\mathbf{I}_{\gamma_j}\left(t_j\right) \equiv \frac{1}{\Theta^2\overline{\sigma}_s^2}\mathbf{H}_{\gamma_j}\left(t_j\right) = \frac{1}{\Theta^2\overline{\sigma}_s^2}\mathbf{h}_{\gamma_j}\left(t_j\right) \otimes \mathbf{h}_{\gamma_j}\left(t_j\right) \tag{48}$$

To determine the information content in the aggregate of measurements collected over the scan interval, we utilize a limit process and integration first introduced by Hamilton and Melbourne [33] and then extended by Curkendall and McReynolds [34]. The information matrix for a pass of data collected in the interval $pT$ is the sum of the matrices for each data point at time $t_j$ with $t_j \in pT$ and can be expressed as

$$\mathbf{I}_\gamma^\Sigma \equiv \frac{1}{\Theta^2\overline{\sigma}_s^2}\sum_{j=0}^{pn_y}\mathbf{h}_\gamma\left(t_j\right) \otimes \mathbf{h}_\gamma\left(t_j\right) \tag{49}$$

where the summation is in time and covers the set of asteroids that conform to the observation schedule identified in Eq. (45). Considering that the interval between images $h_\gamma$ is small relative to the observation period $T$, a useful approximation is to replace the sum over the individual measurements with an integration via taking the limit of $h_\gamma$ as it approaches zero to obtain the following formal result

$$
\begin{aligned}
\mathbf{I}_{\gamma}^{\Sigma} &\equiv \sum_{j=1}^{pn_y} \frac{\mathbf{h}_{\gamma_j}(t_j)\otimes\mathbf{h}_{\gamma_j}(t_j)}{\Theta^2\overline{\sigma}_s^2}\frac{h_\gamma}{h_\gamma} \\
&= \frac{1}{\Theta^2\overline{\sigma}_s^2 h_\gamma}\sum_{j=1}^{pn_y}\mathbf{h}_{\gamma_j}\left(t_j\right)\otimes\mathbf{h}_{\gamma_j}\left(t_j\right)h_\gamma \\
&\cong \frac{1}{\Theta^2\overline{\sigma}_s^2 h_\gamma}\lim_{h_\gamma\to 0}\sum_{j=1}^{pn_y}\mathbf{h}_{\gamma_j}\left(t_j\right)\otimes\mathbf{h}_{\gamma_j}\left(t_j\right)h_\gamma \\
&= \frac{pn_y}{\Theta^2\overline{\sigma}_s^2}\left(\frac{1}{p}\int_0^p\mathbf{h}_\gamma\left(\bar{t}\right)\otimes\mathbf{h}_\gamma\left(\bar{t}\right)d\bar{t}\right)
\end{aligned}
\tag{50}
$$

where, in the integral, we changed the integration variable to $\bar{t}\equiv t/T$ and use the fact that $n_\gamma = t/T$. The observation mapping identified in Eq. (46) is now utilized. Finally, note that the integration limit is now the number of days $p$ that observations are to be collected.

The aggregate information matrix $\mathbf{I}_{\gamma}^{\Sigma}$ defined in Eq. (50) can be block partitioned into the following $2\times 2$ submatrices

$$
\mathbf{I}_{\gamma}^{\Sigma} \equiv \begin{bmatrix} \left(\mathbf{I}_{\gamma}^{\Sigma}\right)_{\mathbf{rr}} & \left(\mathbf{I}_{\gamma}^{\Sigma}\right)_{\mathbf{r\Delta r}} \\ \left(\mathbf{I}_{\gamma}^{\Sigma}\right)_{\mathbf{r\Delta r}} & \left(\mathbf{I}_{\gamma}^{\Sigma}\right)_{\mathbf{\Delta r\Delta r}} \end{bmatrix}
\tag{51}
$$

For the optical tracking case, there is sufficient information in a first-order analysis[3] (i.e., retain only the zeroth-order terms) to obtain representative orbit determination uncertainties, using Eq. (38) and substituting into Eq. (50) yields the following integrals

$$
\left(\mathbf{I}_{\gamma}^{\Sigma}\right)_{\mathbf{rr}} = \frac{pn_\lambda}{\Theta^2\overline{\sigma}_s^2 a^2}\left(\frac{1}{p}\int_0^p \frac{1}{\left(1+\langle\overline{a}_A\rangle^2-2\langle\overline{a}_A\rangle\cos\alpha\right)^2}\begin{bmatrix}\langle\overline{a}_A\rangle\sin\alpha^2 & \langle\overline{a}_A\rangle\sin\alpha\left(1-\langle\overline{a}_A\rangle\cos\alpha\right) \\ \langle\overline{a}_A\rangle\sin\alpha\left(1-\langle\overline{a}_A\rangle\cos\alpha\right) & \left(1-\langle\overline{a}_A\rangle\cos\alpha\right)^2\end{bmatrix}d\bar{t}\right)+\mathbf{O}(\varepsilon)
\tag{52}
$$

$$
\left(\mathbf{I}_{\gamma}^{\Sigma}\right)_{\mathbf{r\Delta r}} = \left(\mathbf{I}_{\gamma}^{\Sigma}\right)_{\mathbf{\Delta rr}}\frac{pn_\lambda}{\Theta^2\overline{\sigma}_s^2 a^2}\left(\frac{1}{p}\int_0^p \frac{\bar{t}}{\left(1+\langle\overline{a}_A\rangle^2-2\langle\overline{a}_A\rangle\cos\alpha\right)^2}\begin{bmatrix}\langle\overline{a}_A\rangle\sin\alpha^2 & \langle\overline{a}_A\rangle\sin\alpha\left(1-\langle\overline{a}_A\rangle\cos\alpha\right) \\ \langle\overline{a}_A\rangle\sin\ \alpha\left(1-\langle\overline{a}_A\rangle\cos\alpha\right) & \left(1-\langle\overline{a}_A\rangle\cos\alpha\right)^2\end{bmatrix}d\bar{t}\right)+\mathbf{O}(\varepsilon)
\tag{53}
$$

$$
\left(\mathbf{I}_{\gamma}^{\Sigma}\right)_{\mathbf{\Delta r\Delta r}} = \frac{pn_\lambda}{\Theta^2\overline{\sigma}_s^2 a^2}\left(\frac{1}{p}\int_0^p \frac{\bar{t}^2}{\left(1+\langle\overline{a}_A\rangle^2-2\langle\overline{a}_A\rangle\cos\alpha\right)^2}\begin{bmatrix}\langle\overline{a}_A\rangle\sin\alpha^2 & \langle\overline{a}_A\rangle\sin\alpha\left(1-\langle\overline{a}_A\rangle\cos\alpha\right) \\ \langle\overline{a}_A\rangle\sin\ \alpha\left(1-\langle\overline{a}_A\rangle\cos\alpha\right) & \left(1-\langle\overline{a}_A\rangle\cos\alpha\right)^2\end{bmatrix}d\bar{t}\right)+\mathbf{O}(\varepsilon)
\tag{54}
$$

[3] For each of the data types, we will examine an asymptotic expansion of their associated information matrices with the objective to expand to the lowest order that achieves meaningful analytic insights into the measurements' information content. For optical navigation, this is a first order analysis (i.e., retain zeroth-order terms); for pulsar navigation, this is a second order analysis (retain up to first order); and, as will be seen for ranging, a numerical approach is required that retains high order terms (beyond 4th order).

where we made the expression for $\delta r^0_{A_j}$ explicit, replaced $a_{A_j}$ with $\langle \overline{a}_A \rangle$, and factored out the spacecraft semi-major axis $a$. In Eqs. (51)–(53), we have separated the coefficient and integral factors in the following manner:

1. The $1/a^2$ factor in the coefficient results from use of the normalized average distance to the asteroids $\langle \overline{a}_A \rangle$. This is used to change the scale of the measurement uncertainty from an angle to a distance. For instance, an image taken of an asteroid using the high-end camera in Table 1 would yield an equivalent distance uncertainty (i.e., the factor $\theta\overline{\sigma}_2 a$) of 563 km for a spacecraft at Mars distances when imaging main-belt asteroids, and 58,500 km at Neptune distances when imaging KBOs.
2. The number of observation days $p$ has been explicitly introduced in the coefficient and the integral has been normalized by $p$ (more on this next). The total number of measurements $pn_\gamma$ taken over the $p$-day interval now appears (recall that $n_\gamma$ is the number of pictures taken per day), and it proves convenient to introduce an aggregate measurement uncertainty for the collection of optical images taken and expressed in the form

$$\overline{\sigma}_\gamma\left(pn_\gamma\right) \equiv \frac{\Theta\overline{\sigma}_s a}{\sqrt{pn_\gamma}} \tag{55}$$

   The effective noise strength decreases according to $1/\sqrt{pn_\gamma}$ as the total number of images increases. This is a standard result for data reduction using Gaussian distributed random variables, and is a theme that will recur when analyzing the pulsar TOA and the ranging data. Unlike the pulsar TOA measurements that require significant integration periods or ranging that also requires integration, if the asteroids are bright enough, then the image integration time is a fraction of a second with the interval between different scenes limited only by gimbal speed. Indeed, imaging the same scene in a burst is a standard technique to reduce image noise (and in the case of the KBOs co-adding images to increase the signal). Selecting a short interval between images $h_\gamma$ yields a larger number of pictures, an image interval of 60 s is reasonable and results in a set of 1440 images per day. With this selection for the number of images per day, $\overline{\sigma}_\gamma(1400) = 14.8$km when imaging main-belt asteroids at Mars distances, and 1541.6 km when imaging KBOs at Neptune distances.
3. Dividing the integrals (rescaled with normalized time $\bar{t}$) by $p$ yields the average geometric information content produced by the observation sequence. Recall, that if the measurement technique where able to directly observe the state being estimated (i.e., $\delta\mathbf{x}(t_0)$ then the uncertainty in the estimate of state vector components would be given by Eq. (55) (assuming Gaussian random variables). However, the images are not direct measures of the state vector, rather the vector is observed via the measurement equation (given formally in Eq. (29)) with data collected over $p$-days. The geometric information content in the complete set of images is specified by the three average integrals given in Eqs. (51)–(53). Later we will relate these to the position dilution of precision (PDOP) that is frequently utilized in GNSS applications.

All of the integrals can be found analytically; however, with the exception of Eq. (52), their resulting forms are complex and do not reveal significant insights. Using the tracking scenario specified by Eq. (46), an analytical expression for the integral in Eq. (52) is obtained explicitly and the integrals in Eq. (53) and (53) will be evaluated numerically. The solution for Eq. (52) can be found using standard Weirstrass tangent half-angle substitutions, but the result has spurious discontinuities. These can be removed via a rectification transformation developed by Jeffery [35] to yield a continuous form with the result

$$\left(\mathbf{I}_{\gamma}^{\Sigma}\right)_{\mathbf{rr}} \cong \begin{cases} \frac{pn_{\gamma}}{\Theta^2\overline{\sigma}_s^2 a^2}\begin{bmatrix} \frac{1}{2\left(\left\langle\overline{a}_A\right\rangle/a\right)^2-2} & 0 \\ 0 & \frac{1}{2\left(\left\langle\overline{a}_A\right\rangle/a\right)^2-2} \end{bmatrix} + \mathbf{O}(\varepsilon), a < \langle a_A \rangle \\ \frac{pn_{\gamma}}{\Theta^2\overline{\sigma}_s^2 a^2}\begin{bmatrix} \frac{1}{2\left(a/\left\langle\overline{a}_A\right\rangle\right)^2-2} & 0 \\ 0 & \frac{2\left(a/\left\langle\overline{a}_A\right\rangle\right)^2-1}{2\left(a/\left\langle\overline{a}_A\right\rangle\right)^2-2} \end{bmatrix} + \mathbf{O}(\varepsilon), a > \langle a_A \rangle \end{cases} . \tag{56}$$

The top row result applies to the scenarios examined in this research where the imaged asteroids are farther from the Sun than the spacecraft being navigated (i.e. main belt asteroids at Mars distances and KBOs at Neptune distances). The bottom row would apply to imaging asteroids that are closer to the Sun than the spacecraft, as might be the case for a spacecraft enroute to Jupiter imaging main belt asteroids.

Later in the high-fidelity analysis, our focus will be on a quantitative assessment of the full navigation problem – trajectory uncertainties/errors, delivery uncertainties/errors, and associated maneuvers; however, our current qualitative analysis focus is only on the position component uncertainties, that is $\left(\mathbf{P}_{\gamma}^{\Sigma}\right)_{\mathbf{rr}}$ and not the full $4\times4$ inverse $\mathbf{P}_{\gamma}^{\Sigma}$. However, a simple inversion of just the $2\times2$ position information matrix $\left(\mathbf{I}_{\gamma}^{\Sigma}\right)_{\mathbf{rr}}$ would yield optimistic results, an accurate position uncertainty estimates require use of all the $2\times2$ information sub-matrices (i.e., velocity information and its correlation with the position information) contained in Eq. (51). Obtaining $\left(\mathbf{P}_{\gamma}^{\Sigma}\right)_{\mathbf{rr}}$, can simplified by taking advantage of the block diagonal structure in Eq. (51) using Schur complements. As given in Ref. [36], a partitioned matrix $\mathbf{M}$ with the square block structure

$$\mathbf{M} = \begin{bmatrix} \mathbf{A} & \mathbf{B} \\ \mathbf{C} & \mathbf{D} \end{bmatrix} \tag{57}$$

can be inverted using

$$\mathbf{M^{-1}} = \begin{bmatrix} (\mathbf{M/D})^{-1} & -(\mathbf{M/D})^{-1}\mathbf{BD}^{-1} \\ -\mathbf{D}^{-1}\mathbf{C}(\mathbf{M/D})^{-1} & \mathbf{D}^{-1} + \mathbf{D}^{-1}\mathbf{C}(\mathbf{M/D})^{-1}\mathbf{BD}^{-1} \end{bmatrix} \tag{58}$$

with $\mathbf{M/D} \equiv \mathbf{A} - \mathbf{CD}^{-1}\mathbf{B}$ defined as the Schur complement of the nonsingular matrix $\mathbf{D}$ in the partitioned matrix $\mathbf{M}$. In our problem, this can be applied to Eq. (51) to find the uncertainty in the estimate of the *initial* position $\left(\mathbf{P}_{\gamma}^{\Sigma}\right)_{\mathbf{rr}}$ via inverting the Schur complement $\mathbf{I}_{\gamma}^{\Sigma}/\left(\mathbf{I}_{\gamma}^{\Sigma}\right)_{\mathbf{\Delta r \Delta r}}$ as follows

$$\mathbf{P}_{\mathbf{rr}}^{\gamma} = \left(\mathbf{I}_{\gamma}^{\Sigma}/(\mathbf{I}_{\gamma}{}^{\Sigma})_{\mathbf{\Delta r \Delta r}}\right)^{-1} = \left(\left(\mathbf{I}_{\gamma}^{\Sigma}\right)_{\mathbf{rr}} - \left(\mathbf{I}_{\gamma}^{\Sigma}\right)_{\mathbf{r\Delta r}}\left(\left(\mathbf{I}_{\gamma}^{\Sigma}\right)_{\mathbf{\Delta r \Delta r}}\right)^{-1}\left(\mathbf{I}_{\gamma}^{\Sigma}\right)_{\mathbf{r\Delta r}}\right)^{-1}. \tag{59}$$

Where we have defined $\left(\mathbf{P}_{\gamma}^{\Sigma}\right)_{\mathbf{rr}}$ to simplify notation. Since the matrices in Eq. (59) are $2\times2$, they are inverted most easily using Cramer's rule. Examination of the explicit expressions for the information matrices given in Eqs. (51)–(53) and the observations that followed – including the overall measurement strength given by Eq. (55) and the geometric information content – we can make the following definition

$$\mathbf{G}_{\mathbf{rr}}^{\gamma} \equiv \frac{1}{\overline{\sigma}_{\gamma}\left(pn_{\gamma}\right)}\mathbf{P}_{\mathbf{rr}}^{\gamma}. \tag{60}$$

We call $\mathbf{G}_{\mathbf{rr}}^{\gamma}$ the *average position dilution of precision matrix*. It contains the average over the observation period of all the geometric information in the optical measurement scenario for estimating the position vector $\delta\hat{\mathbf{r}}\left(t_0\right)$ and scales the overall measurement uncertainty up or down for obtaining this estimate. We can define the associated *position dilution of precision metric* (PDOP), related to the PDOP metric often used in GNSS applications, as follows

$$PDOP_{y} \equiv \sqrt{\mathrm{tr}(\mathrm{G}_{\mathrm{rr}}^{\gamma})} \tag{61}$$

where the subscript $\gamma$ identifies this as the PDOP for optical navigation. Hence, the components of $\mathbf{G}_{\mathbf{rr}}^{\gamma}$ are uncertainty scale factors that provide insight into the fundamental geometric sensitivities of the positioning problem and are independent of the particular camera noise, imaging density, and imaging period. Whereas, the noise uncertainty coefficient $\overline{\sigma}_{\gamma}\left(pn_{\gamma}\right)$ captures all the instrument errors as well as the Gaussian reduction effect due to increasing the number of these measurements. Indeed, if a theoretical measurement method could directly measure each individual component of an $n$-dimensional state vector (i.e., $\mathbf{h}_1 = [1,0,\ldots,0]$, $\mathbf{h}_2 = [0,1,\ldots,0],\ldots,\mathbf{h}_{\mathrm{n}} = [0,0,\ldots,1]$), then its average dilution of precision matrix would be identity. That is, if the following information matrix applied for $m$ observations of uncertainty $\sigma$, then the following covariance and PDOP would result

$$\mathbf{I} = \frac{m}{\sigma^2}\left(\frac{1}{m}\sum_{j=1}^{m}\sum_{i=1}^{m}\mathbf{h}_i \otimes \mathbf{h}_i\right) = \frac{m}{\sigma^2}\left(\frac{1}{m}\sum_{j=1}^{m}\mathbf{1}_{nxn}\right) \rightarrow \mathbf{P} = \frac{\sigma^2}{m}\mathbf{1}_{nxn}, \text{ and PDOP} \equiv \sqrt{\text{tr}(\mathbf{G})} = \sqrt{n} \tag{62}$$

where $\mathbf{1}_{nxn}$ is an $n \times n$ identity matrix. The best achievable PDOP would be $\sqrt{n}$. In the present case, we are focused on the 2-d position vector so the best achievable PDOP = $\sqrt{2}$, and the position uncertainties in each component would improve with each additional measurement as $\sigma/\sqrt{m}$. Returning to the orbit problem, using components from the average dilution of precision matrix $\mathbf{G}^{\gamma}_{\mathbf{rr}}$, the associated one-sigma position uncertainties are found using

$$\begin{aligned} \sigma^{\gamma}_{xx} &= \overline{\sigma}_{\gamma}\left(pn_{\gamma}\right)\sqrt{G^{\gamma}_{xx}} \text{ with } \mathbf{G}^{\gamma}_{xx} \equiv \left(\mathbf{G}^{\gamma}_{\mathbf{rr}}\right)_{11} \\ \sigma^{\gamma}_{xx} &= \overline{\sigma}_{\gamma}\left(pn_{\gamma}\right)\sqrt{G^{\gamma}_{xx}} \text{ with } \mathbf{G}^{\gamma}_{xx} \equiv \left(\mathbf{G}^{\gamma}_{\mathbf{rr}}\right)_{22} \end{aligned} \tag{63}$$

The roots of the diagonals of $\mathbf{G}^{\gamma}_{\mathbf{rr}}$ are, effectively, the geometric scale factors on the aggregate one-sigma measurement uncertainty for obtaining the resultant position uncertainties.

Other than the appearance of the spacecraft semi-major axis, Eq. (56) shows that the upper right position information matrix is independent of the geometry. On the other hand, the other matrices do depend on the geometric relationship between the right ascension of the initial asteroid imaged $\alpha_0$ and the initial velocity vector with the result that $\mathbf{G}^{\gamma}_{\mathbf{rr}}$ can be evaluated as a function of $\alpha_0$. Examining Fig. 1 and the initial assumptions, it can be ascertained that

$$\hat{\mathbf{r}}_{A_0} \bullet \hat{\mathbf{v}}_0 = \sin\alpha_0, \tag{64}$$

which provides a functional relationship between the relative orientation of the *initial* asteroid location and the initial velocity vector. Evaluating Eq. (60), the average dilution of precision matrix $\mathbf{G}^{\gamma}_{\mathbf{rr}}$, as a function of $\alpha_0$ for the Mars distance case at $a = 1.5$ AU with one-day of imaging (i.e. $p = 1$) yields $\left(\sqrt{G^{\gamma}_{xx}}, \sqrt{G^{\gamma}_{yy}}\right)$ (i.e., geometric scale factors on the aggregate measurement uncertainty) and the correlation between them as shown in Fig. 7. The overall PDOP for one day of imaging, seven days of imaging, and fourteen days of imaging are shown in Fig. 8.

For the Neptune distance case at $a = 30$ AU with one-day of imaging (i.e., $p = 1$) KBOs yields $\left(\sqrt{G^{\gamma}_{xx}}, \sqrt{G^{\gamma}_{yy}}\right)$ and associated correlation in Fig. 9. The associated PDOP for one-day, seven-days, and fourteen-days of imaging are shown in Fig. 10.

Examining the one-day imaging results in Fig. 7 for Mars case and Fig. 9 for the Neptune case yields the following bounds

$$\begin{aligned} 2.7 \le \left(\sqrt{G^{\gamma}_{xx}}, \sqrt{G^{\gamma}_{yy}}\right) \le 8.2, \quad a = 1.5 \text{ AU} \\ 1.5 \le \left(\sqrt{G^{\gamma}_{xx}}, \sqrt{G^{\gamma}_{yy}}\right) \le 6.7, \quad a = 30 \text{ AU} \end{aligned} \tag{65}$$

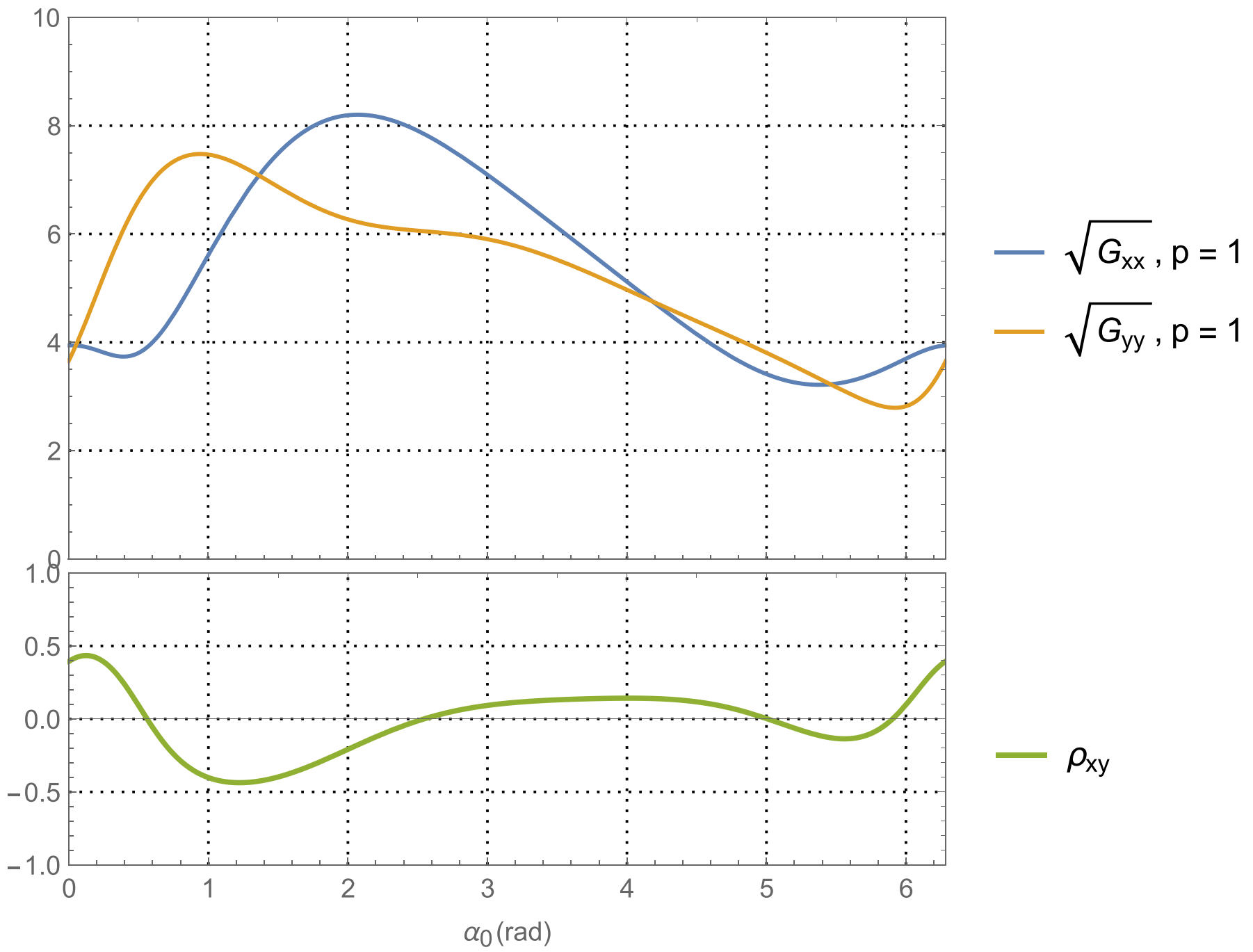

**Fig. 7** Position component uncertainty geometric scale factors for Mars distances (upper) and the correlation between components (lower) using optical imaging of the main belt asteroids and a full 360° scan over one-day

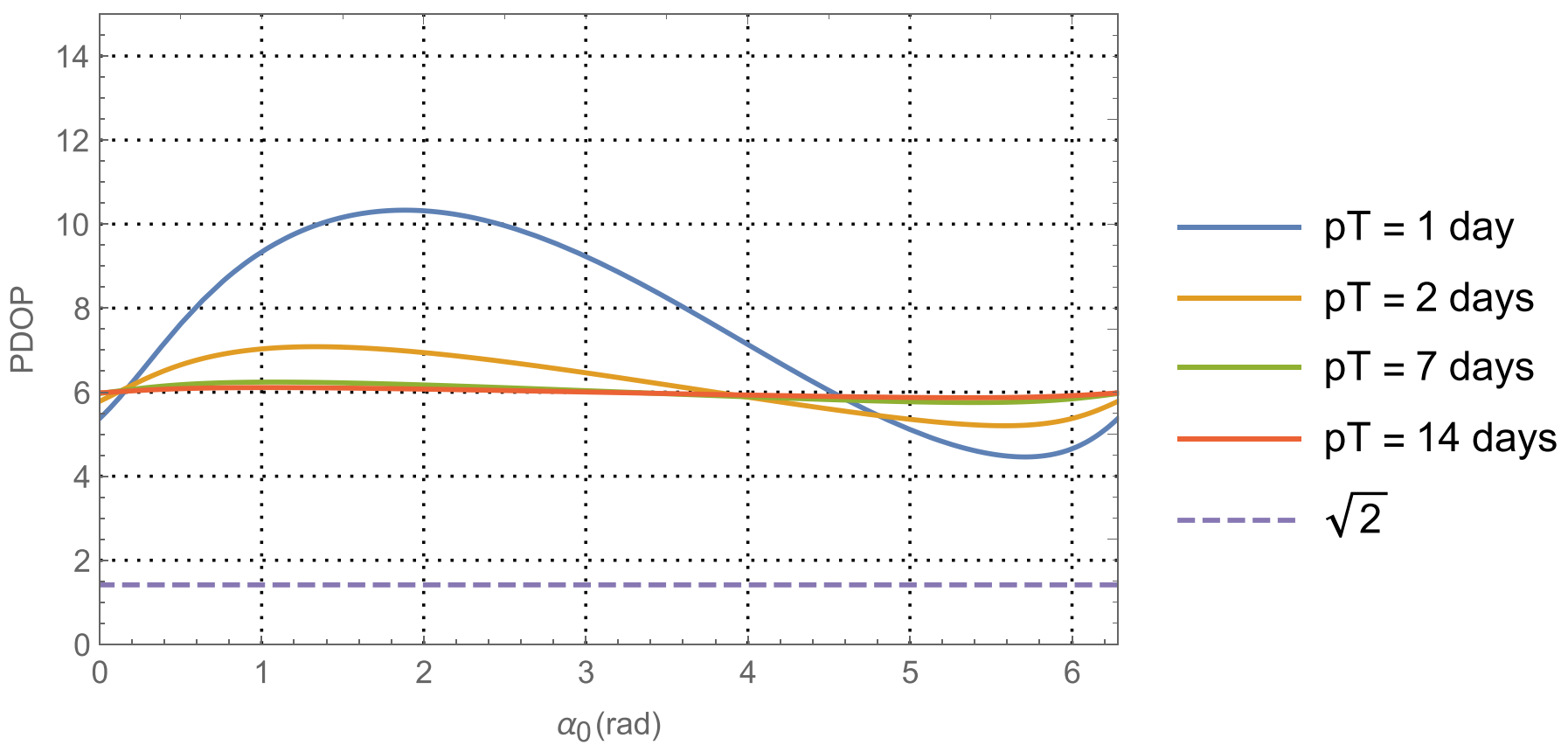

**Fig. 8** PDOP for Mars distances using one-day of imaging, seven days, and fourteen days. For reference, the theoretical PDOP obtained with direct observation of the position components is shown as the dashed line at $\sqrt{2}$

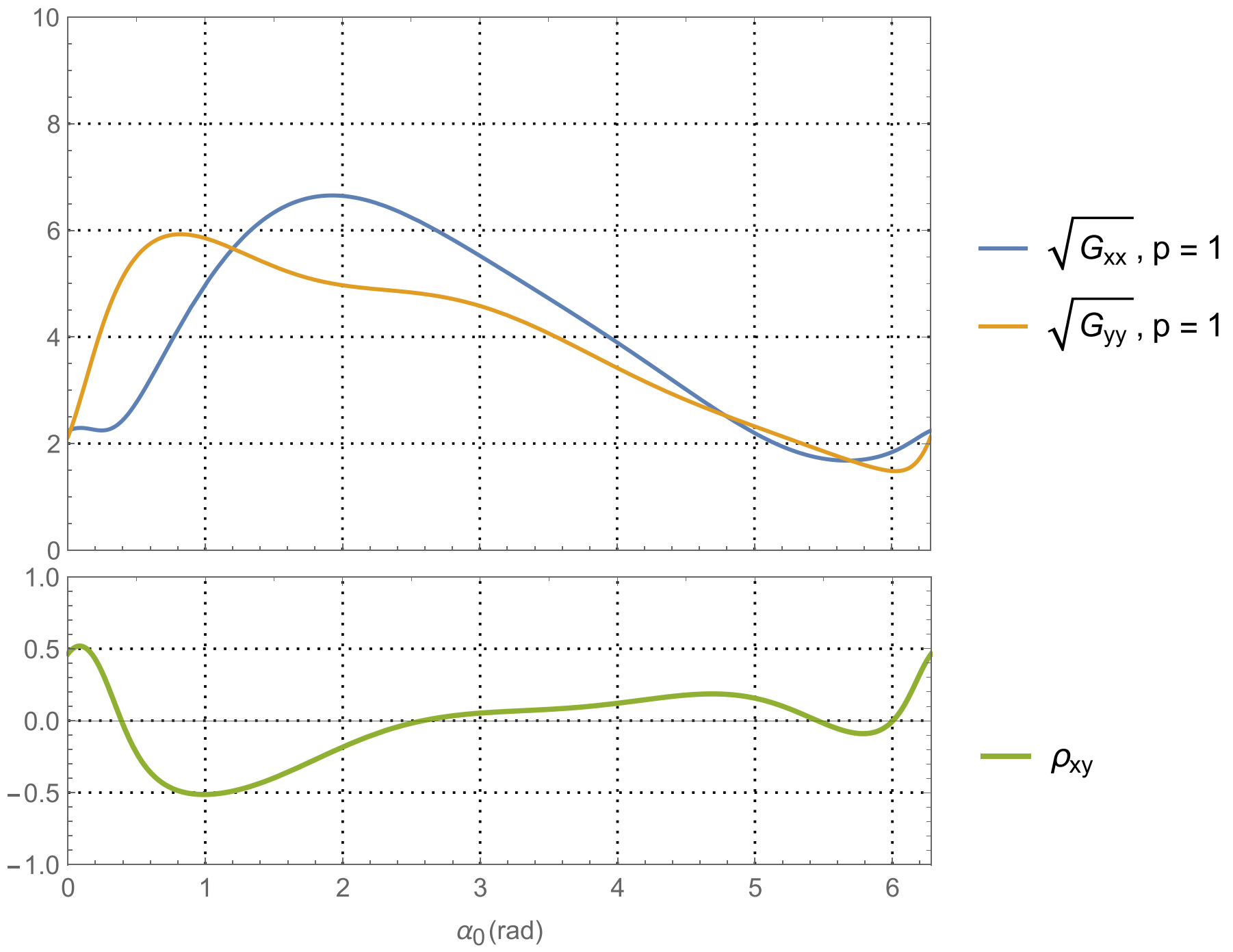

**Fig. 9** Position component uncertainty geometric scale factors for Neptune distances (upper) and the correlation between components (lower) using optical imaging of KBOs at an average heliocentric distance of 40 AU with a full 360° scan over one day

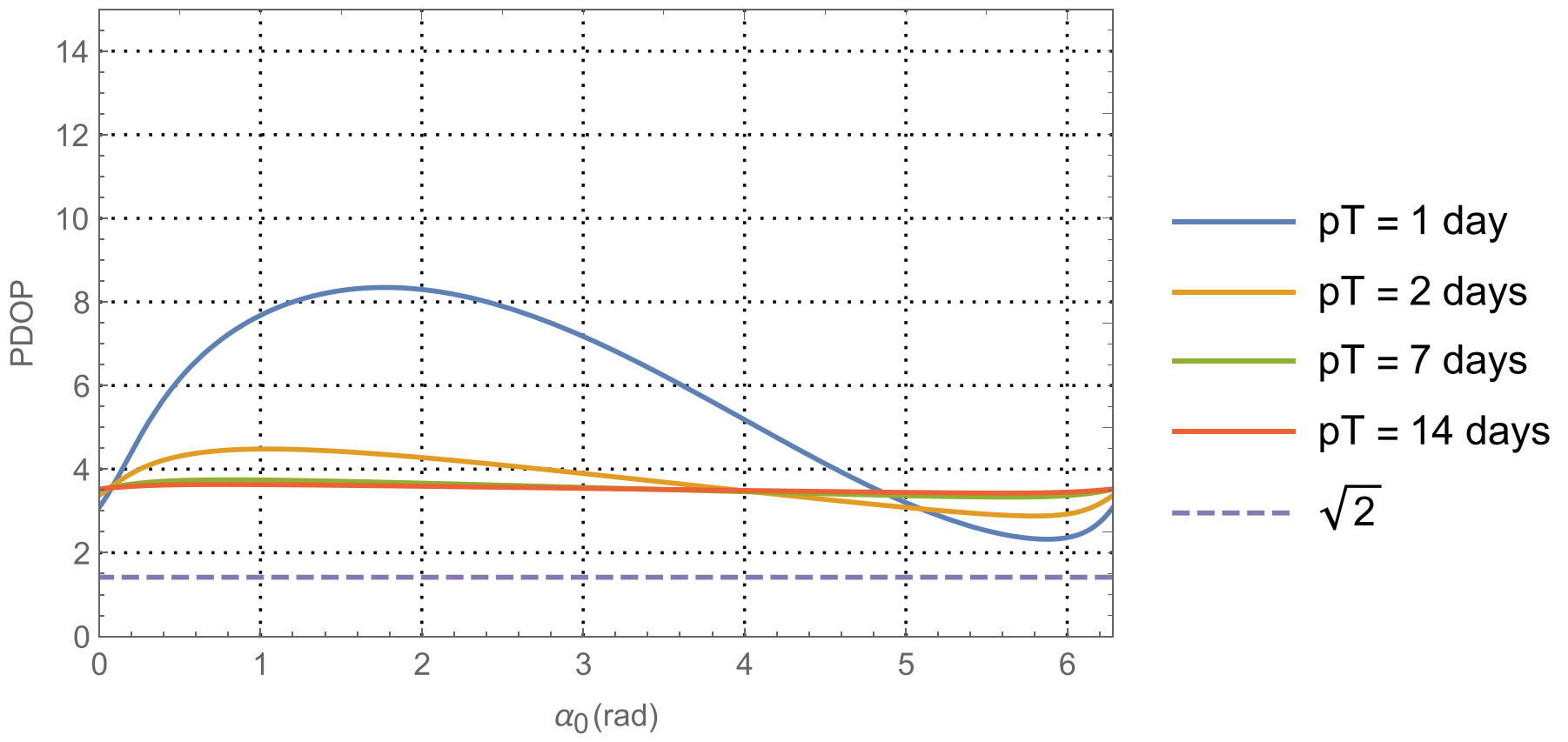

**Fig. 10** PDOP for Neptune distances using one-day of imaging, seven days, and fourteen days of KBOs (rather than main belt asteroids). For reference, the theoretical PDOP obtained with direct observation of the position components is shown as the dashed line at $\sqrt{2}$

Presumably the choice of gimbal orientation for the onboard camera is a free parameter, in which case an imaging campaign would select the initial orientation that gives the best positioning result. As seen in Fig. 7 for the Mars case, this occurs at $\alpha_0 \cong 5.5$ rad with $\sqrt{G^{\gamma}_{xx}} = \sqrt{G^{\gamma}_{yy}} \cong 3.2$, and at $\alpha_0 \cong 5.7$ rad with  in Fig. 9 for the Neptune case. Assuming use of the high-end camera with a 1-min imaging interval yields the following 1-sigma position component uncertainties

$$\text{Mars case (a = 1.5 AU)}: \quad \left(\sigma^{\gamma}_{xx}, \sigma^{\gamma}_{xx}\right) \cong (3.2 \times 14.8\ \text{km}, 3.2 \times 14.8\ \text{km}) \cong (47\text{km}, 47\text{km}) \tag{66}$$

and

$$\text{Neptune case } (a = 30\ \text{AU}): \quad \left(\sigma^{\gamma}_{xx}, \sigma^{\gamma}_{xx}\right) \cong (1.7 \times 1541.6\ \text{km}, 1.7 \times 1541.6\ \text{km}) \cong (2621\ \text{km}, 2621\ \text{km}) \tag{67}$$

Turning our attention to the PDOP results for the Mars case in Fig. 8 and the Neptune case in Fig. 10, the average ‘dilution of precision’ over the selected time span appears to asymptotically approach constant values that are independent of the initial asteroid angle. For Mars, this approaches a PDOP of ~6, and for Neptune a PDOP of ~3.5. While at the same time, the aggregate measurement uncertainty is reduced relative to the value at one day by a factor of $1/\sqrt{7} \cong 0.38$ (at 7-days) and $1/\sqrt{7} \cong 0.27$ (at 14-days). What these two observations illustrate is geometric scale factors improve with more imaging; however, after a few days the scale factors hit a limit at which point the solution uncertainty becomes information-content limited and geometry no longer is able to improve the uncertainties. While it is tempting to use the 7-day or 14-day factors to determine the component position uncertainties, the results would be overly optimistic as they do not include the necessary and realistic model effects that are needed to obtain accurate solution uncertainties on these time scales. This will be done with the higher fidelity simulations conducted and examined in Part 2 of this paper. Indeed, even the 1-day results are optimistic, but they do yield *representative order of magnitude* estimates that will allow us to make qualitative comparisons with the other measurement techniques.

## Pulsar TOA Information Content

We turn now to the pulsar time of arrival measurement, the linearized measurement equation for observing the $k^{\text{th}}$-pulsar's (and, in our 2-d problem, located at $\beta_k$) time-of-arrival (TOA) at time $t_k$ provides information that can be used to estimate $\delta\mathbf{x}_0$ and is found using

$$\delta(c\tau)_k \equiv \mathbf{h}_\tau\left(t_k\right)\delta\mathbf{x}_0 + v_{\tau_k} = \frac{\partial(c\tau)_k}{\partial\mathbf{x}_0}\delta\mathbf{x}_0 + v_{\tau_k} \tag{68}$$

The measurement sensitivity (or gradient) $\mathbf{h}_\tau(t_k)$ is obtained using Eq. (26), Eq. (7), and the following chain rule relationship

$$\mathbf{h}_\tau\left(t_k\right) = \frac{\partial(c\tau)_k}{\partial\mathbf{r}_k}\frac{\partial\mathbf{r}_k}{\partial\mathbf{x}_0} = \frac{\partial(c\tau)_k}{\partial\mathbf{x}_0}\left[\mathbf{\Phi}_{\mathbf{rr}}\left(t_k, t_0\right) \vdots \mathbf{\Phi}_{\mathbf{r\Delta r}}\left(t_k, t_0\right)\right] \tag{69}$$

The partial of an individual pulsar measurement with respect to position vector at time $t_k$ is found by differentiating Eq. (26) and takes the form

$$\frac{\partial (c\tau)_k}{\partial \mathbf{r}_0} = \left(\hat{\mathbf{n}}_k^p\right)^T \tag{70}$$

Because the partials in Eq. (70) are exact, and the zeroth-order term of the STM is actually good to second order, a pulsar TOA expansion that retains only leading terms yields a result that is good to second order. Using $\mathbf{\Phi}^{(1)}\left(t_k, t_0\right)$, Eq. (69) can be assembled into an approximate linearized measurement sensitivity vector as follows

$$\mathbf{h}_\tau\left(t_k\right) \equiv \left[\cos\beta_k, \sin\beta_k\right]\begin{bmatrix} 1 & 0 & \frac{t_k}{T} & 0 \\ 0 & 1 & 0 & \frac{t_k}{T} \end{bmatrix} + \mathbf{O}\left(\varepsilon^2\right) = \left[\cos\beta_k, \sin\beta_k, \frac{t_k}{T}\sin\beta_k, \frac{t_k}{T}\cos\beta_k\right] + \mathbf{O}\left(\varepsilon^2\right) \tag{71}$$

At each observation time $t_k$ a pulsar located at $\beta_k$ is observed that is usually different (but not necessarily) from the pulsar observed at time $t_{k-1}$. In this analysis, we assume that contiguous observations are uniformly spaced in time such that $t_k = kh_\tau$ where $h_\tau$ is the time interval between each observation and the integration time for each pulsar observation.

As demonstrated by NASA's SEXTANT experiment and documented by Ray [3], pulsar measurement precision improves with increasing $h_\tau$ and is proportional to $1/\sqrt{h_\tau}$ on the short time scales that are relevant for navigation. Furthermore, the measurement precision improves with increasing the area $A$ of the X-ray detector and is proportional to $1/\sqrt{A}$. The SEXTANT experiment obtained pulsar TOA measurement uncertainties that ranged from ~2 km to ~35 km (1σ) for one hour integration times. These uncertainty results depend primarily on the source pulsar stability and the specifics of the SEXTANT detector, which uses 56 individual X-ray telescopes for an effective aggregate collection area of 1800 cm$^2$. Given these values and relationship to $h_\tau$, the following empirical uncertainty model can be used for measuring the $k$th-pulsar's TOA

$$\sigma_{\tau_k} = \frac{S_{\tau_k}}{\sqrt{Ah_\tau}}. \tag{72}$$

For $h_\tau$ measured in hours and $A$ in centimeters, the pulsars selected by the SEXTANT experiment yielded $S_{\tau_k}$coefficients (converted to values of distance rather than time) that ranged in value from 82.5 km$\left(\text{cm}\sqrt{\text{hr}}\right)$ to 1443.1 km$\left(\text{cm}\sqrt{\text{hr}}\right)$. To expand on this, a pulsar catalog of likely 'navigation grade' pulsars has been aggregated from Refs. [3, 4, 37] and listed in Table 2. Those sources that were used in the SEXTANT flight experiment and characterized by Ray [3] are noted by a superscript $^{\text{S}}$ at the end of their name. Pulsar locations are from the SIMBAD Astronomical Database and presented in J2000 ecliptic coordinates and sorted on the right ascension of the ascending node (RA). The associated location uncertainty $\sigma_\beta$ values are from Shemar [38]. Finally, when available, their stability characteristics have been listed and converted into the $S_{\tau_k}$ figure of merit as specified in Eq. (72).

**Table 2** Pulsar Catalog ('name' S identifies those pulsars used in the SEXTANT experiment)

| Pulsar | RA (deg) | $RA_i - RA_{i-1}$ | Dec (deg) | $s_\tau \mathrm{km}\left(\mathrm{cm}\sqrt{\mathrm{hr}}\right)$ | $\sigma_\beta$ (mas) |
|---|---|---|---|---|---|
| PSR J0030+0451$^S$ | 8.91 | 20.10 | 1.45 | 431.2 | 21.35 |
| PSR J0218+4232$^S$ | 47.05 | 38.14 | 27.01 | 200.3 | 31.12 |
| PSR J0437–4715$^S$ | 50.47 | 3.42 | −67.87 | 697.1 | 0.05 |
| PSR B0531+21 (M1 - Crab Pulsar)$^S$ | 84.10 | 33.63 | −1.29 | 28.9 | 3.43 |
| PSR J0751+1807 | 116.33 | 32.24 | −2.81 | | 7.23 |
| PSR J1012+5307$^S$ | 133.36 | 17.03 | 38.76 | 1474.7 | 0.48 |
| PSR B0833–45 | 153.37 | 20.01 | −60.36 | 778.9 | 0.37 |
| PSR J1024–0719 | 160.73 | 7.36 | −16.04 | 2,791,358.2 | 0.67 |
| PSR B1055–52 | 195.77 | 35.04 | −52.39 | | 209.51 |
| PSR B1509-58$^S$ | 243.89 | 48.12 | −39.40 | 1672.3 | 1216.4 |
| PSR B1821–24 (J1824-2452A)$^S$ | 275.56 | 31.67 | −1.55 | 41.1 | 6.01 |
| PSR B0540–69 | 301.60 | 26.03 | −86.66 | 573.9 | 69.07 |
| PSR B1937+21$^S$ | 301.99 | 0.39 | 42.33 | 62.1 | 0.04 |
| PSR J2124–3358 | 312.74 | 10.75 | −17.82 | | 0.79 |
| PSR J2214+3000 | 348.81 | 36.07 | 37.71 | | |

This sort on the pulsar's right ascension of ascending node illustrates their distribution in the ecliptic plane. For this selection of pulsars, the average difference in right ascension between neighboring pulsars is 24° and the associated standard deviation is 14°. A representative statistic is the mean value of $s_\tau$ in Table 2 associated with the pulsars that were used by the SEXTANT mission. It has a value of $\langle s_\tau \rangle = 576\mathrm{km}\left(\mathrm{cm}\sqrt{\mathrm{hr}}\right)$. The associated mean position uncertainty for these selected pulsars is $\langle \sigma_\beta \rangle = 160$ mas (more on the effect of the pulsar position uncertainty later). At present, this represents the set of pulsars' that the community has found useful for navigation. If X-Ray pulsar navigation finds wider application, it would be reasonable to anticipate that a focused campaign to discover and document other stable source X-Ray signals might occur. As noted in the prior section on optical navigation, the recent ESA Gaia mission to catalog and characterize millions of optical sources will be a significant aid in documenting solar system objects. In anticipation of a similar investment for pulsar timing, we conjecture that a more uniform distribution of source signals could become available that would allow a space-based navigation scheme to point to a geometrically diverse set of stable pulsars and find a suitable signal to track. Assuming this facilitates our parametric investigation of the geometries induced for pulsar-based navigation. As with the previous analytical investigation on optical imaging, we will assume that a suitable source signal is available at any value of right ascension of ascending node (i.e., RA in Table 2), and that the detector can scan a full 360° over the tracking period $T$ (recall that a 'keep-out' zone near the Sun is being neglected for the Part 1 analysis). This yields the following hypothetical 'best-case' discrete tracking scenario relationship between consecutive pulsar measurements

$$\beta_k = \beta_{k-1} + \Delta\beta = \beta_{k-1} + \frac{2\pi}{n_\tau} \rightarrow \beta_k = \beta_0 + \frac{2\pi}{n_\tau}k = \beta_0 + 2\pi\frac{t_k}{T} \text{ (full 360° scan)} \tag{73}$$

where the times between measurements conform to $t_k = kh_\tau$ and $T = n_\tau h_\tau$. Analogous to the optical case, it is useful to transition from a discrete measurement model to a continuous one – doing so will facilitate the use of integrals (vs discrete sums) and simplify the analysis of the information matrix. We do so by assuming the integration time is small relative one-day arcs $h_\tau$ $T$, which allows the use of the following

$$\beta(t) = \beta_0 + \frac{2\pi}{T}t, \quad h_\tau \ll T \tag{74}$$

Note that the angle $\beta_0$ to the first pulsar now represents an initial phase angle between the start of pulsar tracking and the spacecraft initial velocity $\mathbf{v}_0$, as with the optical case, and conforms to the relation

$$\hat{\mathbf{n}}_{\beta_1} \bullet \hat{\mathbf{v}}_0 = \sin\beta_1 \;\; \hat{\mathbf{n}}_{\beta_1} \bullet \hat{\mathbf{v}}_0 = \sin\beta_0 \tag{75}$$

To facilitate the qualitative analysis, we will select a constant source signal stability using $\langle s_\tau \rangle$, the mean value of the pulsars used for the SEXTANT mission in Table 2. Using this value in Eq. (72) yields a generic measurement uncertainty

$$\sigma_\tau(h_\tau) = \sqrt{\frac{\langle s_\tau \rangle^2}{Ah_\tau}} \tag{76}$$

with the functional dependence on $h_\tau$ explicitly called out.

The information matrix $\mathbf{I}_\tau(t_k)$ in a single pulsar TOA measurement at $t_k$ from pulsar $P_k$ is computed using

$$\mathbf{I}_\tau(t_k) \equiv \frac{\mathbf{h}(t_k) \otimes \mathbf{h}(t_k)}{c^2\sigma_\tau^2(h_\tau)} \tag{77}$$

where, to facilitate the current analytical analysis, we have replaced the measurement uncertainty resulting from each individual pulsar $P_k$ with the median uncertainty as expressed in Eq. (76). In the later high-fidelity analysis, we will use the uncertainties that are specific to the pulsar being observed as listed in Table 2 to obtain a more accurate estimate of the trajectory uncertainties. Note that $\sigma_\tau(h_\tau)$ is independent of the absolute time $t_k$; however, this can change when other measurement errors, such as those from a clock or a pulsar template that changes with time, are introduced. Using the additive property of information matrices, the aggregate information content in a sequence of $pn_\tau$ pulsar time of flight measurements, with $n_\tau$ measurements per day taken over a total of $p$ days, is computed using

$$\mathbf{I}_\tau^\Sigma = \sum_{k=1}^{pn_\tau} \frac{\boldsymbol{h}_\tau(t_k) \otimes \boldsymbol{h}_\tau(t_k)}{\sigma_\tau^2\,(h_\tau)} \tag{78}$$

Similar to the optical case, we can approximate the sum over the individual measurements with an integration via taking the limit as $h_\tau$ approaches zero to obtain the following formal result

$$\begin{aligned}\mathbf{I}_\tau^\Sigma &\equiv \sum_{k=1}^{pn_\tau} \frac{\mathbf{h}_\tau(t_k)\otimes\mathbf{h}_\tau(t_k)}{\sigma_\tau^2(h_\tau)} \\ &= \sum_{k=1}^{pn_\tau} \mathbf{h}_\tau(t_k)\otimes\mathbf{h}_\tau(t_k)\frac{Ah_\tau}{\langle S_\tau\rangle^2} \\ &\cong \frac{A}{\langle S_\tau\rangle^2}\left(\lim_{h_\tau\to 0}\sum_{k=1}^{pn_\tau}\mathbf{h}_\tau(t_k)\otimes\mathbf{h}_\tau(t_k)h_\tau\right) \\ &= \frac{1}{\sigma_\tau^2(pT)}\left(\frac{1}{p}\right)\int_0^p h_\tau(\bar{t})\otimes\mathbf{h}_\tau(\bar{t})d\bar{t}\end{aligned} \tag{79}$$

where $\sigma_\tau(pT) = \sqrt{\langle S_\tau\rangle^2/ApT}$ and $\bar{t} \equiv t/T$. The aggregate information matrix $\mathbf{I}_\tau^\Sigma$ defined in Eq. (79) can be block partitioned into the following $2\times2$ submatrices

$$\mathbf{I}_\tau^\Sigma \equiv \begin{bmatrix} \left(\mathbf{I}_\tau^\Sigma\right)_{\mathbf{rr}} & \left(\mathbf{I}_\tau^\Sigma\right)_{\mathbf{r\Delta r}} \\ \left(\mathbf{I}_\tau^\Sigma\right)_{\mathbf{r\Delta r}} & \left(\mathbf{I}_\tau^\Sigma\right)_{\mathbf{\Delta r\Delta r}} \end{bmatrix} \tag{80}$$

where

$$\left(\mathbf{I}_\tau^\Sigma\right)_{\mathbf{rr}} = \frac{1}{\sigma_\tau^2(pT)}\left(\frac{1}{p}\int_0^p \begin{bmatrix} \cos^2\beta & \cos\beta\sin\beta \\ \cos\beta\sin\beta & \sin^2\beta \end{bmatrix} d\bar{t}\right) + \mathbf{O}\left(\varepsilon^2\right) \tag{81}$$

$$\left(\mathbf{I}_\tau^\Sigma\right)_{\mathbf{r\Delta r}} = \left(\mathbf{I}_\tau^\Sigma\right)_{\mathbf{\Delta rr}} = \frac{1}{\sigma_\tau^2(pT)}\left(\frac{1}{p}\int_0^p \begin{bmatrix} \bar{t}\cos^2\beta & \bar{t}\cos\beta\sin\beta \\ \bar{t}\cos\beta\sin\beta & \bar{t}\sin^2\beta \end{bmatrix} dt\right) + \mathbf{O}\left(\varepsilon^2\right) \tag{82}$$

and

$$\left(\mathbf{I}_\tau^\Sigma\right)_{\mathbf{\Delta r\Delta r}} = \frac{1}{\sigma_\tau^2(pT)}\left(\frac{1}{p}\int_0^p \begin{bmatrix} \bar{t}^2\cos^2\beta & \bar{t}^2\cos\beta\sin\beta \\ \bar{t}^2\cos\beta\sin\beta & \bar{t}^2\sin^2\beta \end{bmatrix} dt\right) + \mathbf{O}\left(\varepsilon^2\right) \tag{83}$$

Utilizing the time dependent behavior for $\beta$ as specified in Eq. (74), Eqs. (81), (82), and (83) can be evaluated with the following results

$$\left(\mathbf{I}_\tau^\Sigma\right)_{\mathbf{rr}} = \frac{1}{\sigma_\tau^2(pT)}\begin{bmatrix} \frac{1}{2} & 0 \\ 0 & \frac{1}{2} \end{bmatrix} + \mathbf{O}\left(\varepsilon^2\right) \tag{84}$$

$$\left(\mathbf{I}_\tau^\Sigma\right)_{\mathbf{r\Delta r}} = \left(\mathbf{I}_\tau^\Sigma\right)_{\mathbf{\Delta rr}} = \frac{1}{4}\frac{1}{\sigma_\tau^2(pT)}\begin{pmatrix} p + \frac{\sin^2\beta_0}{2\pi} & \frac{-\cos 2\beta_0}{2\pi} \\ -\frac{\cos 2\beta_0}{2\pi} & p - \frac{\sin 2\beta_0}{2\pi} \end{pmatrix} + \mathbf{O}\left(\varepsilon^2\right) \tag{85}$$

and

$$\left(\mathbf{I}_{\tau}^{\Sigma}\right)_{\Delta\mathbf{r}\Delta\mathbf{r}} = \frac{1}{16\pi^2}\frac{1}{\sigma_{\tau}^2(pT)}\begin{bmatrix} \frac{8\pi^2p^2}{3} + 2\pi\, p\sin 2\beta_0 + \cos 2\beta_0 & -2p\pi\cos 2\beta_0 + \sin 2\beta_0 \\ -2p\pi\cos 2\beta_0 + \sin 2\beta_0 & \frac{8\pi^2p^2}{3} + 2\pi\, p\sin 2\beta_0 + \cos 2\beta_0 \end{bmatrix} + \mathbf{O}\left(\varepsilon^2\right) \quad (86)$$

As with the optical analysis, we will find the position covariance by inverting the Schur complement for the velocity-only information sub matrix in Eq. (80). That is, we seek

$$\mathbf{P}_{\mathbf{rr}}^{\tau} = \left(\mathbf{I}_{\tau}^{\Sigma}/\left(\mathbf{I}_{\tau}^{\Sigma}\right)_{\Delta\mathbf{r}\Delta\mathbf{r}}\right)^{-1} = \left(\left(\mathbf{I}_{\tau}^{\Sigma}\right)_{\mathbf{rr}} - \left(\mathbf{I}_{\tau}^{\Sigma}\right)_{\mathbf{r}\Delta\mathbf{r}}\left(\left(\mathbf{I}_{\tau}^{\Sigma}\right)_{\mathbf{r}\Delta\mathbf{r}}\right)^{-1}\left(\mathbf{I}_{\tau}^{\Sigma}\right)_{\mathbf{r}\Delta\mathbf{r}}\right)^{-1} \quad (87)$$

Substituting Eqs. (84), (85), and Eq. (86) into Eq. (87), and simplifying leads to the following expressions for the uncertainty in the estimate of the spacecraft *initial* position good to second order with $p$-days of pulsar TOA tracking

$$\mathbf{P}_{\mathbf{rr}}^{\tau} = \sigma_{\tau}^2(pT)\mathbf{G}_{\mathbf{rr}}^{\tau} \quad (88)$$

where the average dilution of precision matrix takes the form

$$\mathbf{G}_{\mathbf{rr}}^{\tau} = \begin{bmatrix} \frac{4\left(32\pi^4p^4-24\pi^2p^2-18+9\left(1-4\pi^2p^2\right)\cos 2\beta_0-6\pi\left(3-4\pi^2p^2\right)\sin 2\beta_0\right)}{16\pi^2p^4-24\pi^2p^2-27} & \frac{24\pi p\left(3-4\pi^2p^2\right)\cos 2\beta_0+36\left(1-4\pi^2p^2\right)\sin 2\beta_0}{16\pi^4p^4-24\pi^2p^2-27} \\ \frac{24\pi^2\left(3-4\pi^2p^2\right)\cos 2\beta_0+36\left(1-4\pi^2p^2\right)\sin 2\beta_0}{16\pi^4p^4-24\pi^2p^2-27} & \frac{4\left(32\pi^4p^4-24\pi^2p^2-18-9\left(1-4\pi^2p^2\right)\cos 2\beta_0+6\pi\left(3-4\pi^2p^2\right)\sin 2\beta_0\right)}{16\pi^2p^4-24\pi^2p^2-27} \end{bmatrix} + \mathbf{O}\left(\varepsilon^2\right) \quad (89)$$

Examination of Eq. (89) shows that as $p$ increases, the terms with $p^4$ dominate and, in taking the limit, the following results

$$\lim_{p\to\infty}\mathbf{G}_{\mathbf{rr}}^{\tau} = \begin{bmatrix} 8 & 0 \\ 0 & 8 \end{bmatrix} \to PDOP_{\tau} = 4 \text{ (limiting case for pulsar TOA)} \quad (90)$$

Recall that the components of $\mathbf{G}_{\mathbf{rr}}^{\tau}$ yield uncertainty scale factors that are independent of the particular pulsar detector characteristics, source noise, and pulsar catalog density; rather, they provide insight into the fundamental geometric sensitivities of the positioning problem. Examination of Eq. (84) shows that the upper right position information matrix is constant, thus independent of geometry. The other matrices, Eqs. (85) and (86), depend on $\beta_0$ that relates, via Eq. (75), the initial pulsar location relative to initial velocity vector. Similar to optical imaging, the choice of gimbal orientation for the onboard pulsar detector is a free parameter, so a detection campaign would select the initial pulsar position $\beta_0$ that gives the best positioning result. As seen in Fig. 11 the position component uncertainty geometric scale factors $\left(\sqrt{G_{xx}^{\tau}},\sqrt{G_{yy}^{\tau}}\right)$ oscillate within the interval (2.6,3.4), and at $\pi/2$ intervals they are equal with an approximate value of 3 (for the one-day of tracking case). Compared with the optical case,

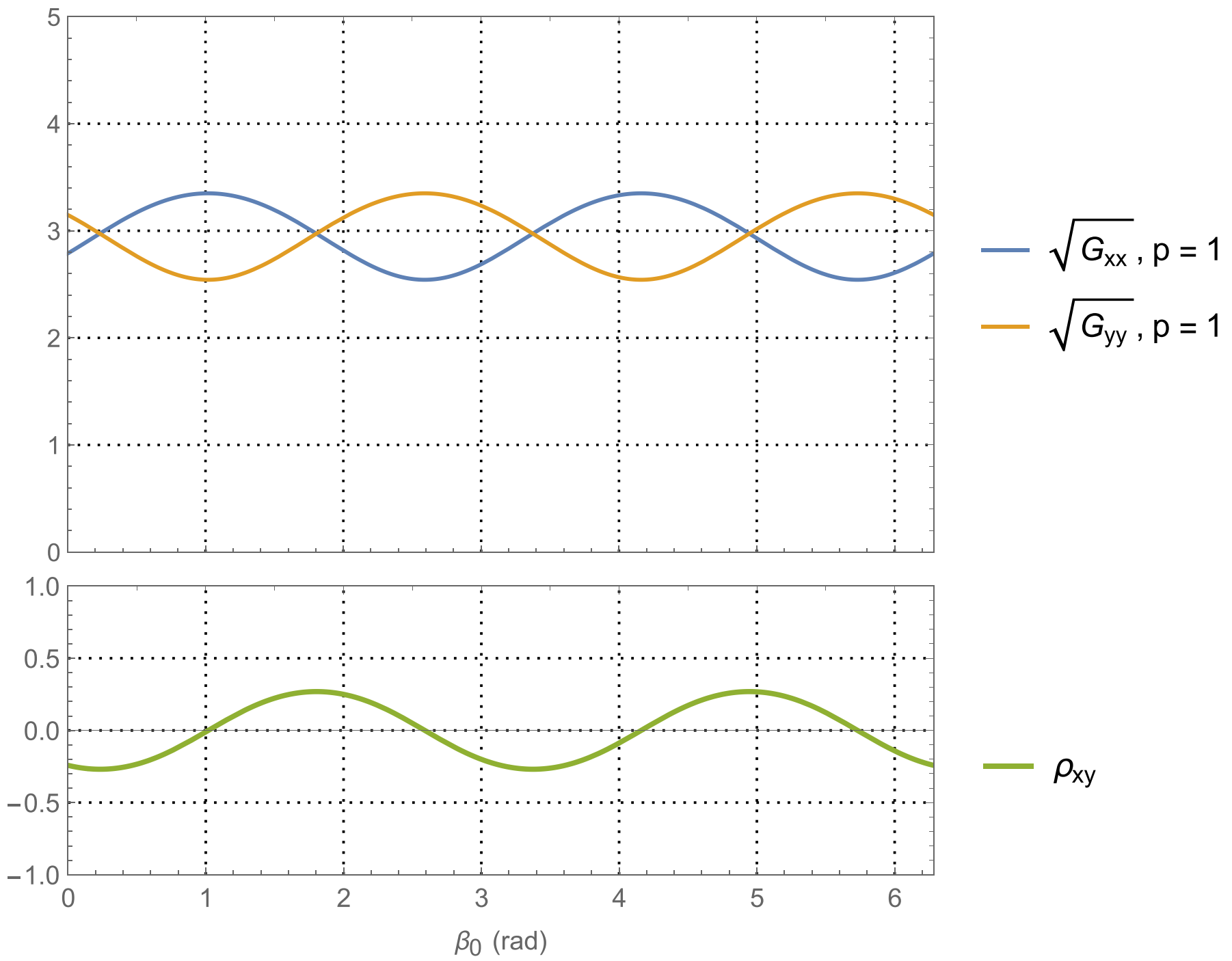

**Fig. 11** Position component uncertainty geometric scale factors for pulsar-based TOA measurements (upper) and the correlation between components (lower) using a full 360° scan over a one-day tracking period

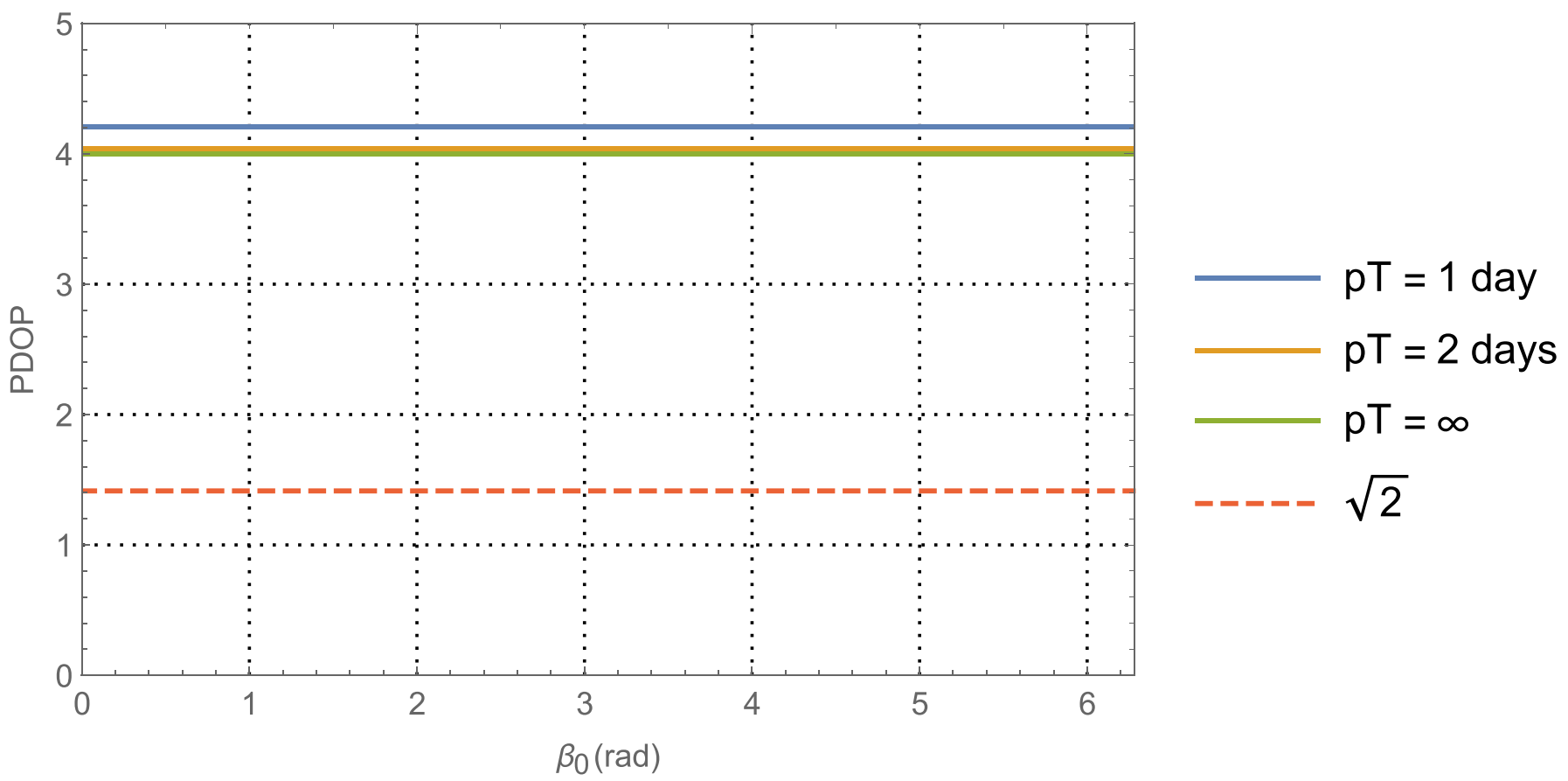

**Fig. 12** PDOP for pulsar TOA using one day of tracking, two days, and in the limit. For reference, the theoretical PDOP obtained with direct observation of the position components is shown as the dashed line at $\sqrt{2}$

the pulsar TOA geometry yields comparably better geometric information that has a more consistent distribution across initial pulsar positions $\beta_0$. The associated PDOP evaluated over $p=1$, $p=2$, and the limit is shown in Fig. 12, where it is clear that the overall dilution of precision is independent of $\beta_0$ and converges quickly to its limit at PDOP = 4 (compare with the optical case that took about 7 days to converge).

Before computing the resultant position uncertainties in Eq. (88) associated with the $\mathbf{G}^{\tau}_{\mathbf{rr}}$ results given in Eq. (89), we need to consider another significant error source in using the pulsar TOA measurements for orbit determination. As with optical imaging, pulsar TOA measurements are sensitive to the location knowledge of the tracked pulsars. This sensitivity can be bounded by

$$\delta(c\tau)_k = \frac{\partial(c\tau)_k}{\partial\beta_k}\delta\beta_k = a\big(\sin nt\cos\beta_k - \cos nt\sin\beta_k\big)\delta\beta_k = -a\sin\beta_k\delta\beta_k + O(\varepsilon) < a\delta\beta_k \tag{91}$$

The preceding result indicates that the measurement uncertainty introduced by a pulsar location uncertainty grows linearly as the spacecraft distance from the solar system barycenter increases. To compare the impact of the pulsar position errors with the impact of the detector noise, we need to determine a set of pulsar TOA detector values that are practical for use in deep space. For instance, the SEXTANT detector in its current configuration would be prohibitive to deploy on most spacecraft heading into deep space, its overall size, weight, and power would need to be reduced. Fortunately, the SEXTANT detector is scalable by eliminating collimators and detectors in the telescope array. Doing so reduces the effective collection area $A$ and increases measurement noise for a given collection interval, but also significantly reduces the instrument's volume to a more practical value. To this end, we have selected $A = 129$ cm (with dimensions of ~11 cm by 11 cm), representing a detector that uses 4 of the 56 SEXTANT X-ray telescope collimators as a candidate detector. Using the mean values $\langle s_\tau \rangle$, $\langle \sigma_\beta \rangle$ and the reduced collecting area with a three-hour integration time leads to the following detector noise uncertainty and pulsar location errors

$$\begin{aligned} \sigma_r(3) &= 29.3\ \text{km}\ \big(h_r = 3\ \text{hr}, A = 129\ \text{cm}^2\big) \\ a\big\langle \sigma_\beta \big\rangle &= 174\ \text{km (Mars case@}a = 1.5\ \text{AU)} \\ a\big\langle \sigma_\beta \big\rangle &= 3491\ \text{km (Neptune case@}a = 30\ \text{AU)} \end{aligned} \tag{92}$$

where it is evident that the pulsar location errors can dominate over the detector noise and is further exacerbated as the spacecraft gets further from the solar system center. With asteroids it is possible to estimate their positions and velocities in conjunction with the spacecraft orbit with reasonable results over short arcs (relative to the asteroid orbit) for navigation purposes as the images are direct observation of the objects that, at large distances, are point sources; however, this must be done carefully so as to not produce optimistic solutions (such as by adding small amounts of process noise in the estimation states to account for unmodeled effects on the asteroids trajectory). In contrast, X-Ray pulsars have more complicated signals with structures not related to the pulsar location that could easily alias into their location estimates if not modeled properly. A more conservative approach for pulsar-based navigation is to consider their location errors when estimating the spacecraft orbit.

We now examine how to augment the qualitative position uncertainties to consider pulsar location error.

As detailed by Tapley, et al. [39], a vector of discrete scalar measurements $\mathbf{y}$ that can be related to the state $\mathbf{x}$ and a vector of consider parameters $\mathbf{c}$ as follows

$$\mathbf{y} = \mathbf{H_x x} + \mathbf{H_c c} + \mathbf{v} \tag{93}$$

with $\mathbf{R} = E[\mathbf{v}\mathbf{v}^T]$. The state covariance $\mathbf{P}_\mathbf{x}^\mathbf{c}$ of a batch filter that includes the effects of the consider parameters can then be expressed as

$$\mathbf{P}_\mathbf{x}^\mathbf{c} = \mathbf{P_x} + \left(\mathbf{P_x H_x^T R^{-1} H_c}\right)\mathbf{P_c}\left(\mathbf{P_x H_x^T R^{-1} H_c}\right)^\mathrm{T} \tag{94}$$

where $\mathbf{P_c}$ is the covariance of the consider parameters and $\mathbf{P_x}$ is the typical state uncertainty obtained from filtering the measurement vector $\mathbf{y}$ (i.e., in the present case this is $\mathbf{P}_\mathbf{rr}^\tau$). Note that the result in Eq. (94) applies when no a priori state uncertainty is present. In applying this result to the present problem, we will focus on the case with one-day of tracking (i.e., $p=1$) in which $n_\tau$ measurements are collected and, as posed by Eq. (73), $n_\tau$ different pulsars are tracked with an integration time $h_\tau$. As such, the following associations apply

$$\begin{aligned}
&\mathbf{H_x} = \begin{bmatrix} \mathbf{h}_\tau(t_1) \\ \vdots \\ \mathbf{h}_\tau(t_{n_\tau}) \end{bmatrix}, \boldsymbol{R} = \begin{bmatrix} \sigma_\tau^2(h_\tau) & 0 & 0 \\ 0 & \ddots & 0 \\ 0 & 0 & \sigma_\tau^2(h_\tau) \end{bmatrix} = \sigma_\tau^2(h_\tau)\mathbf{1}_{n_\tau \times n_\tau}, \\
&\mathbf{c} = \begin{bmatrix} \delta\beta_1 \\ \vdots \\ \delta\beta_{n_t} \end{bmatrix}, \mathbf{H_c} = \begin{bmatrix} a\sin\beta_1 & 0 & 0 \\ 0 & \ddots & 0 \\ 0 & 0 & a\sin\beta_{n_\tau} \end{bmatrix}, \mathbf{P_c} = \begin{bmatrix} \sigma_\beta^2 & 0 & 0 \\ 0 & \ddots & 0 \\ 0 & 0 & \sigma_\beta^2 \end{bmatrix} = \sigma_\beta^2\mathbf{1}_{n_\tau \times n_\tau} \\
&\mathbf{P_x} = \mathbf{P}_\mathbf{rr}^\tau = \sigma_\tau^2(T)\mathbf{G}_\mathbf{rr}^\tau, \left(\mathbf{P}_\mathbf{rr}^\tau\right)^\mathbf{c} \equiv \mathbf{P}_\mathbf{x}^\mathbf{c}
\end{aligned} \tag{95}$$

The $\mathbf{H_c P_c H_c^T}$ part of Eq. (94) is expressed as

$$\mathbf{H_c P_c H_c^T} = a^2\sigma_\beta^2 \begin{bmatrix} \sin^2\beta_1 & 0 & 0 \\ 0 & \ddots & 0 \\ 0 & 0 & \sin^2\beta_{n_\tau} \end{bmatrix} \tag{96}$$

In this form, Eq. (94) becomes exceedingly complex to evaluate analytically because it yields a position dependent entry for every measurement. Since our objective is to bound the effect of the pulsar location error, we can simplify the analysis by replacing the matrix in Eq. (96) with the 2-norm of the matrix (a closer bound then the Frobenius norm) to produce the following upper bound

$$\mathbf{H_c P_c H_c^T} \le a^2\sigma_\beta^2 \left\| \begin{bmatrix} \sin^2\beta_1 & 0 & 0 \\ 0 & \ddots & 0 \\ 0 & 0 & \sin^2\beta_{n_\tau} \end{bmatrix} \right\|_2 \mathbf{1}_{n_\tau \times n_\tau} = a^2\sigma_\beta^2 \max\left(\sin^2\beta_k\right)\mathbf{1}_{n_\tau \times n_\tau} \le a^2\sigma_\beta^2\mathbf{1}_{n_\tau \times n_\tau} \tag{97}$$

where we also made use of the fact that $\left\|\mathbf{1}_{n_\tau \times n_\tau}\right\|_2 = 1$ to ensure a consistent bound on the spectral radius of $\mathbf{H_c P_c H_c^T}$. Using the following observation that

$$\mathbf{H}_{\mathbf{x}}^{\mathrm{T}}\mathbf{H}_{\mathbf{x}} = \sum_{k=1}^{n_\tau} \mathbf{h}_\tau\left(t_k\right) \otimes \mathbf{h}_\tau\left(t_k\right) = \left(\mathbf{G}_{\mathbf{rr}}^\tau\right)^{-1} \tag{98}$$

and substituting the associations in Eq. (95), the results of Eq. (97) and Eq. (98), and the fact that $\sigma_\tau^2(T) = \sigma_\tau^2\left(h_\tau\right)/n_\tau$ into Eq. (94) leads to the following expression for the position uncertainty $\left(\mathbf{P}_{\mathbf{rr}}^\tau\right)^{\mathbf{c}}$ that considers the effects of pulsar position errors conforming to the measurement scheme in Eq. (74)

$$\left(\mathbf{P}_{\mathbf{rr}}^\tau\right)^{\mathbf{c}} = \mathbf{P}_{\mathbf{rr}}^\tau + \frac{1}{n_\tau^2} a^2 \sigma_\beta^2 \mathbf{G}_{\mathbf{rr}}^\tau = \left(\sigma_\tau^2(T) + \frac{a^2\sigma_\beta^2}{n_\tau^2}\right)\mathbf{G}_{\mathbf{rr}}^\tau \tag{99}$$

We see that the effect of considering the pulsar position errors is to increase the measurement variance by the additive amount $a^2\sigma_\beta^2/n_\tau^2$ yielding an effective uncertainty $\overline{\sigma}_\tau\left(T, n_\tau\right)$ for the tracking over the period $T$ of

$$\sigma_\tau\left(T, n_\tau\right) = \sqrt{\frac{\langle s_\tau\rangle^2}{AT} + \frac{a^2\left\langle\sigma_\beta\right\rangle^2}{n_\tau^2}} \tag{100}$$

where, again, we have used the mean values for the pulsar noise and location errors. The effect of considering the pulsar location errors (without estimating them) is to increase the overall uncertainty. Since $\left\langle s_\tau\right\rangle$ is not a function of distance from the SSB while $a\left\langle\sigma_\beta\right\rangle$ is, Eq. (100) illustrates that pulsar location errors begin to dominate as the spacecraft distance increases; however, when this begins depends on the relative magnitudes of $\left\langle s_\tau\right\rangle$ and $\left\langle\sigma_\beta\right\rangle$. Also note, if the pulsars were to be tracked more than once during the period $T$, then the structure of the matrices in Eqs. (95) would change (becoming more complex) but would not qualitatively change the result that pulsar location errors add to the overall error.

In order to calculate specific representative values, we select 8 pulsars to track – the same ones as used for the SEXTANT mission – and over a 24-h period equates to $h_\tau = 3$ hrs. Using the smaller X-ray detector with 4 collimators, collecting measurements for the full period $T$ to compute $\overline{\sigma}_\tau^2\left(T, n_\tau\right)$, and selecting $\max\left(\sqrt{G_{xx}^\tau(\beta_0)}, \sqrt{G_{yy}^\tau(\beta_0)}\right)$ from Fig. 11 leads to the following 1-sigma position component uncertainties for the Mars and Neptune cases

$$\text{Mars case } (a = 1.5 \text{ AU}) : \left(\sigma_{xx}^\tau, \sigma_{yy}^\tau\right) \cong (72 \text{ km}, 72 \text{ km}) \tag{101}$$

$$\text{Neptune case } (a = 30 \text{ AU}) : \left(\sigma_{xx}^\tau, \sigma_{yy}^\tau\right) \cong (1310 \text{ km}, 1310 \text{ km}) \tag{102}$$

Of course, these results are highly dependent on the selected pulsars, which amongst the population in Table 2 can have a lot of variability. For comparison, four pulsars (one in each quadrant or close to another quadrant) have been selected to provide other representative statistics. The selected ones include J0437 + 4715 in Q1,[4] B0833–45 in Q2, B1821–24 close to Q3, and B1937 + 21 in Q4 and yield an average signal stability statistic $\langle s_\tau \rangle$ of 397 $\mathrm{km}\left(\mathrm{cm}\sqrt{\mathrm{hr}}\right)$ and an average location uncertainty $\langle \sigma_\beta \rangle$ of 1.62 mas. Using these values, the resulting 1-sigma position component uncertainties for the Mars and Neptune cases become

$$\text{Mars case } (a = 1.5 \text{ AU}) : \left(\sigma_{xx}^\tau, \sigma_{yy}^\tau\right) \cong (21 \text{ km}, 21 \text{ km}) \quad \text{(best 4 pulsars)} \tag{103}$$

$$\text{Neptune case } (a = 30 \text{ AU}) : \left(\sigma_{xx}^\tau, \sigma_{yy}^\tau\right) \cong (34 \text{ km}, 34 \text{ km}) \quad \text{(best 4 pulsars)} \tag{104}$$

where it is obvious that the effect of the pulsar location error has been minimized by this selection.

The preceding results indicate the pulsar data type can be used for absolute positioning during cruise phase orbit determination throughout the solar system with relatively uniform results. However, they also show that the solution uncertainties increase with the distance from the solar system barycenter; but this effect can be mitigated with a judicious selection of pulsars with lower position uncertainties. Another key factor into usability of the pulsar TOA data for navigation is the variability in the available catalog can result in significant variability in the solution performance. Ideally, a more extensive catalog of well surveyed pulsars will be needed to ensure that reliable, high performing navigation can be obtained. Finally, it will be seen in the high-fidelity quantitative analysis, that the pulsar data type is insensitive to the location of the spacecraft relative to any object that it might be navigating towards. That is, target relative trajectory estimates using pulsar TOA measurements do not improve, as would be needed for any precision navigation, as a spacecraft approaches a body of interest for flybys, orbit insertions, or landings. In contrast, optical data is by design an excellent target relative measurement, and to a lesser degree Earth-based radiometric data.

## Range Information Content

An analytic asymptotic analysis of the range tracking case does not prove to be as straightforward as for the prior two cases. For the optical imaging case, a first order analysis proved sufficient to obtain analytically tractable estimates of the position uncertainties, for pulsar TOAs a second order analysis was sufficient; however, for the range case, a fifth-order analytic expansion would be required. In Appendix 2: Asymptotic Analysis of the Radio Information Matrix, an argument is presented for why a fifth-order expansion is necessary. Since fifth-order analytic expressions would not be very illuminating, we instead analyze our simplified example using direct numerical integration of the nonlinear range measurement sensitivity equations to obtain bounds on the position uncertainties.

[4] While the PSR B0531 + 21 (M1 - Crab Pulsar) has lower noise figures and position knowledge, it is also prone to sporadic noise glitches. As a result, J0218–4232 was selected as more representative of a stable population.

We seek the partials of the slant range $\rho_i$ from the Earth station $i$ at time $t$ with respect to the initial state vector $\mathbf{x}_0$ at time $t_0$. The measurement sensitivity (or gradient) vectorcan be obtained using the following chain rule expression

$$\mathbf{h}_{\rho_i}(t) = \frac{\partial \rho_i}{\partial \mathbf{r}} \frac{\partial \mathbf{r}}{\partial \mathbf{x}_0} = \frac{\partial \rho_i}{\partial \mathbf{r}} \left[ \boldsymbol{\Phi}_{\mathbf{rr}}\left(t, t_0\right) \vdots \boldsymbol{\Phi}_{\mathbf{r\Delta r}}\left(t, t_0\right) \right] \tag{105}$$

The partial for the slant range is readily obtained as

$$\frac{\partial \rho_i}{\partial \mathbf{r}} = \frac{\partial \rho_i(t)}{\partial \mathbf{r}} = \frac{\boldsymbol{\rho}_i(t)}{\rho_i(t)} = \hat{\boldsymbol{\rho}}_i(t) \tag{106}$$

which is the slant range unit vector. Applying the assumptions in the 2-d example leads to the following explicit equation for the slant range unit vector

$$\begin{aligned}\hat{\boldsymbol{\rho}}_i(t) &= \frac{\boldsymbol{\rho}_i(t)}{\rho_i(t)} = \frac{\left(\boldsymbol{r}_i(t) - \boldsymbol{r}_{s_i}(t)\right)}{\sqrt{\left(\boldsymbol{r}_i(t) - \boldsymbol{r}_{s_i}(t)\right) \bullet \left(\boldsymbol{r}_i(t) - \boldsymbol{r}_{s_i}(t)\right)}} \\ &= \frac{\left(a \cos nt - R_\oplus \cos\left(\omega_\oplus t + \phi_{i,0}\right) - a_\oplus \cos\left(n_\oplus t + \xi_0\right)\right)\hat{\mathbf{x}} + \left(a \sin nt - R_\oplus \sin\left(\omega_\oplus t + \phi_{i,0}\right) - a_\oplus \sin\left(n_\oplus t + \xi_0\right)\right)\hat{\mathbf{y}}}{\sqrt{\left(a \cos nt - R_\oplus \cos\left(\omega_\oplus t + \phi_{i,0}\right) - a_\oplus \cos\left(n_\oplus t + \xi_0\right)\right)^2 + \left(a \sin nt - R_\oplus \sin\left(\omega_\oplus t + \phi_{i,0}\right) - a_\oplus \sin\left(n_\oplus t + \xi_0\right)\right)^2}}\end{aligned} \tag{107}$$

Since the result in Eq. (107) has not been asymptotically expanded, the logical choice for the STM would be to use the exact expressions in Eq. (8) and (9) to formulate an expression for $\mathbf{h}_{\rho_i}$; however, numerical experiments with the third order expansion of the STM have shown that the loss in accuracy is minimal when using the approximation in Eq. (13) and with a significant improvement in processing time for the numerical integrals. Therefore, a complete expression for the $\mathbf{h}_{\rho_i}(t)$ can be written (in column form for readability) as

$$\mathbf{h}_{\rho_i}(t) = \frac{\begin{bmatrix} \left(1 + n^2 t^2\right)\left(a \cos nt - R_\oplus \cos\left(\omega_\oplus t + \phi_{i,0}\right) - a_\oplus \cos\left(n_\oplus t + \xi_0\right)\right) \\ \left(1 - \frac{n^2 t^2}{2}\right)\left(a \sin nt - R_\oplus \sin\left(\omega_\oplus t + \phi_{i,0}\right) - a_\oplus \sin\left(n_\oplus t + \xi_0\right)\right) \\ \left(\frac{t}{T} + \frac{n^2 t^3}{3T}\right)\left(a \cos nt - R_\oplus \cos\left(\omega_\oplus t + \phi_{i,0}\right) - a_\oplus \cos\left(n_\oplus t + \xi_0\right)\right) \\ \left(\frac{t}{T} - \frac{n^2 t^3}{6T}\right)\left(a \sin nt - R_\oplus \sin\left(\omega_\oplus t + \phi_{i,0}\right) - a_\oplus \sin\left(n_\oplus t + \xi_0\right)\right) \end{bmatrix}^{\mathrm{T}}}{\sqrt{\left(a \cos nt - R_\oplus \cos\left(\omega_\oplus t + \phi_{i,0}\right) - a_\oplus \cos\left(n_\oplus t + \xi_0\right)\right)^2 + \left(a \sin nt - R_\oplus \sin\left(\omega_\oplus t + \phi_{i,0}\right) - a_\oplus \sin\left(n_\oplus t + \xi_0\right)\right)^2}} \tag{108}$$

It is necessary to consider the geometry of the Earth tracking stations and when it is geometrically possible to track the spacecraft. The typical pass of an Earth station from rise-to-set can vary by hours and ranges from 6 h up to 11. Over the course of a day, continuous coverage would transition between three different DSN complexes (Goldstone, Madrid, and Canberra). In our *simplified* 2-d planar geometry, we determine a constraint between the Earth angle $\xi$ and $\phi$ the station angle when the Earth 'rises' and 'sets' in the station's antenna field of coverage and is documented in Appendix 1: Two-dimensional Earth Station Viewing Geometry. From this analysis, the key Earth station viewing angle constraints at rise (defined to be the spacecraft ascending above the horizon) are given to first order by

$$\sin\phi_{0,0} = -\frac{a - a_{\oplus}\cos\xi_0}{\sqrt{a_{\oplus}^2 + a^2 - 2a_{\oplus}a\cos\xi_0}}, \cos\phi_{0,0} = -\frac{a_{\oplus}\sin\xi_0}{\sqrt{a_{\oplus}^2 + a^2 - 2a_{\oplus}a\cos\xi_0}} \tag{109}$$

where the indices on the angles $\xi_0$ and $\phi_{0,0}$ indicate the simplifying assumption that the first Earth station begins tracking at the initial time. Since we are assuming a spherical Earth and rise/set defined by the horizon, the first station sets after the Earth has rotated 180 degrees. In this case, the length of the tracking pass is 12 h or half of the observation period $T$. The second station is assumed to be located so that it comes into view immediately after the first one sets, thus the following geometric constraint applies

$$\phi_{0,1} = \phi_{0,0} - \pi \tag{110}$$

Thus, for this analysis, it is sufficient to assume there are only two Earth tracking stations providing continuous coverage. Combining the results in Eq. (110) and (109) constrains the initial tracking station location to be a function of the initial Earth location, that is $\phi_{0,0}=\phi_{0,0}(\xi_0)$ and reduces Eq. (108) to be functions of time that are parameterized by only the initial Earth angle $\xi_0$ (again, effects due to conjunction with the Sun are neglected in the Part 1 analysis). Using these observations, we can now proceed with the analysis. The associated linearized instantaneous range measurement equation can be represented formally as

$$\delta\rho_i(t) = \mathbf{h}_{\rho_i}(\mathrm{t})\delta\mathbf{x}_0 + v_{s_i} \tag{111}$$

As demonstrated by the recent NASA New Horizons mission to Pluto, deep space PN-range measurements can obtain sub-meter noise levels with integration times as short as 30 s at distances that are 10s of AUs [40]. For a given received signal power, the noise in the range measurement scales as $\rho_i/\sqrt{h_\rho}$ where $h_\rho$ is the range measurement integration time [15]. Range integration times are 10's of seconds and yield a significant number of measurements in any given observation period $T$. For the current analysis, a 1 m ($1\sigma$) ranging noise strength at 1 AU with a 60 s integration time is selected as representative and leads to the following noise strength expression

$$\sigma_\rho = 0.001\left(\frac{\rho}{1\mathrm{AU}}\right)\sqrt{\frac{60\ \mathrm{sec}}{h_\rho}}\ \mathrm{km} \cong s_\rho\frac{\rho_0}{\sqrt{h_\rho}}\ \text{with}\ s_\rho = 8.6\times10^{-13}\sqrt{\mathrm{hr}} \tag{112}$$

For the problems being investigated, the maximum slant range for the Mars case is $\rho < 2.5$ AU and for the Neptune $\rho < 31$ AU case. At these distances, the factor $s_\rho \rho_0$ take the following values

$$\begin{aligned} s_\rho \rho_0 &= 0.0003\ \mathrm{km}\sqrt{\mathrm{hr}} \quad \text{(Mars)} \\ s_\rho \rho_0 &= 0.004\ \mathrm{km}\sqrt{\mathrm{hr}} \quad \text{(Neptune)} \end{aligned} \tag{113}$$

and, as a function of integration time $h_\rho$, we define the measurement uncertainty as

$$\sigma_\rho\left(h_\rho\right) = \sqrt{\frac{s_\rho^2 \rho_0^2}{h_\rho}} \tag{114}$$

The information matrix for a single instantaneous range measurement at time $t$ from tracking station $i$ is formulated using

$$\mathbf{I}_{\rho_i}(t) \equiv \frac{1}{\sigma_\rho^2\left(h_\rho\right)} \mathbf{h}_{\rho_i}(t) \otimes \mathbf{h}_{\rho_i}(t) \tag{115}$$

The information matrix for the aggregate of data that is collected at a rate of $n_\rho$ measurements per day over a total of $p$ days from the set of Earth stations $\{E_i | i \in \{0, \cdots, n_E - 1\}\}$ can be formulated similarly to the prior cases. To facilitate this, we restrict the analysis to two tracking stations $\{E_0, E_1\}$ that are separated according to Eq. (110). The timeline at each station that is actively tracking and taking measurements at the discrete times $t_{m,\,i}$ (with the interval between consecutive measurements being $h_\rho$) over the $p$ days is represented as the following sets

$$\begin{aligned} E_0 : E_0 &= \left\{ t_{m,0} \middle| t_{m,0} \in \left(0, \tfrac{T}{2}\right), \left(T, \tfrac{3T}{2}\right), \cdots, \left((p-1)T, (p-1)T + \tfrac{T}{2}\right) \right\} \\ E_1 : E_1 &= \left\{ t_{m,1} \middle| t_{m,1} \in \left(0, \tfrac{T}{2}, T\right), \left(\tfrac{3T}{2}, 2T\right), \cdots, \left((p-1)T + \tfrac{T}{2}, pT + \tfrac{T}{2}\right) \right\} \end{aligned} \tag{116}$$

Nondimensionalizing the time via the substitution $\bar{t} \equiv t/T$ yields

$$\begin{aligned} E_0 &: \left\{ \bar{t}_{m,0} \middle| \bar{t}_{m,0} \in \left[ (j-1), (j-1) + \tfrac{1}{2}; j = 1, \cdots, p \right] \right\} \\ E_1 &: \left\{ \bar{t}_{m,1} \middle| \bar{t}_{m,1} \in \left[ (j-1) + \tfrac{1}{2}, j; j = 1, \cdots, p \right] \right\} \end{aligned} \tag{117}$$

Using the preceding observations about the timeline of measurements, the aggregate information matrix can be expressed as

$$\mathbf{I}_\rho^\Sigma \equiv \sum_{i=0}^{1} \sum_{E_i} \frac{\mathbf{h}_{\rho_i}\left(t_m^i\right) \otimes \mathbf{h}_{\rho_i}\left(t_m^i\right)}{\sigma_\rho^2\left(h_\rho\right)} \tag{118}$$

We now approximate the sum over the individual measurements with integrations via taking the limit as $h_\rho$ approaches zero, with the result

$$
\begin{aligned}
\mathbf{I}_\rho^\Sigma &\equiv \sum_{\mathrm{i}=0}^{1}\sum_{E_i}\frac{\mathbf{h}_{\rho_i}(t_{m,i})\otimes\mathbf{h}_{\rho_i}(t_{m,i})}{\sigma_\rho^2(h_\rho)} \\
&= \sum_{\mathrm{i}=0}^{\mathbf{1}}\sum_{E_i}\mathbf{h}_{\rho_i}(t_{m,i})\otimes\mathbf{h}_{\rho_i}(t_{m,i})\frac{h_\rho}{(s_\rho\rho_0)^2} \\
&\cong \frac{1}{(s_\rho\rho_0)^2}\sum_{i=0}^{1}\left(\lim_{h_\rho\to 0}\sum_{E_i}\mathbf{h}_{\rho_i}(t_{m,i})\otimes\mathbf{h}_{\rho_i}(t_{m,i})h_\rho\right) \\
&= \frac{1}{\sigma_\rho^2(pT)}\left[\frac{1}{p}\sum_{j-1}^{p}\left(\int_{j-1}^{(j-1)+\frac{1}{2}}\mathbf{h}_{\rho_0}(\bar{t})d\bar{t}+\int_{(j-1)+\frac{1}{2}}^{j}\mathbf{h}_{\rho_1}(\bar{t})\otimes\mathbf{h}_{\rho_i}(\bar{t})d\bar{t}\right)\right]
\end{aligned}
\tag{119}
$$

where $\sigma_\rho(pT) \equiv \sqrt{s_\rho^2\rho_0^2/pT}$. Analytic results for Eq. (119) turn out to be exceptionally difficult to obtain. Inverting Eq. (119) to compute covariance estimates requires the information matrix $\mathbf{I}_\rho^\Sigma$ to be full rank (in the present case that is 4). An asymptotic expansion through third order (i.e., obtaining explicit expressions at zeroth, first, and second order) is required to achieve a full rank analytic expression. Since the zeroth order expansion is not full rank an analytic expression for the inversion requires a Laurent series expansion (see Ref. [41] for details). Only after expanding to third order does the Laurent expansion yield a full rank invertible matrix. Furthermore, for the covariance to be correct the inversion must yield a positive definite symmetric matrix, to obtain numerical results that conform to this requires an asymptotic expansion through fifth order. There is little to be gained with developing complex fifth-order analytic expressions, so the better course is to numerically integrate Eq. (119) to obtain curves similar to those found previously for optical imaging and pulsar TOA measurements. A numerical analysis of the eigenvalues of the various asymptotic expansions of the information matrix $\mathbf{I}_\rho^\Sigma$ reveals the behaviors that make it difficult to analytically expand Eq. (119) and is presented in Appendix 2: Asymptotic Analysis of the Radio Information Matrix.

As with the prior analyses, the aggregate information matrix $\mathbf{I}_\rho^\Sigma$ can now be partitioned into the following submatrices

$$
\mathbf{I}_\rho^\Sigma \equiv \begin{bmatrix} \left(\mathbf{I}_\rho^\Sigma\right)_{\mathbf{rr}} & \left(\mathbf{I}_\rho^\Sigma\right)_{\mathbf{r}\Delta\mathbf{r}} \\ \left(\mathbf{I}_\rho^\Sigma\right)_{\mathbf{r}\Delta\mathbf{r}} & \left(\mathbf{I}_\rho^\Sigma\right)_{\Delta\mathbf{r}\Delta\mathbf{r}} \end{bmatrix} \tag{120}
$$

We find the position covariance by inverting the Schur complement for the velocity-only information sub matrix in Eq. (120) using

$$
\mathbf{P}_{\mathbf{rr}}^\rho = \left(\mathbf{I}_\rho^\Sigma/\left(\mathbf{I}_\rho^\Sigma\right)_{\Delta\mathbf{r}\Delta\mathbf{r}}\right)^{-1} = \left(\left(\mathbf{I}_\rho^\Sigma\right)_{\mathbf{rr}} - \left(\mathbf{I}_\rho^\Sigma\right)_{\mathbf{r}\Delta\mathbf{r}}\left(\left(\mathbf{I}_\rho^\Sigma\right)_{\Delta\mathbf{r}\Delta\mathbf{r}}\right)^{-\mathbf{1}}\left(\mathbf{I}_\rho^\Sigma\right)_{\mathbf{r}\Delta\mathbf{r}}\right)^{-1} \tag{121}
$$

As done in the prior cases, we make the following association for the average geometric dilution of precision matrix $\mathbf{G}_{\mathbf{rr}}^\rho$ and the position covariance in Eq. (121) as follows

$$\mathbf{P}_{\mathbf{rr}}^{\rho} = \sigma_{\rho}^{2}(pT)\mathbf{G}_{\mathbf{rr}}^{\rho} \tag{122}$$

The components of $\mathbf{G}_{\mathbf{rr}}^{\rho}$ yield uncertainty scale factors that are independent of the particular range noise characteristics (i.e., $s_{\rho}\rho_0$) and provide insight into the fundamental geometric sensitivities of the positioning problem. Note that the integration over time in Eq. (119) eliminates time as an independent variable leaving the covariance components to be functions of only the initial Earth angle $\xi_0$. However, unlike optical imaging and the pulsar TOA, the initial Earth angle $\xi_0$ is not a free variable that can be selected at will. This angle is set by the Earth's relative location to the spacecraft, which is determined by the initial epoch. That is, while the initial asteroid angle $\alpha_0$ or the initial pulsar angle $\beta_0$ can be freely selected (via use of a gimbal), the initial Earth angle $\xi_0$ is dictated by the current positions of the solar system bodies associated with the initial epoch. The position compontent uncertainty geometric scale factors $\left(\sqrt{G_{xx}^{\rho}}, \sqrt{G_{yy}^{\rho}}\right)$ for the Mars case are shown in Fig. 13.

One of the most notable observations is the geometric scale factors are up to five orders magnitude larger than the prior results seen with optical and or pulsar tracking. However, this is ammeloriated by the fact that the range measurement uncertainty is about seven orders of magnitude smaller than both the optical and pulsar tracking systems. This results in position uncertainties that are, even for the worst

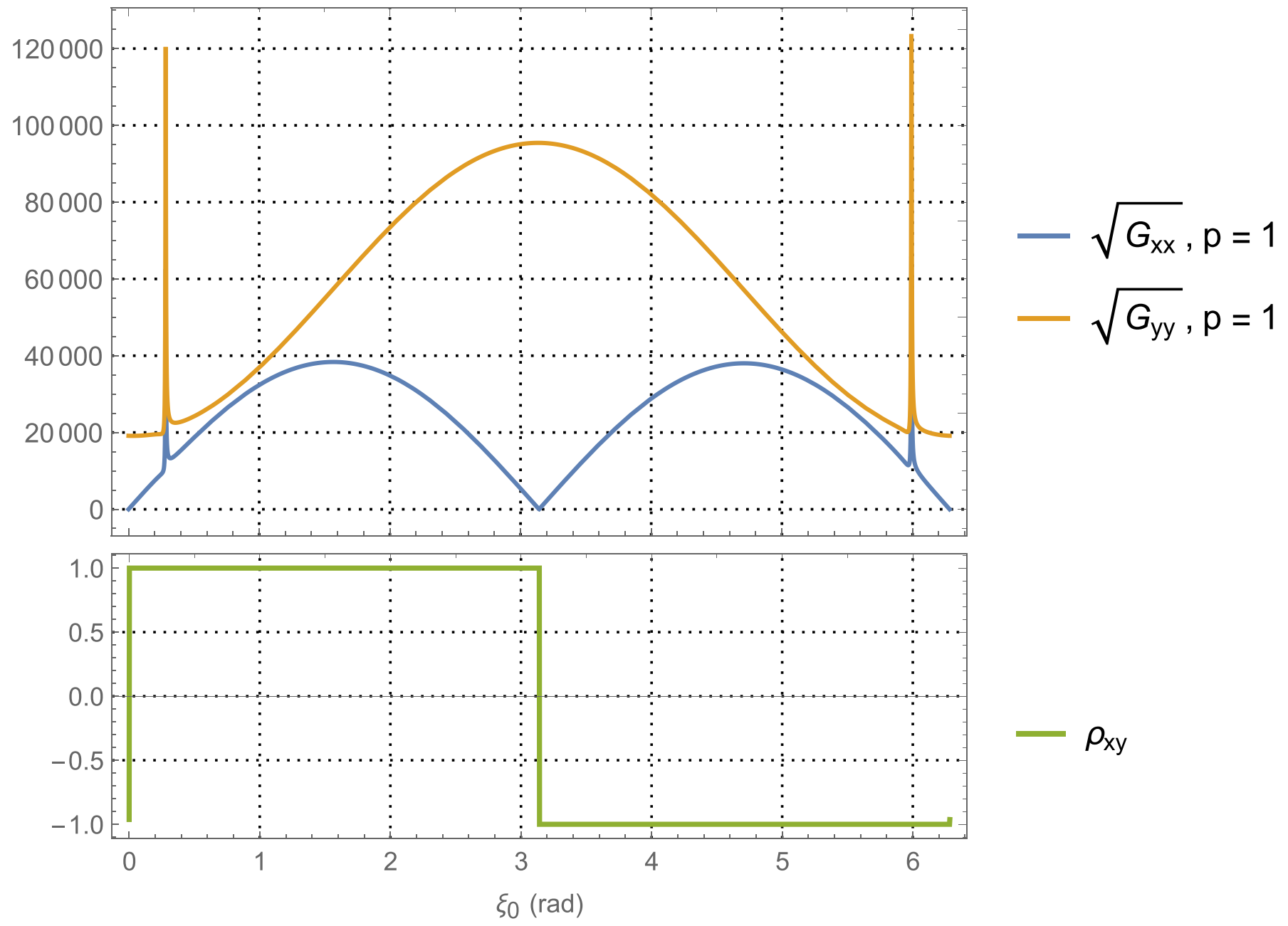

**Fig. 13** Position component uncertainty geometric scale factors using ranging measurements at Mars distances and a one-day tracking period (upper) and the correlation between components (lower) as a function of the initial Earth angle

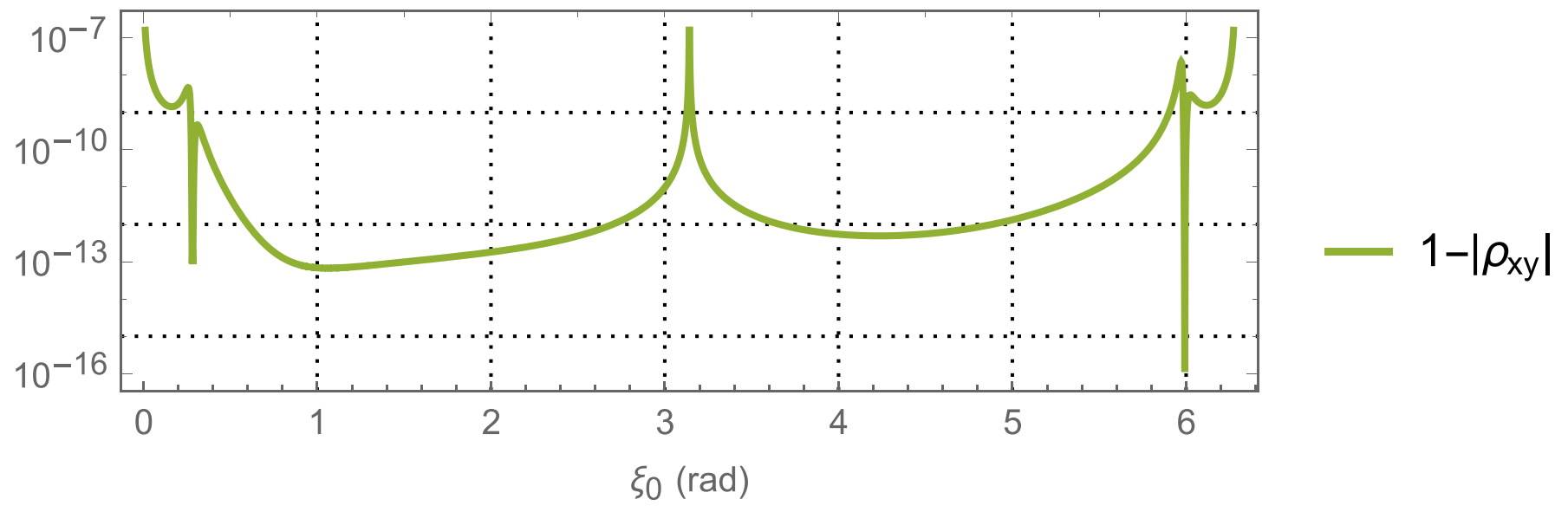

**Fig. 14** Log scale plot of $1-|\rho_{xy}|$ showing the deviation from one of the correlation between the two position component uncertainties when using range tracking at Mars distances

performing initial Earth angles, much smaller than with the prior measurement cases. Indeed, this phenomenal measurement precision is a fundamental reason that radiometric tracking has been the basis for deep space navigation since its first use in the early 1960's. Using the ranging uncertainty factor for Mars distances in Eq. (113) and a one-day tracking period yields the following position uncertainty bounds

$$\text{Mars case}: \begin{cases} 0.2\text{ m} < \sigma_{xx}^{\gamma} < 2.5\text{ km}, \\ 1\text{ km} < \sigma_{yy}^{\gamma} < 6\text{ km}\left(8\text{ km at the}'\text{spikes}'\right) \end{cases} \tag{123}$$

As expected, these results illustrate that the ranging line of sight direction (at $\xi_0=0$ this is the x-axis) is consistently better determined than the orthogonal direction. Examination of the correlations shown in Fig. 13 are an indication of the poor numerical conditioning in the ranging scenario; however, they remain sufficient to maintain a proper covariance structure. To get a closer look at the behavior of the correlation near the value of one, the quantity $1-|\rho_{xy}|$ is plotted on a log scale and shown in Fig. 14 where it can be seen that the correlation is close to one, but never exactly one. As compared to the pulsar TOA and optical imaging results, a single days worth of tracking for a spacecraft at Mars distances is well determined using ranging data. In practice, these estimates are further improved by augmenting the range with Doppler data. Indeed, the phase measurements used to derive the Doppler can be measured at approximately 5 mm (at X-band), which is three orders of magnitude more precise than range. The addition of this precision data improves velocity estimates tremendously and ultimately the position estimates.

In Fig. 15 for the Mars case, we plot the PDOP results for various periods of ranging from one day to fourteen days. The relative improvement is significant, resulting in almost an order of magnitude reduction in the scale factor. Furthermore, the observability issues seen at the spikes actually 'invert' at seven days producing the smallest PDOP values. This suggests that the numerical difficulties with inverting the information matrices are significantly reduced. However, the reader is cautioned to not intepret that after fourteen days the *current* state position uncertainty (vs the initial state uncertainties found presently) would reduce commensurately. Current state uncertainties are found by propagating the initial condition unceratinty forward plus any stochastic errors present using the state transition matrix $\mathbf{\Phi}(t, t_0)$, which is unbounded as $t$ increases. The forward propagation coupled with the continuing measurement collection (for any observatio scheme – range data, optical imaging, or pulsar TOA) typically yields state uncertainties that achieve some form of balanced result. The high-fidelity simulations in the sequel will better illustrate these characteristic than the present analytical analysis.

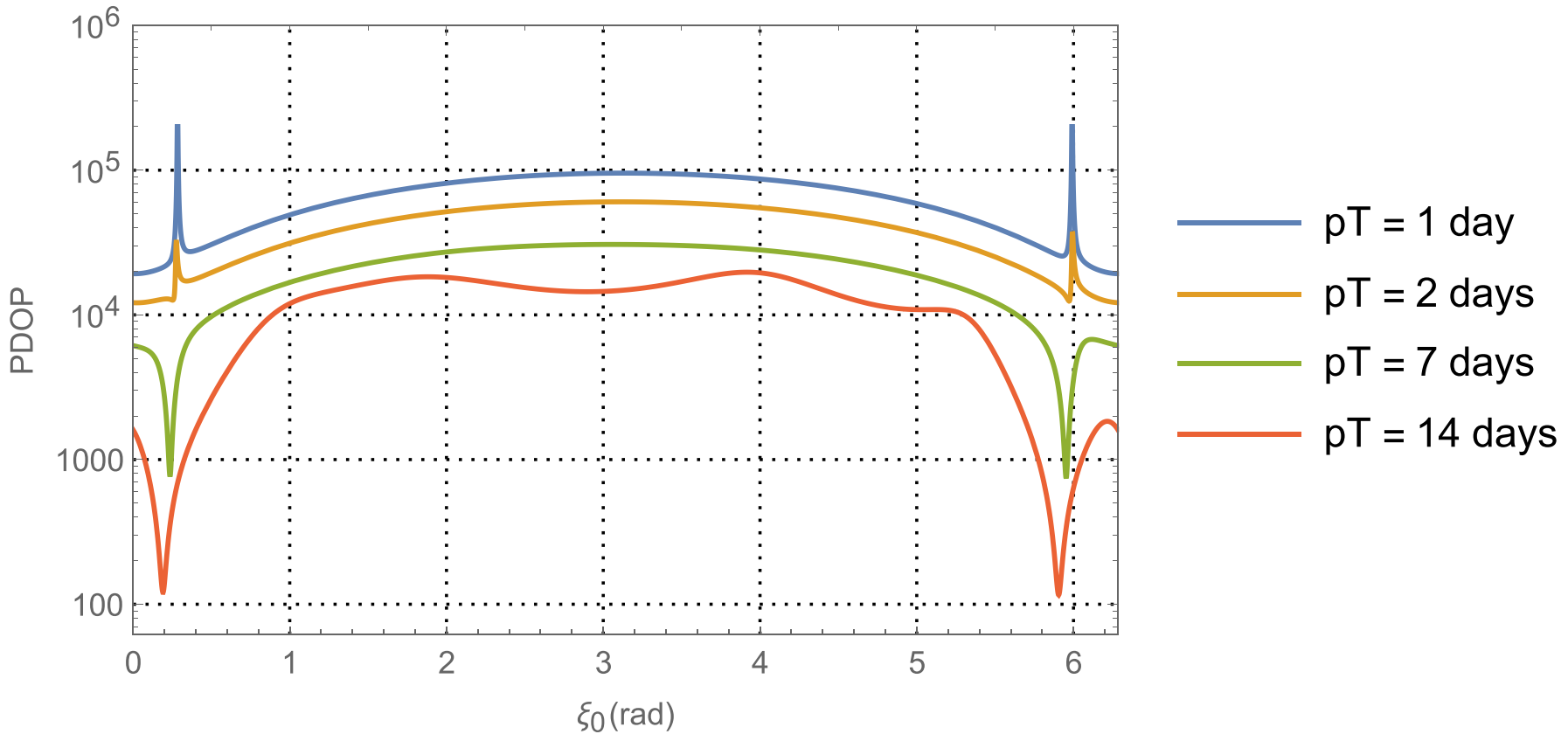

**Fig. 15** PDOP using range measurements for the Mars case with one day, two days, seven days, and fourteen days of tracking

We now extend our results to the Neptune case with initial position component uncertainties shown (on log scales) in Fig. 16, the line-of-sight direction degrades somewhat as compared to the Mars case (even the 'spikes' are similar in magnitude but shifted in initial Earth angle $\xi_0$); however, the orthogonal direction has degraded by an order of magnitude. Using the ranging uncertainty factor for Neptune distances in Eq. (113) and a 24 h tracking period yields the following position uncertainty bounds

$$\text{Neptune case}: \begin{cases} 2\ \text{m} < \sigma^{\gamma}_{xx} < 40\ \text{km}\ \left(600\ \text{km at the}'\text{spikes}'\right) \\ 820\ \text{km} < \sigma^{\gamma}_{yy} < 980\ \text{km}\ \left(3200\ \text{km at the}'\text{spikes}'\right) \end{cases} \tag{124}$$

The deviation of the correlation away from one, shown in Fig. 16, illustrates that the system retains full rank; however, the conditioning is an order of magnitude worse near the 'spikes' than with the Mars case. The PDOP results for the Neptune case are shown in Fig. 17, where similar to the Mars case, there is about an order magnitude improvement with a fourteen- day tracking period relative to a one-day period.

### Summary of the Analytical Comparison between Pulsar TOA, Optical Imaging, and Ranging

The preceding analysis results for the three separate measurement scenarios are summarized in Table 3 for the case with one day of tracking. It is important to understand their context, they represent idealized scenarios when the error is primarily from the measurement error and, for pulsar TOA directly and optical indirectly, source location errors. These simplifications necessarily result in optimistic solutions as they do not include the myriad of other error sources that impact solutionw quality. But what can be inferred from them are the relative geometric and measurement system strengths. For instance, range data has the smallest measurement noise, but the worst geometric strength (as measured by $\sqrt{G_{xx}}$ and$\sqrt{G_{yy}}$); while pulsar TOA has the best geometric strength and, for the bulk of the observed pulsars, the worst measurement noise. To obtain more accurate results one needs to consider more error sources in the context of a realistic trajectory towards a destination of interest. Nonetheless, these results do provide a measure with which to compare the relative performance when in deep space cruise using the different tracking methods. Indeed, one conclusion might be that using pulsar TOA with the most stable small set of pulsars yields the best performing orbit determination. However, consider the PDOP behaviors for ranging and pulsar TOA after 14 days of tracking, the ranging PDOP improves by about an order of magnitude while the pulsar result is already near its best levels; hence, even at Neptune distances, OD from ranging or pulsar TOA measurements are at commensurate levels. Furthermore, one might conclude that optical cannot compete with the other choices; however, optical imaging is the only target relative measurement in the set so as the spacecraft nears its destination optical imaging would begin to provide superior results relative to the others. This hasn't been captured by the analytic analysis, but will be characterized in Part 2 with high-fidelity modeling using realistic errors and approaching a target destination.

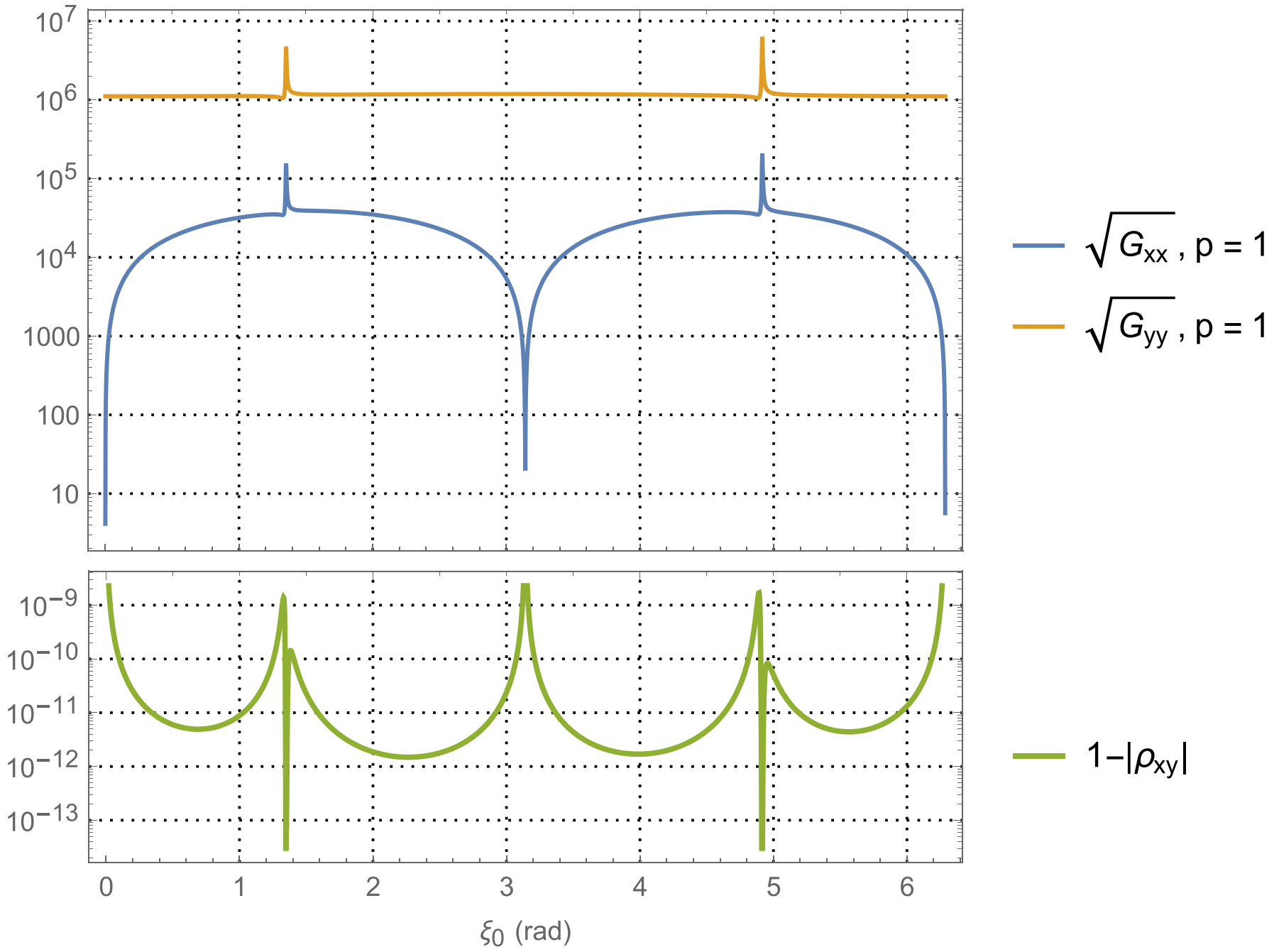

**Fig. 16** Normalized position component uncertainties for ranging measurements at Neptune distances using a one-day tracking period (upper) and the correlation deviation away from 1 between components (lower) as a function of the initial Earth angle

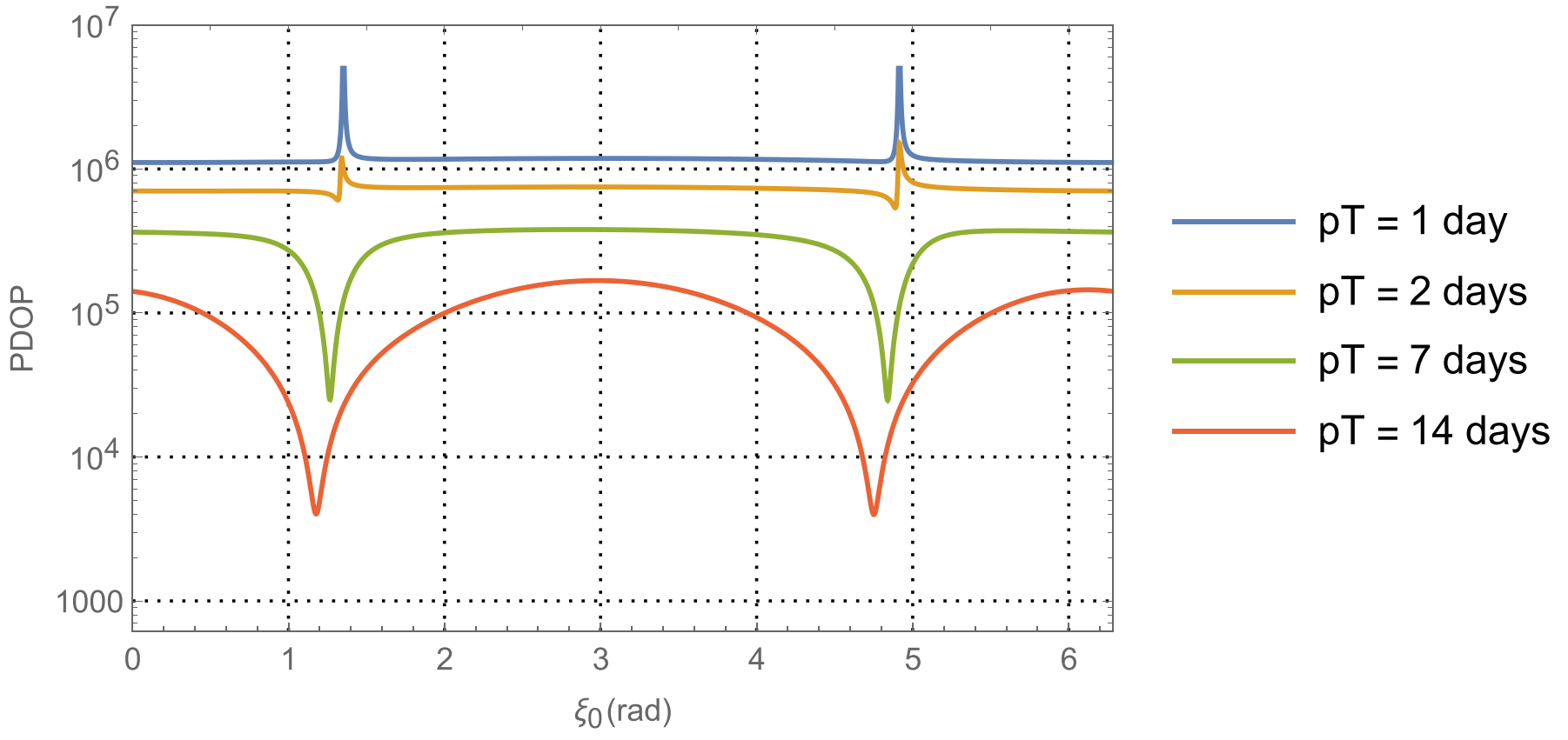

**Fig. 17** PDOP using range measurements for the Neptune case with one day, two days, seven days, and fourteen days of tracking

**Table 3** Comparing system measurement uncertainties, geometric scale factors, and analytic initial position uncertainties estimated using one day of tracking use on of optical imaging, pulsar TOA, or range data at Mars or Neptune distances

| Distances | Measurement Scenario | Measurement Uncertainty at 1-day $\sigma(T)$ km | Position Uncertainty Scale Factors $\left(\sqrt{G_{xx}}, \sqrt{G_{yy}}\right)$ | Initial Position Component Uncertainties at 1-day $(\sigma_{xx}, \sigma_{yy})$ km |
|---|---|---|---|---|
| Mars | Optical/Main Belt Images | 14.8 | (3.2, 3.2) | (47, 47) |
| | Pulsar/SEXTANT | 24.1 | 3 | (72, 72) |
| | Pulsar/Best 4 | 7.1 | 3 | (21, 21) |
| | Ranging | $6.1\times10^{-5}$ | $3 < \sqrt{G_{xx}} < 41,000$<br>$16,300 < \sqrt{G_{yy}} < 98,000$ | $0.0002<\sigma_{xx}<2.5$<br>$1<\sigma_{yy}<6$ |
| Neptune | Optical/KBO Images | 1541.6 | (1.7, 1.7) | (2621, 2621) |
| | Pulsar/ SEXTANT | 436.5 | 3 | (1310, 1310) |
| | Pulsar/Best 4 | 11.4 | 3 | (34, 34) |
| | Ranging | $8.2\times10^{-4}$ | $3 < \sqrt{G_{xx}} < 49,000$<br>$1,004,290 < \sqrt{G_{yy}} < 1,200,250$ | $0.0002<\sigma_{xx}<2.5$<br>$820<\sigma_{yy}<980$ |

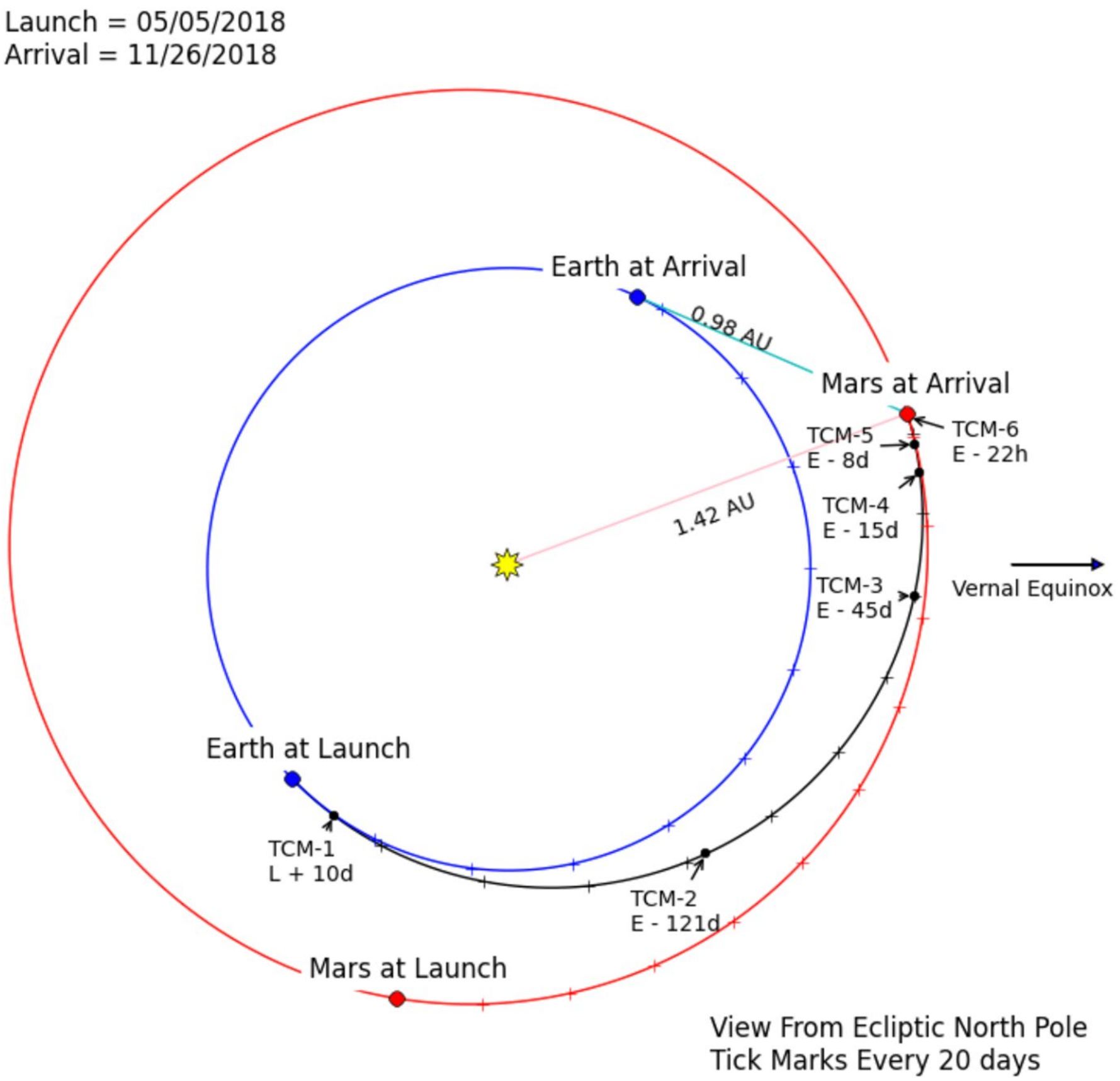

**Fig. 18** InSight trajectory for launch on May 5, 2018 from Abilleira. [23] The period of investigation in this study is the last 45 days beginning after TCM-3 and ending at entry on Nov 26, 2018

## Part 2: High-Fidelity, Quantitative Analysis of Mars Cruise, Approach, and Entry Navigation

The preceding analytical analysis of range, optical imaging, and pulsar TOA measurements in Part 1 provided rough order of magnitude estimates for the relative capabilities of each data type using idealized assumptions. We now turn to a high-fidelity analysis that considers realistic errors (both dynamic and measurement) in the context of a dynamic deep space navigation problem. This will be a full dimensional problem with 3-d position and 3-d velocity estimation plus estimation/consideration of other model parameters. We select the late-cruise, approach, and entry navigation of the recent Mars InSight lander as a representative (and extensible) use case for quantitatively assessing the relative merits of navigation using onboard uplink radiometric tracking, optical imaging, and/or pulsar TOAs. We examine this case because it is a mission that demonstrates the state of the art in deep space navigation, and there is a wealth of recent data for use in validating our results. Our truth simulations use the models and error assumptions that the InSight navigation team utilized in their navigation system design as documented by Abilleira [23] with the exception that the reaction control system (RCS) small forces values that were used in the current study came from earlier InSight models. The nominal InSight trajectory for launch on May 5, 2018 is shown in Fig. 18. We have selected the last 45 days (after TCM-3) as the period for our investigation since this represents the most dynamic phase of the trajectory where specific entry flight path angle constraints need to be achieved. In particular, InSight's delivery and knowledge requirements were:

1. Delivery entry flight path angle (EFPA) should be $-12° \pm 0.21°$ (3-σ),
2. Knowledge of the delivered EFPA must be within $\pm 0.15°$ (3-σ).

Using simulated truth trajectories and measurement data derived from the InSight models, we will be able to compare the absolute and relative navigation capabilities that the three data types are able to provide individually and in selected combinations. We will augment the radiometric, optical, and pulsar TOA data models presented in Part I with the most significant errors that impact their accuracy and, thus, their utility for navigation. We start with selecting the navigation filter approach and design that is appropriate for the current study.

### Filter Algorithm Selection and Filter Design

Because our current focus is on comparing data types and *not* the model reductions needed for an actual onboard application, we will use the highest-fidelity models and the best calibration data assumptions that are available for the analysis. These are the same assumptions that a ground-based navigation team would use, in the present case for InSight navigation. The navigation filter will also be optimally tuned for all the model errors that are present, use all available calibration data, and not require expanded filter compensation for reduced-order models (as would be the case for a true onboard implementation). For comparison, a recent paper by Ely, et al. [42] thoroughly examined the onboard case with model simplifications and reduced calibration modeling (i.e., sub-optimal with filter compensations) that are sufficient for an autonomous navigation system with radio and/or optical data that supports this Mars cruise, approach, and entry problem. In the present paper, use of optimal modeling with matched filter process noise (i.e., the same that would be used by a ground-based navigation team) will allow for better identification of the relative merits of one-way radiometric, optical imaging, and pulsar TOA data without 'clouding' the interpretation of the results that reduced-order models and calibration data predicts would introduce.

The state-of-art navigation filter method, with decades of successful use for deep space navigation, is the batch sequential linearized Kalman filter (LKF) with iteration using fixed-interval backwards smoothing. [43, 44] Note, iteration is defined as incorporating the LKF solution for an initial state (based on processing an arc of data in a forward pass) into the reference initial state, repropagating the trajectory and variational equations, and then reprocessing the data arc to determine a new trajectory solution. The process repeats to convergence or after a specified number of iterations. To ensure numerical stability, the filter must be implemented in a factorized form such as UD or in an upper-triangular Square Root Information Array (SRIF). These methods are well known and have been thoroughly documented in Bierman [43] and implemented in Monte [45] that has been used for this analysis and operationally for many NASA missions. It should be noted that the arrival of multiple measurements, at differing times, and with gaps requires modifications to the typical Kalman filtering algorithms, which has been documented in Ely [46] and utilized in this study.

## Spacecraft Dynamic Model and Solar System Modeling Considerations

We now examine the spacecraft dynamic perturbations that impact the overall n-body solar system gravity forces affecting the motion of the spacecraft. These model effects will apply to all the navigation scenarios that will be examined. Additionally, we discuss the solar system modeling including ephemeris assumptions for the planets and relevant moons and asteroids prior to discussing models for each data type.

### Solar Pressure and Unbalanced Reaction Control System Effects

The pre-launch high-fidelity InSight shape model consists of many components that require their own reflectivity parameters and nominal orientations.[5] The shape model has five components representing the solar arrays, the launch vehicle adapter on the cruise stage, the cruise stage outer ring, and the backshell (two components). The InSight cruise configuration is illustrated in Fig. 19. The models include diffuse and specular reflectivity coefficients for each component. For the period being investigated, InSight maintains its orientation such that the normal to the solar arrays is oriented towards the Sun while still allowing access to the Earth via the MGA.

[5] During actual flight, the InSight navigation team switched the SRP model from a component model to one based on spherical harmonics.

Because the InSight cruise stage maintains attitude with a three-axis unbalanced thruster control system, there are numerous (almost continuous) small translational forces perturbing the trajectory resulting from this attitude maintenance. [23] Like SRP, these small forces impact not only the trajectory, requiring periodic trajectory correction maneuvers, but the trajectory's associated knowledge errors as these small forces degrade overall trajectory estimation and prediction. The InSight project, via experience with the Mars' Phoenix mission that used a similar spacecraft, generated a model with aggregate bias accelerations in each spacecraft body fixed axis direction. The initial bias values were uncertain as well as the associated thrust direction. On top of these initial uncertainties, the acceleration magnitudes varied stochastically. The truth acceleration profile used in this study is an earlier version used by the InSight navigation team. Details of these accelerations and the other spacecraft truth models are listed in Table 4.

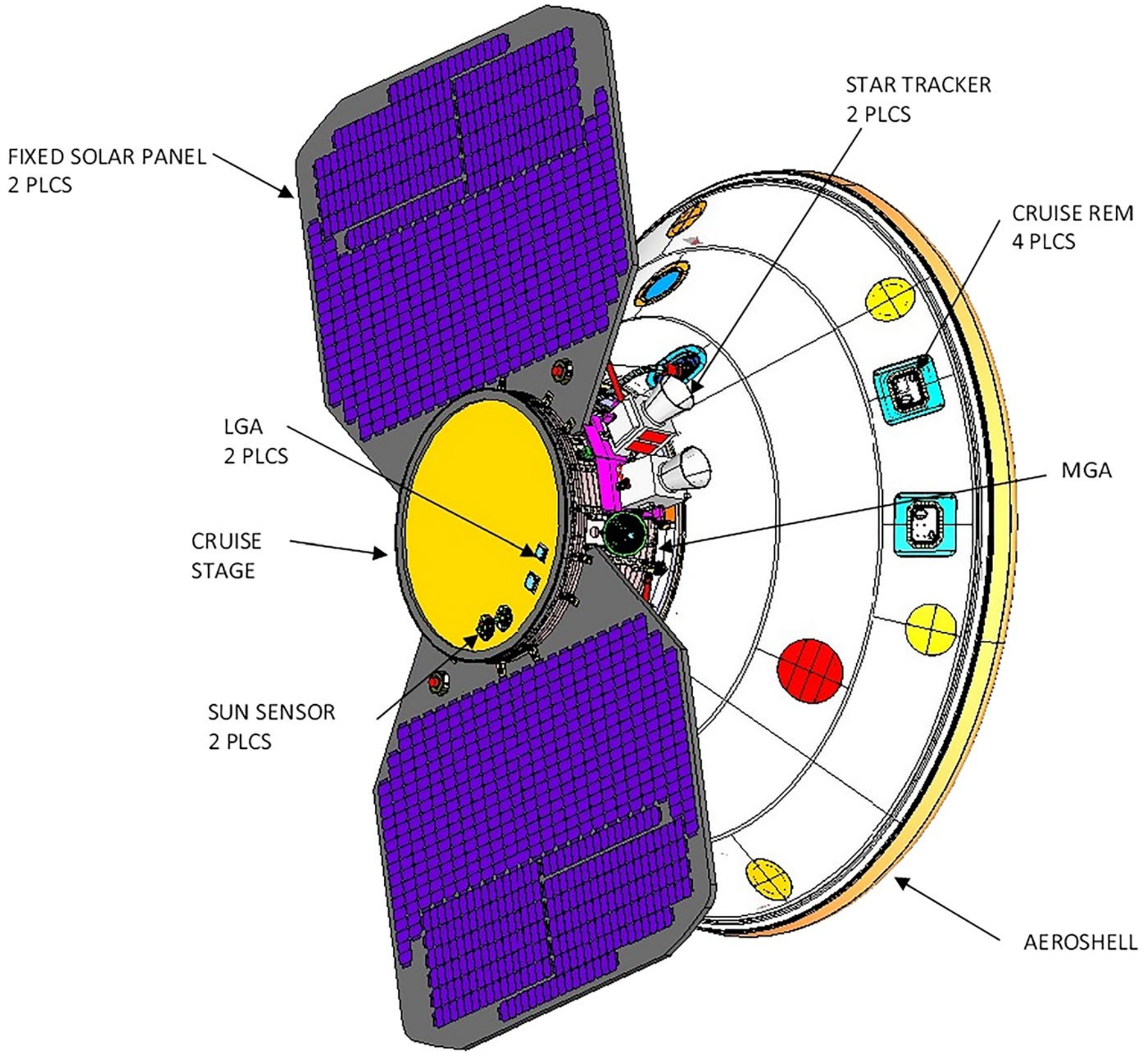

**Fig. 19** InSight Cruise Configuration. [23]

**Table 4** Truth and Filter Dynamic Models for Mars Approach Spacecraft

| Parameter | Reference Model/Comments | Injected Truth Error (1σ) | Filter Model (1σ) |
|---|---|---|---|
| Initial Position | Targeted for each simulated truth trajectory to achieve Mars entry conditions. | 100 km each component | 1000 km each component |
| Initial Velocity | Targeted for each simulated truth trajectory to achieve Mars entry conditions. | 1 cm/s each component | 1 m/s each component |
| Solar Pressure Model | s/c with 5 shape components and separate reflective properties | Scale Factor (SF) Bias = 10%<br>SF ECRV $\sigma$ = 3% with $\tau$ = 7 days<br>$h$ = 1-day | Estimate bias and stochastic at truth error levels |
| Small Forces | S/C x-accel = 6.6 x $10^{-12}$ km/s$^2$<br>S/C y-accel = 8.8 x $10^{-14}$ km/s$^2$<br>S/C z-accel = −3.9 x $10^{-14}$ km/s$^2$ | Accel SF Bias = 3%<br>Accel Pointing Bias = 3°<br>Accel SF White Noise = 5%<br>$h$ = 1-day | Estimate biases and stochastic at truth error levels |
| Trajectory Correction Maneuvers | Realized delta-V varies with each trajectory. | 8.95 mm/s magnitude additive bias<br>0.667% magnitude SF bias<br>13.5 mm/s direction additive bias<br>0.00472 rad pointing bias | Estimate biases at truth error levels |

ECRV = Exponentially Correlated Random Variable or, equivalently, 1st – order Gauss-Markov Process, $\tau$ = correlation time, $h$ = discrete noise interval between realized value, or the interval for applying process noise in the filter (i.e., the batch length).

### Trajectory Correction Maneuver (TCM) Targeting and Errors

In the period being investigated (i.e., the last 45 days prior to entry) there are three maneuver opportunities: TCM-4 at E-15 days, TCM-5 at E-8 days, and TCM-6, the final maneuver, at E-22 h. These maneuvers are needed to correct for all of the trajectory perturbations that result from modeling errors with solar pressure, small forces, orbit knowledge errors and execution errors propagating from prior TCMs. Since the current studies focus is on orbit determination and not on the most efficient targeting method, the selected targeting algorithm used for determining the TCMs that guide the separate truth trajectories to achieving the entry condition is a robust, capable nonlinear least squares optimization algorithm developed by Hanson and Krogh. [47] For simplicity, the targeting is fixed-time, which usually results in higher delta-V costs (more on this later when discussing EFPA results).

### Other System Models

The other solar system models that an onboard navigation system must use when propagating trajectories, computing observables, and other relevant quantities (such as Mars entry target conditions) include planet, satellite, and asteroid ephemerides and their associated body properties (like reduced mass). Of course, the knowledge of these quantities is imperfect and, as with the other errors discussed previously, they have been injected into the truth simulation. Note that the imaging campaign with the OpNav camera includes 37 different main belt asteroids. These were arrived at using the heuristic asteroid selection algorithm developed by Broschart [17] as being an optimal set for this trajectory that factors brightness, camera quality, and location plus associated uncertainties. Additionally, Phobos and Deimos, Mars' moons, are imaged – this becomes especially valuable as the spacecraft nears Mars since these images provide strong target-relative information. Table 5 has a list of the models, errors, and asteroids. For the pulsar TOA measurements, the relevant pulsar information is identified in Table 2 including the measurement noise (per pulsar) and pulsar location error. For both the optical data and the pulsar TOA data, the relevant body location errors are considered by the filter.

## Higher Fidelity Model Considerations for the Radiometric, OpNav, and Pulsar TOA Data

### One-Way CPF-Phase Model and Campaign

The one-way charged-particle–free phase or CPF-phase is a suitable data type for onboard use because it eliminates charged-particle effects from the Sun and the Earth's ionosphere on the measured path lengths (one of the most significant error sources impacting radiometric tracking data). Furthermore, when derived using a high precision clock such as DSAC, it retains the same level of accuracy as its ground-based two-way counterparts. It is formed using a linear combination of the traditional one-way uplink range $R(t)$ and total count phase $\Phi(t)$

**Table 5** Truth and Filter Models for Solar System Bodies

| Parameter | Reference Model/Comments | Injected Error (1σ) | Filter Model (1σ) |
|---|---|---|---|
| Earth/Mars Ephemeris | JPL DE430 | DE430 correlated covariance for Earth and Mars bias error | Consider at truth error levels |
| Mars GM | DE430 value | 2.8 x $10^{-4}$ $km^3/s^2$ | Estimate bias at truth error levels |
| Phobos/Deimos Ephemeris | JPL MAR097 | 0.5 km in each position component | Consider at truth error levels |
| Asteroid Ephemeris | PDS asteroid database for the following asteroids: 2,000,258, 2,004,483, 2,000,140, 2,000,269, 2,001,550, 2,002,577, 2,005,142, 2,002,839, 2,001,432, 2,001,946, 2,000,030, 2,001,627, 2,000,070, 2,000,043, 2,000,172, 2,000,173, 2,000,694, 2,000,951, 2,025,916, 2,001,987, 2,000,198, 2,001,224, 2,000,204, 2,001,235, 2,000,852, 2,000,598, 2,000,475, 2,000,606, 2,000,352, 2,000,353, 2,002,406, 2,001,006, 2,000,112, 2,000,498, 2,000,115, 2,001,147, 2,000,253 | PDS asteroid database correlated covariance for each asteroid | Consider at truth error levels |
| Pulsars | See Table 2 | See Table 2 for both measurement noise figures and location errors | Consider pulsar location at truth errors levels |

$$\Phi_{\text{CPF}}(t) \equiv \frac{R(t) + \Phi(\text{t})}{2} \tag{125}$$

For the error effects included in the current analysis, a useful model for the CPF-phase can be represented as

$$\Phi_{\text{CPF}}(t) = p(t) + c\left[x_R(t) - x_T(t-\tau)\right] + T(t) - \frac{N}{2} + \frac{b_R(t)}{2} + \frac{v_p(t) + \varepsilon_p(t)}{2} \tag{126}$$

and, since this is an onboard measurement, each measured CPF-phase value will be time-tagged with the onboard clock's value of time, represented as

$$C(t) = t + x_R(t) \tag{127}$$

In Eq. (126), $\rho(t)$is the geometric path length between the receiving antenna's phase center and transmitting antenna's phase center, $x_R$ is the receiver clock's time deviation from $t$, $x_T$ is the transmitter clock's time deviation from $\boldsymbol{t}$, $T(t)$ is the troposphere delay, $N$ is the total count phase ambiguity, $b_T(t)$ is the range bias (which can change on a per pass basis), and $v_\rho(t)$ and $\varepsilon_\rho(t)$ are the range and phase measurement noise, respectively. Other error sources may be present, such as multipath, but are not considered in the present analysis. Note that because the time-tag is sensitive to the onboard clock phase error $x_R$, the measured CPF-phase value will be senstive to clock errors directly, via Eq. (126), and indirectly, via Eq. (127). Correct measurement partials need to account for this.

Thorough discussions of this data type, associated errors, and its use for autonomous onboard navigation are presented in Ely [7, 48]. The geometric path length $\rho(t)$ is computed accurately using an iterative algorithm, as documented in Moyer [49]. This computation includes the full n-body, relativistic gravity effects on path length and times, precision Earth orientation data and Earth station locations, and accurate spacecraft trajectory modeling that accounts for the significant forces affecting the spacecraft's motion (multi-component shape models with reflectivity coefficients for solar radiation pressure, small forces from an unbalanced reaction control system, maneuvering errors).

Note that by construction, neither charged-particle delays from the ionosphere nor solar corona plasma effects are present in Eq. (126), which eliminates the need to calibrate for these at the expense of an increased overall noise relative to the phase by itself (DSN-based range measurement noise typically is 1 to 3 m (1-σ) while the phase noise is 5 mm (1-σ)). The effects of the phase ambiguity term can be minimized to the level of the range error via using the range measurement value at the beginning of a pass to calibrate for the ambiguity.

Selecting a clock such as DSAC that has a stability on par with ground-based atomic clocks enables simple clock offset and rate bias estimation in the navigation filter after some form of initial clock calibration activity with the ground has occurred. This will be assumed for all the data types studied in this paper. For other choices of clocks (such as a USO or a CSAC), this would not be the case and specific stochastic process filter compensations would be required with periodic clock recalibrations needed. After correlating for the spacecraft clock, a conservative bound for the range bias uncertainty (including charge particle delays at X-band) is 3 m (1-sigma).

The CPF-phase measurement models including needed calibration modeling of related Earth parameters and associated filter models are delineated in Table 6.

Turning to the tracking campaign, traditionally in the late cruise near 45 days prior to entry of a Mars lander, DSN tracking is increased to continuous 24/7 support. That is, as each of the DSN stations rotates into view of the lander, a DSN antenna at the complex will collect two-way Doppler and range data between the station and the spacecraft. Typically, only one antenna is in view at a time. However, when two DSN antennas at different complexes are in view of the spacecraft, this data can be augmented with Delta Differenced One-Way Range (DDOR) data (providing 'plane of sky' information to complement the 'line of sight' data from range and Doppler).

For our onboard navigation case, we replace the two-way range and Doppler with uplink-only one-way CPF-phase and start with the continuous level of support, but then consider the impact of reducing the DSN support to two hours of transmission from one station per day (vs 24/7 support). Since DDOR is a ground-based data type, it is not considered – instead we will examine the onboard optical data and/or the pulsar TOA data as a complementary data type. We include only uplink one-way CPF-phase tracking from the in view DSN stations (using a traditional 10° elevation mask). The selected DSN antennas include DSS-15 at Goldstone, California, DSS-45 at Madrid, Spain, and DSS-65 at Canberra, Australia for the case with continuous tracking.

### Onboard OpNav Imaging Models and Campaign

For the current study, we have selected a gimballed, high-end camera system (later we pair this with the CPF-phase measurements) with the imaging and measurement noise characteristics identified in Table 1. When the camera images an asteroid, it will be projected onto its 2-d focal plane that is then digitally sampled by an array of detectors. This results in a set of sample (horizontal) and line (vertical) coordinates for the center of figure of all the detectable asteroids (and other bright bodies) in the camera's field of view. The sample/line coordinates of an asteroid, coupled with asteroid ephemeris knowledge, and camera pointing can be used to determine the spacecraft position relative to the imaged bodies; thus, using it for spacecraft navigation. The optical measurement model adapted for use is documented by Owen [50] and is part of JPL's operational Monte navigation software system. [45, 51, 52] The use of the gimbal enables the spacecraft to point the camera without changing attitude, thus enabling the spacecraft to image as needed without interrupting other operations. Prior operational experience with optical data has shown that, in most cases, the image pointing direction can be ascertained using the star background to a micro radian accuracy. This has been assumed in our simulation with pointing errors injected at this level. Use of this approach also implies that separate attitude information is not required to process the image; however, in the event that orientation determination is not possible an explicit attitude interface to the spacecraft would be needed. Also, for this study we are only imaging bodies that appear as point sources in the focal plane (i.e., they are distant enough that the center of figure can easily be determined within a pixel). The selected asteroids are identified in Table 5 with ephemeris errors injected in the simulated truth trajectories and, in the filter, they are considered. Phobos and Deimos trajectory errors are also injected and the filter estimates theses trajectory offsets as biases.

**Table 6** Truth and Filter Models for CPF-phase, and Other Associated Earth-centric Model Effects

| Parameter | Reference Model/Comments | Injected Error (1σ) | Filter Model (1σ) |
|---|---|---|---|
| One-way phase noise | | White noise at 4.24 mm | Same |
| One-way range noise | | White noise at 1 m | Same |
| One-way range Bias | | 2 m bias applied at start of each pass | Estimate as a Per-Pass Stochastic at truth error levels |
| Clock phase Bias | Calibrated DSAC | $1\times10^{-6}$ s (assumes clock calibration) | Estimate biases at truth error levels |
| Clock frequency bias | Calibrated DSAC | $1\times10^{-14}$ (assumes clock calibration) | Estimate biases at truth error levels |
| Clock stochastic frequency | DSAC stochastic model | White Frequency (WF) Noise Sequence with AD = $3\times10^{-15}$ @ 1-day | Not Estimated |
| Earth troposphere delays | High precision daily and long period calibration data | Daily Dry Delay ECRV $\sigma=1$ cm with $\tau=6$ h<br>Daily Wet Delay ECRV $\sigma=1$ cm with $\tau=6$ h $h=1$ h | Estimate stochastic at truth error levels<br>Estimate stochastic at truth error levels |
| Earth UT1 errors | High precision daily EOP calibration data | ECRV $\sigma=5\times10^{-5}$ with $\tau=6$ hrs, and $h=1$ h | Estimate stochastic at truth error levels |
| Earth (X, Y) pole motion errors | High precision daily EOP calibration data | ECRV $\sigma=1.6\times10^{-9}$ radians with $\tau=2$ days, and $h=1$ h | Estimate stochastic at truth error levels |
| Station Locations | Surveyed DSN station locations | Bias error from sampling fully correlated 2003 covariance | Estimate biases at truth error levels |

For the gimbaled camera's imaging campaign, there are four separate cadences that have been programmed. Beginning at E-45 days, 9 to 10 different asteroids are imaged every 5 days and Phobos and Deimos are imaged every day. At E-12 days, the asteroid imaging frequency increases to 10 every day while Phobos and Deimos also continue to be imaged daily. Then at E-8 days, the frequency increases to every 6 h. Finally, at E-2 days it is every hour until entry. The images of one target are taken in bursts of 5 images as aid in reducing measurement noise.

### Pulsar TOA Modeling and Campaign

High fidelity pulsar TOA observables using photon arrival timing, counting, and folding models suitable for the Earth orbiting SEXTANT experiment have been developed in Winternitz [4]. This work identifies some of the most significant error sources effecting the precision and accuracy of the pulsar TOA measurement beyond the pulsar source signal noise, and include:

1. Pulsar timing noise and glitches,
2. Uncertainties between the radio and x-ray model phases,
3. Astrometric errors including pulsar location and proper motion,
4. Mismatches in the source flux rates,
5. Reference clock errors,
6. Earth ephemeris error impacts on the source template signature.

Similar observation models using photon arrival timing, counting, and folding have also been examined by Emadzadeh [37] and Chen [5]. These references develop processing algorithms that account for the fact that the folding period on a moving spacecraft is Doppler shifted and determining the proper period is dependent on having accurate spacecraft velocity knowledge a priori. In the present study, we will focus on two significant error sources (in the list above) that impact the accuracy of the measurement beyond the fundamental pulsar source signal noise, these include the previously discussed pulsar location errors and the reference clock errors. By construction, the pulsar TOA model in Eq. (26) already includes sensitivity to pulsar location errors, but the addition of local clock errors do change the form of the model to the following

$$\left(\widetilde{c\tau}\right)(t) = \mathbf{r}(t)^T \hat{\mathbf{n}}_{P_k} + c x_R(t) + v_\tau(t) \tag{128}$$

where the onboard reference clock now appears directly, as was the case with CPF-phase, and the measured pulsar TOA value will be time-tagged with the local clock's reading of time $C(t)$. Note that in the present analysis, we are not considering proper motion errors so the unit vector to the observed pulsars $\hat{\mathbf{n}}_{P_k}$ remain constant but have unknown bias errors that have the $1-\sigma$ uncertainties given in Table 2.

The pulsars selected for the analysis are the same as those observed by SEXTANT and are identified in Table 2 with a subscript 'S' and a subset of these labeled the 'best' 4 (J0437+4715 in Q1,[6] B0833–45 in Q2, B1821–24 close to Q3, and B1937+21). As with the analytical analysis, we assume a gimballed X-ray detector that can rotate its observation focus consecutively and continuously between the selected pulsars. For the current analysis, each pulsar is observed for 30-min at a time to generate a TOA measurement conforming to the model in Eq. (128), and then the detector will move to the next SEXTANT pulsar in the table with a higher right ascension value. Since this is a full 3-d position space problem, the gimbal will also point the detector to the pulsar in all directions including the out-plane coordinate. Once the end of the selected set is reached, the observation schedule repeats. The result is a 45-day simulation of continuous pulsar observations with 30-min spacing between measurements. Finally, the X-ray detector is the same as assumed in the analytical analysis with a detector area of 129 cm$^2$.

## Results

### CPF-Phase Tracking Only

We begin by presenting the Mars approach and entry current state navigation errors and uncertainties when using only CPF-phase data (no OpNav or pulsar TOA data present) when there is a continuous broadcast of uplink carrier and range signals from whichever of DSS-15/45/65 is in view. The Monte Carlo simulation includes 200 realizations of truth trajectories and the associated set of CPF-phase observables. In each realization, the filter model starts with the same nominal reference trajectory, processes that realization's data, and iterates for 3 times. The final iterated current state position errors and associated uncertainties in each position component and associated RSS are presented in Fig. 20. Shown are the 1-σ and 3-σ uncertainty bounds as well as the filter solution error obtained by differencing the filter solutions (+ reference trajectory) with the truth trajectory. We see trajectory uncertainties reach a maximum at 950 h with 1-σ and 3-σ uncertainties of 20 km and 60 km, respectively. Also listed in the results are statistics of the percentage that a filter solution exceeds the 1-σ and the 3-σ bounds. In the present example, this is 6% and 0.0%, respectively, from a population of 649,200 samples. Assuming ergodic processes and statistics that conform to a normal distribution, one expects 1-σ exceedances to be <31.7% of the time and 3-σ exceedances to be <0.3% of the time, which is true for this case. Another observation is as the spacecraft nears Mars, the errors

[6] While the PSR B0531+21 (M1 - Crab Pulsar) has lower noise figures and position knowledge, it is also prone to sporadic noise glitches. As a result, J0218–4232 was selected as more representative of a stable population.

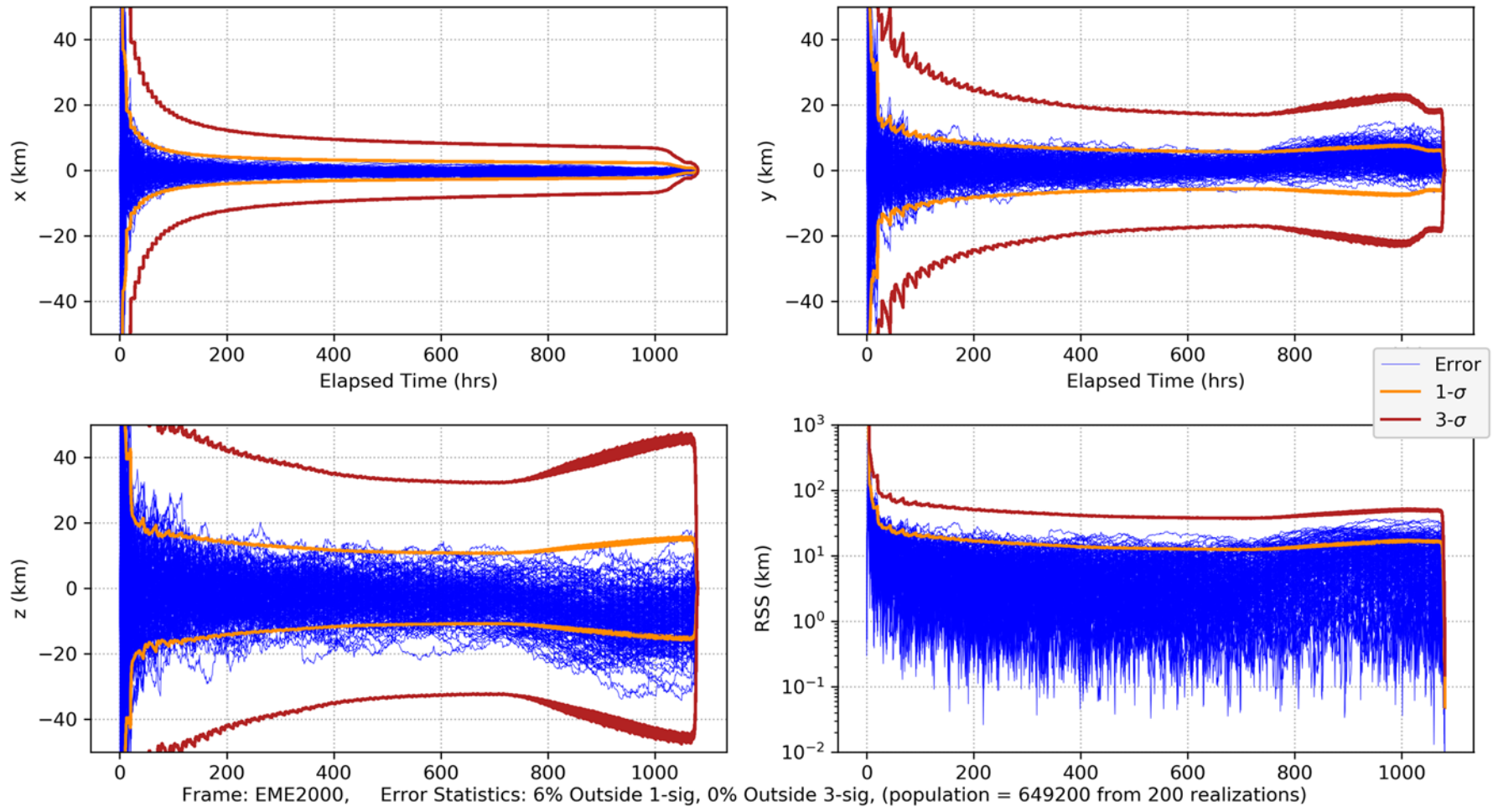

**Fig. 20** Mars approach and entry position errors for 200 realizations and uncertainties (1-sigma dark orange, 3-sigma brickred) with only uplink one-way CPF-phase data from the DSN

grow until the final few hours when the uncertainties (and errors) drop to less than 150 m. The growth in the error/uncertainty is due primarily to execution errors with TCM 4–6 beginning approximately 700 h into the run. This is more clearly evident with the three discrete TCM events seen in the period after 700 h in the velocity error and uncertainty plots shown in Fig. 21.

In Fig. 22 the entry flight path angle (EFPA) is shown as a function of a data cut off (DCO) time (with the solution at that time propagated to the entry condition

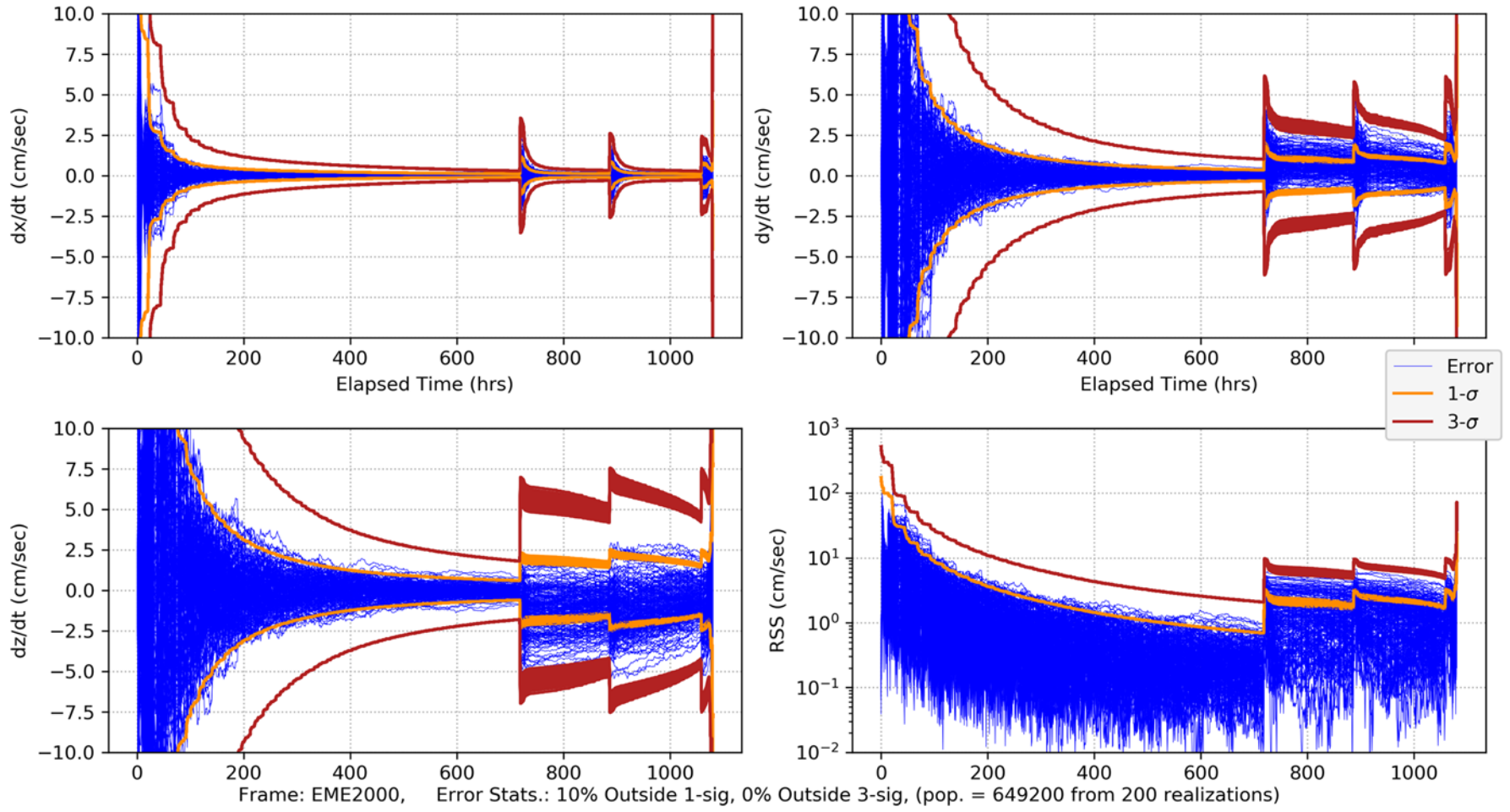

**Fig. 21** Mars approach and entry velocity errors for 200 realizations and uncertainties (1-sigma dark orange, 3-sigma brickred) with only uplink one-way CPF-phase data from the DSN

for computing the EFPA and associated uncertainties). The delivery and knowledge requirements for InSight listed previously (both the DCO time and the associated EFPA uncertainty required at that time) are drawn. At this point it is important to make a distinction between InSight's DCO requirements when using ground-based data processing versus onboard autonomous navigation processing. Because of light time delays and associated ground processing delays, the ground processing DCOs for both delivery and knowledge are set to accommodate these delays; however, a fully autonomous onboard system's DCOs could be set much closer to the key events that dictate the delivery and knowledge errors. In particular, for delivery the onboard navigation solution is needed prior to executing TCM6, and for knowledge the onboard navigation solution is needed at entry. So when evaluating the utility of the measurement types for meeting the EFPA requirements, the distinction of when the DCO applies is different for onboard processing versus ground-based processing as summarized in Table 7.

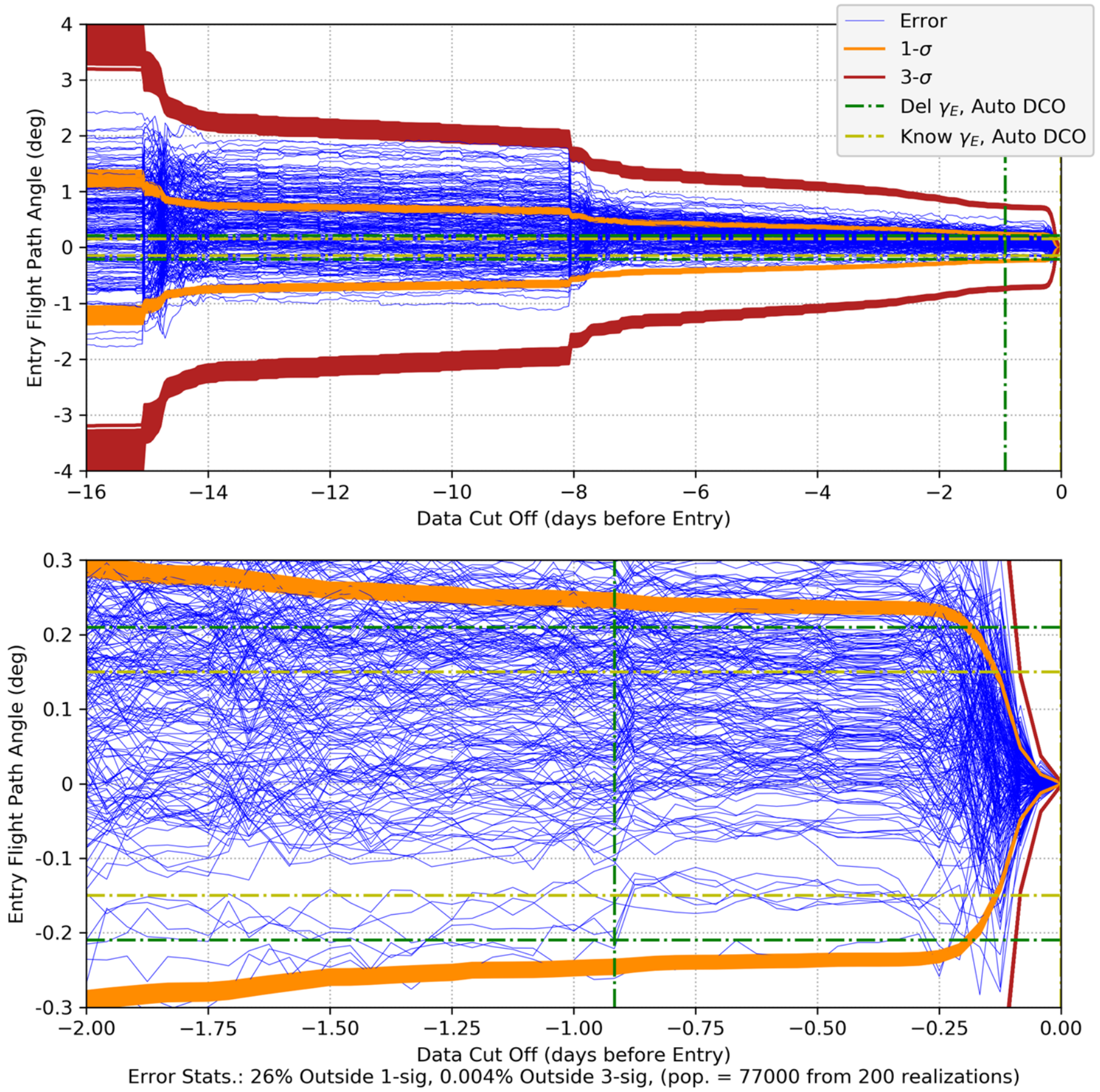

**Fig. 22** Mars entry flight path angle as a function of data cut off prior to entry with CPF-phase only tracking. Top plot spans from entry-16 days to entry. The bottom plot is for the final two days. Note that the vertical green dash-dot line represents the onboard DCO (just before TCM-6) for the delivery EFPA requirement and the vertical yellow dash-dot line represents the DCO (at entry) for the knowledge EFPA

In all of the EFPA figures to be presented, the DCO constraints for the onboard autonomous case have been plotted with green dash-dot vertical lines at the TCM6 time for the delivery DCO, and yellow dash-dot at the entry time (far right of the plot) for the knowledge DCO. The associated 3-σ EFPA uncertainty requirements (denoted as $\gamma_E$ in the legend) for delivery at $\pm 0.21°$ and knowledge at $\pm 0.15°$ are also identified with dash-dot horizontal lines using the same colors, respectively. In Fig. 22, it is clear that navigating with CPF-phase measurements alone does not meet the delivery requirement; however, the knowledge requirement is easily met with the 3-σ EFPA uncertainty bounds achieving values near $\pm 0.01°$, nearly an order of magnitude better than required. Of course, the goal with precision entry knowledge is for an onboard guidance system to fly out delivery errors as an aid for precision landing. This level of state knowledge would enable an onboard navigation system to know its actual flight trajectory sufficiently well, that, when paired with sufficient maneuverability, could return the vehicle to its desired flight path. Reducing the delivery error and associated uncertainty could be accomplished via pairing with another measurement technique. This is the case with ground-based navigation, that always pairs 2-way range and Doppler measurements with double differenced one-way range (DDOR) measurements to achieve acceptable delivery errors for Mars entry. After examining optical and pulsar TOA navigation separately, pairings of CPF-phase with optical and CPF-phase with pulsar TOA measurements will be examined.

### Optical Imaging Only

We turn now to the optical-only case using a gimballed high-end camera that is imaging selected asteroids as well as Phobos and Deimos while enroute to Mars. The position and velocity errors with associated 1-sigma and 3-sigma uncertainties are shown in Fig. 23 and Fig. 24, respectively. Recall with the analytical analysis, the effect of the asteroid ephemeris uncertainty was sufficiently small to not significantly affect the overall uncertainty, in the present analysis this remains true as well. However, it is necessary to explicitly consider the asteroid locations to ensure that the filter maintains solution convergence and that the solution statistics remain consistent with the formal uncertainties. This increases the complexity of the onboard filter, but is still tractable. It is clear that the filter errors are consistent

**Table 7** Delivery and Knowledge Nav State Data Cut-Off Times for Ground-based and Onboard Processing

| | Delivery Nav State Data Cut-Off | Knowledge Nav State Data Cut-Off |
|---|---|---|
| Ground-Based Navigation | TCM6–1 day | Entry – 6 h |
| Onboard Autonomous Navigation | Just before TCM6 | Entry |

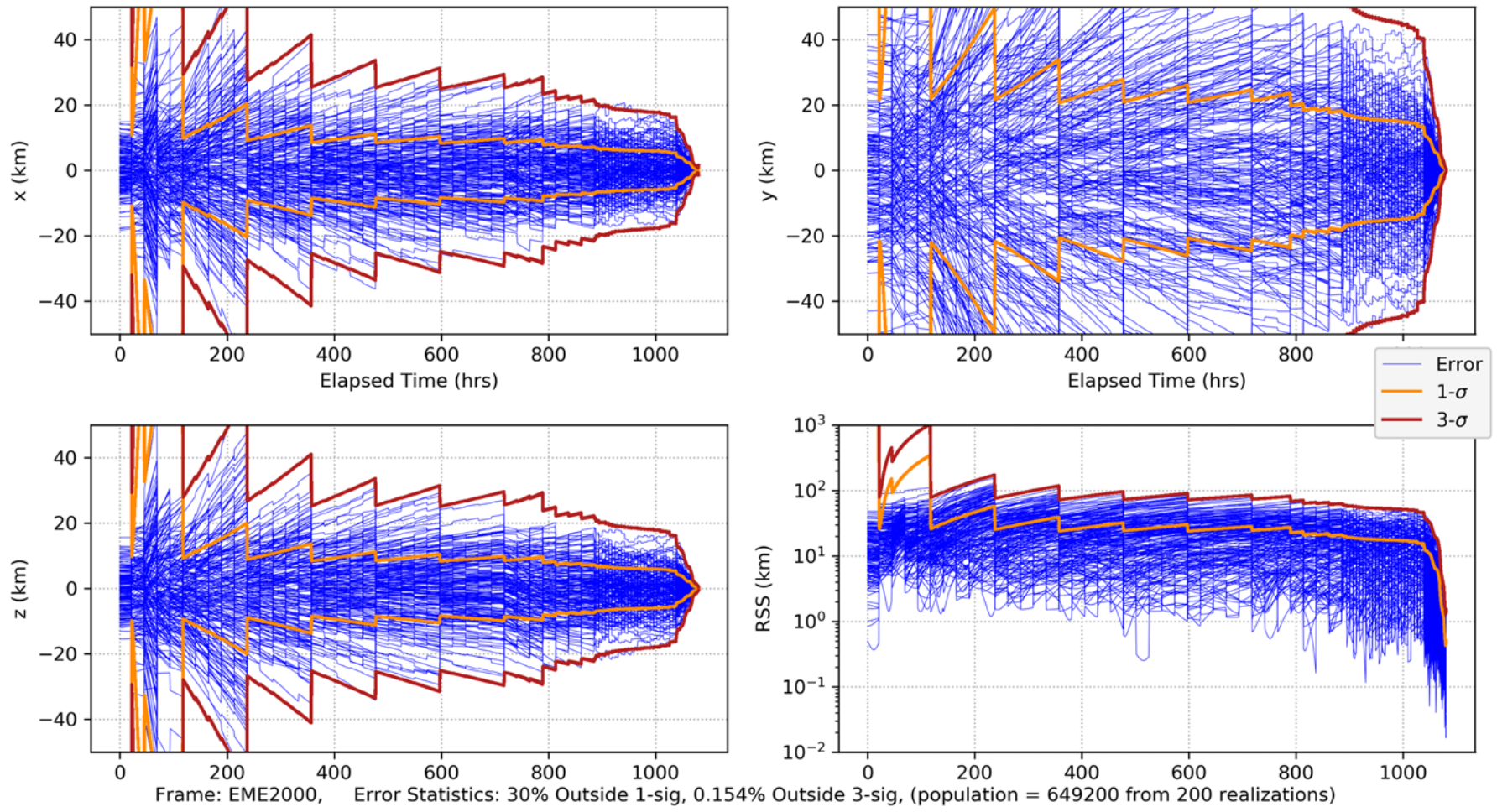

**Fig. 23** Mars approach and entry position errors for 200 realizations and uncertainties (1-sigma dark orange, 3-sigma brickred) with only optical imaging of main belt asteroids and Phobos and Deimos

with the formal uncertainties with exceedances well below the limits imposed by normal distribution assumptions. Comparing the position solutions with those of the CPF-phase only case (Fig. 20), the formal uncertainties for the optical case are 2 to 3 times larger for most of the cruise, but, as is typical of OpNav, improve as the spacecraft nears Mars. Imaging Phobos and Deimos provides direct Mars relative measurements and the results show an expected gradual improvement as the spacecraft nears Mars in the final few days before entry versus the CPF phase that exhibits a more dramatic improvement in the final hours. Indeed, at entry the uncertainties

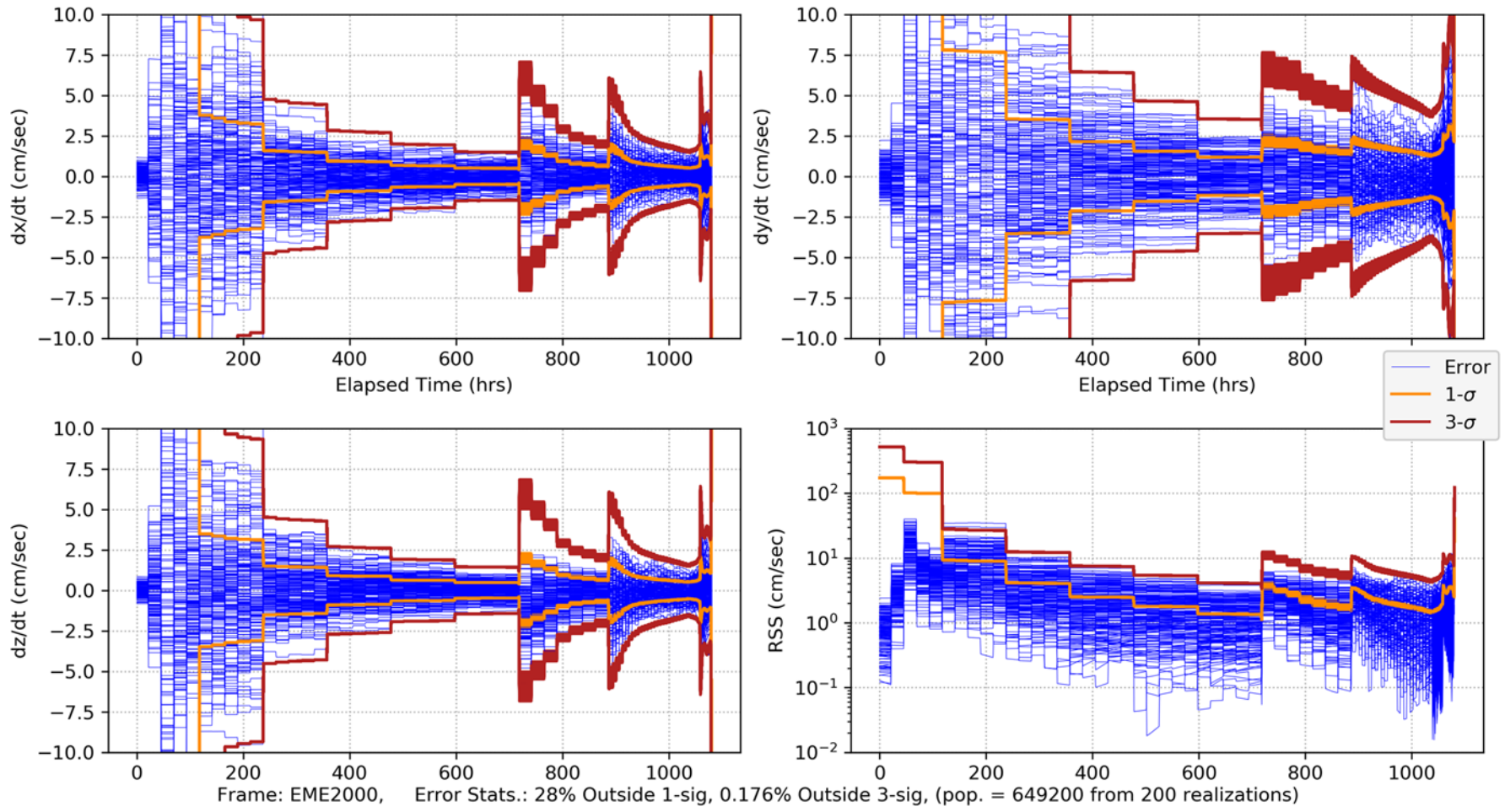

**Fig. 24** Mars approach and entry position errors for 200 realizations and uncertainties (1-sigma dark orange, 3-sigma brickred) with only optical imaging of main belt asteroids and Phobos and Deimos

for OpNav are on the order of ~2 km (3-σ) – compared to <150 m (3-σ) for the CPF-phase case. Comparing the velocity uncertainties, especially in the last 15 days in the period with the TCM 4–6 are performed, the uncertainties are overall about 50% greater than seen with CPF-phase data. This has the potential to increase the overall delta-V needs relative to the CPF-phase case. Nonetheless, a distinct advantage that the optical imaging has is its direct sensitivity to the target body when being imaged. In fact, from an examination of the EFPA results shown in Fig. 25, it is clear that the optical-only case meets both the delivery and knowledge EFPA requirements, and is close to meeting the delivery requirements at E-6 days. This level of approach knowledge early on is important to reducing risk for delivery to Mars with sufficient foreknowledge to affect outcomes or have the ground intervene in the event an anomaly occurs. Furthermore, of the three data types only optical imaging could be used for close approach of bodies where there is little knowledge of the target, as is often the case with small bodies. Imaging provides direct information of the spacecraft relative to the body that can be used to navigate towards it when an accurate ephemeris or body data are not available.

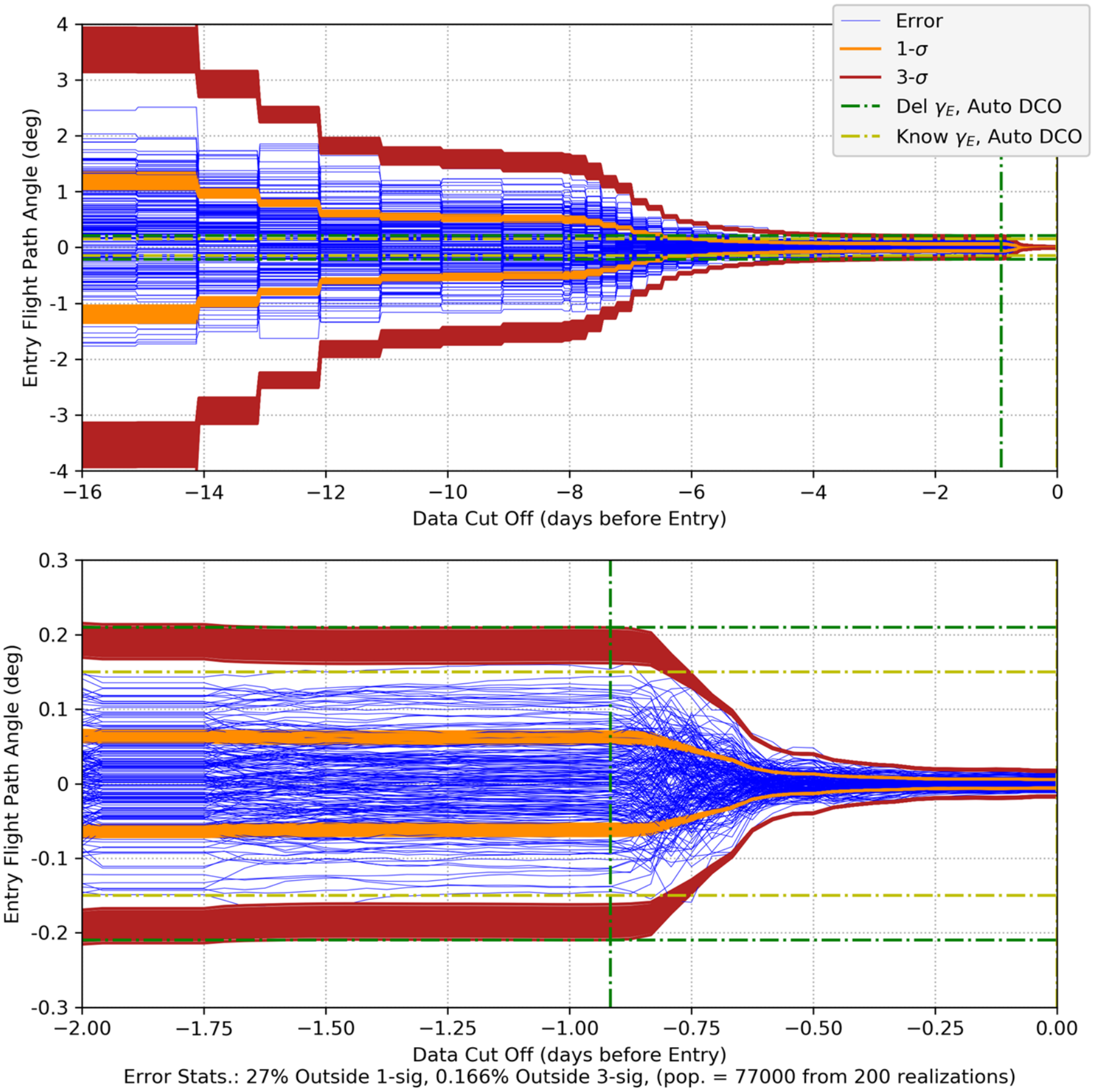

**Fig. 25** Mars entry flight path angle as a function of data cut off prior to entry with optical-only imaging. Top plot spans from entry-16 days to entry. The bottom plot is for the final two days. Note that the vertical green dash-dot line represents the onboard DCO (just before TCM-6) for the delivery EFPA requirement and the vertical yellow dash-dot line represents the DCO (at entry) for the knowledge EFPA

## Pulsar TOA Tracking Only

Next we examine pulsar TOA measurements using a gimballed detector with an area of 129 cm$^2$ and, in this first case, tracking the set of pulsars used by the SEXTANT mission (as identified in Table 2). The position determination results are shown Fig. 26 with the RSS position formal uncertainties reaching steady state values between 70 and 80 km (3-σ) and a 'peak' at entry near 105 km. As done with the analytical analysis, we next restrict the tracking to the 'best 4' pulsars to assess the performance improvement attained because of the lower measurement noise and pulsar location errors. This is shown in Figs. 27 and 28, where the RSS position uncertainties steady state values improve to the range of 50–60 km (3-σ), and the 'peak' at entry is lowered to near 80 km (3-σ). Unlike the results with either CPF phase or optical measurements, there is no improvement in position uncertainties (or velocities) as the spacecraft nears or enters the Martian atmosphere. In fact, there is a distinct growth in the uncertainties by a few tens of kilometers in the last two hours prior to entry. The associated EFPA performance is shown in Fig. 29 where neither the delivery nor knowledge EFPA requirements are met. The delivery EFPA uncertainty is bounded by approximately ± 1.7° (3-σ) – exceeding the requirement by 8 times - and knowledge uncertainty of ± 1.5° (3-σ) – exceeding the requirement by an order of magnitude. Clearly, this presents an issue with pulsar TOA measurements

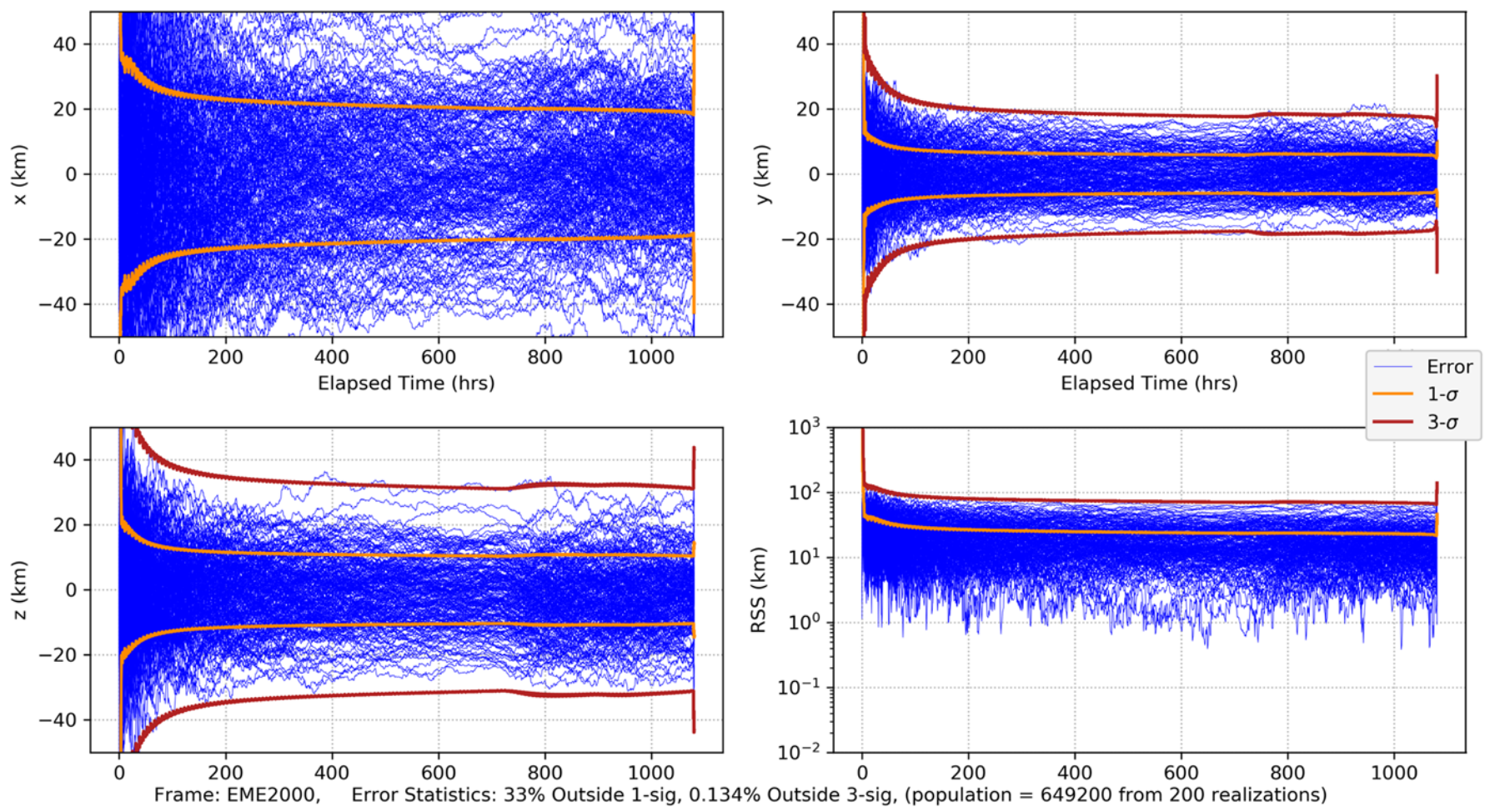

**Fig. 26** Mars approach and entry position errors for 200 realizations and uncertainties (1-sigma dark orange, 3-sigma brickred) for the case with only pulsar TOA tracking using the complete set of SEXTANT pulsars

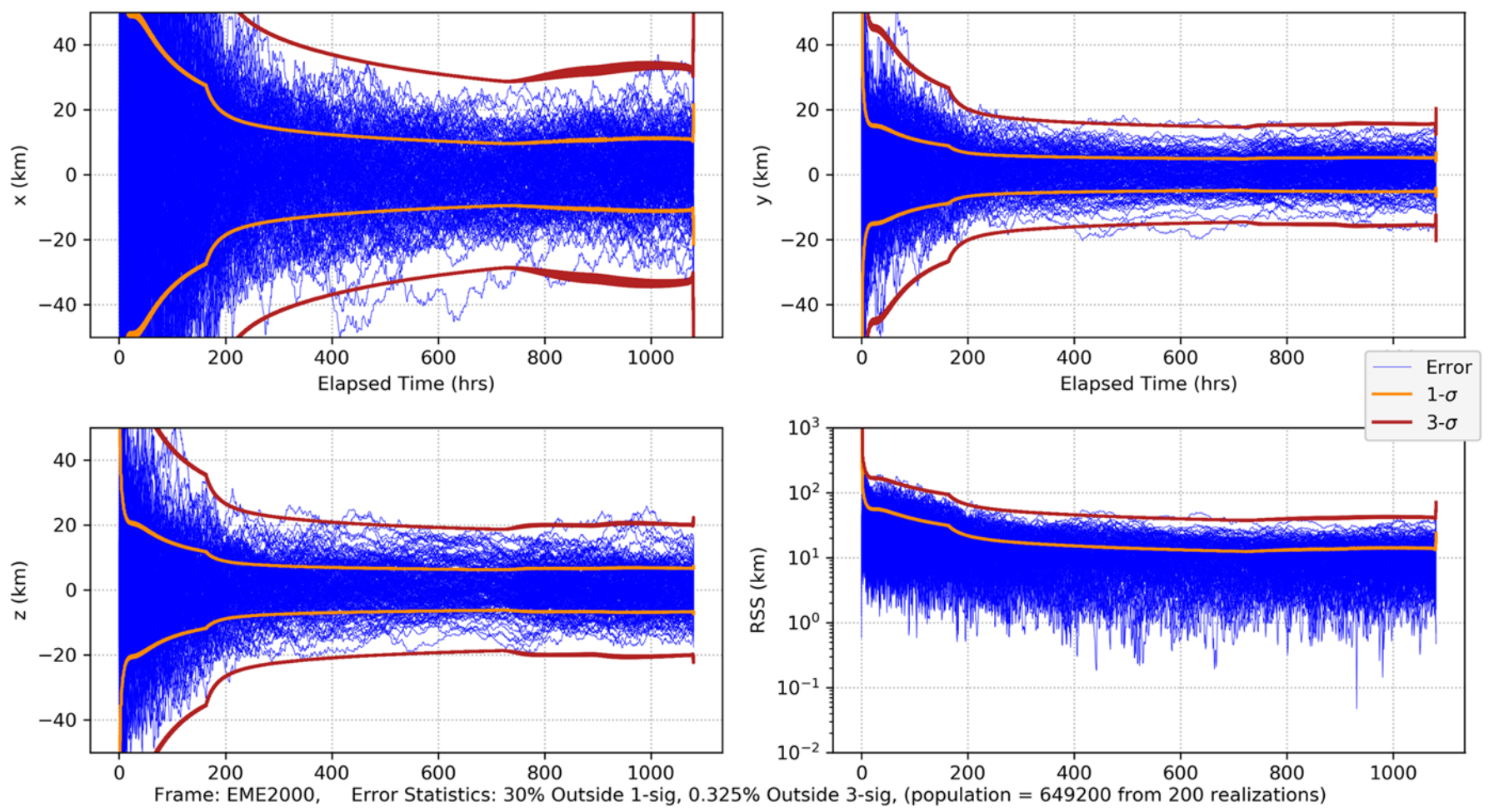

**Fig. 27** Mars approach and entry position errors for 200 realizations and uncertainties (1-sigma dark orange, 3-sigma brickred) with only pulsar TOA tracking using the 'best 4' SEXTANT pulsars

being able to support planetary entry to the levels that are necessary for safe entry of a lander. As with the CPF phase, the pulsar TOA would need to be paired with another measurement type to support Mars lander navigation. A reasonable question to consider is whether these performance levels would be sufficient for Mars orbit insertion, which are typically less stringent than for a lander. Consider the case of the Mars Reconnaissance Orbiter (MRO) that had a periapsis delivery requirement at Mars orbit insertion (MOI) of greater than 200 km and less than 400 km, hence a 200 km range. Even though this case is not for the MRO trajectory, we can use it as a proxy for periapsis altitude delivery, an examination of the state uncertainty at TCM 6 (the delivery DCO) mapped to periapsis yields approximately ± 32 km (3-σ) for a total range of delivery errors at 64 km. This indicates that pulsar TOAs could be used for a Mars orbit insertion in certain cases.

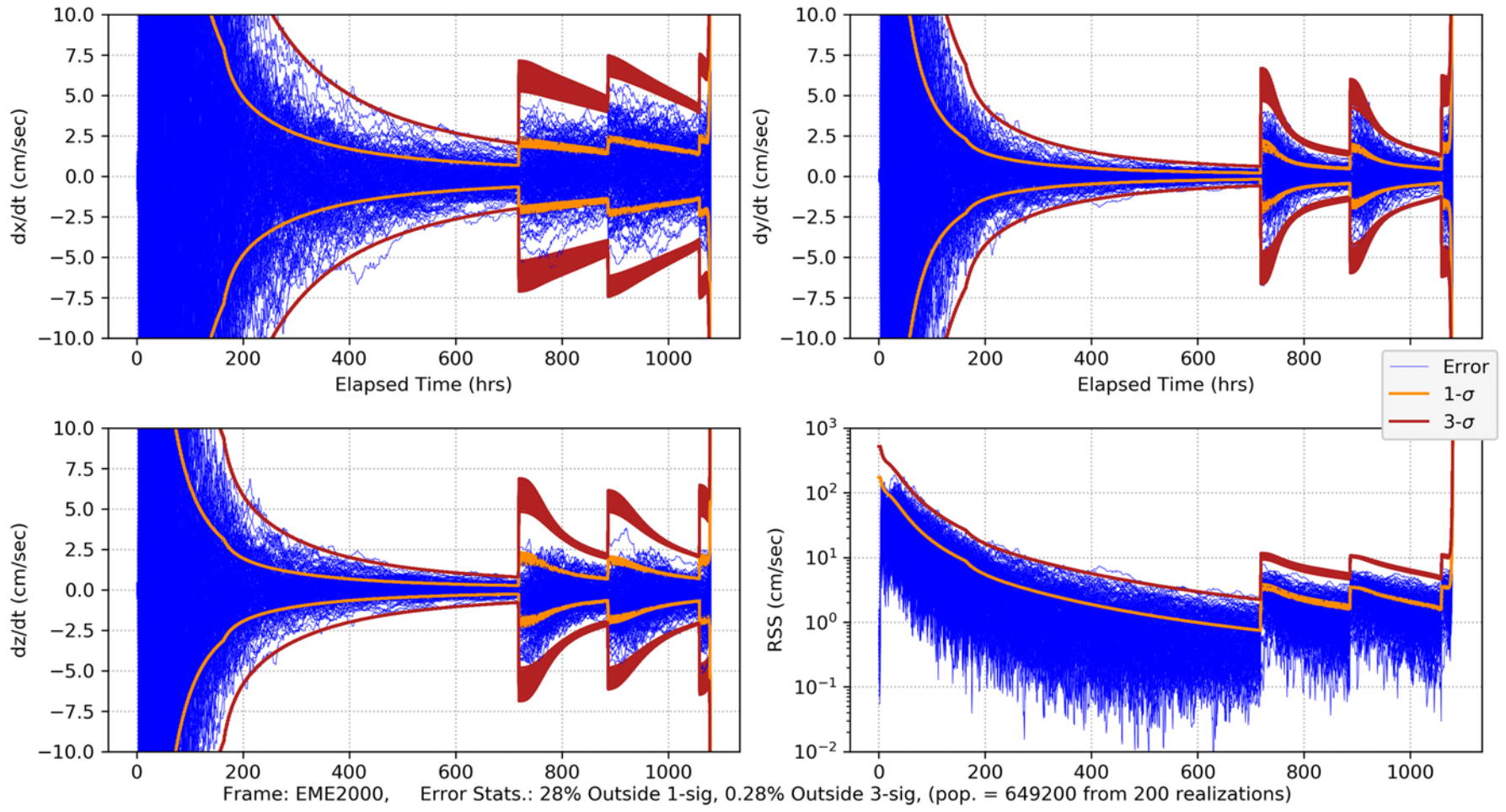

**Fig. 28** Mars approach and entry velocity errors for 200 realizations and uncertainties (1-sigma dark orange, 3-sigma brickred) with only pulsar TOA tracking using the 'best 4' SEXTANT pulsars

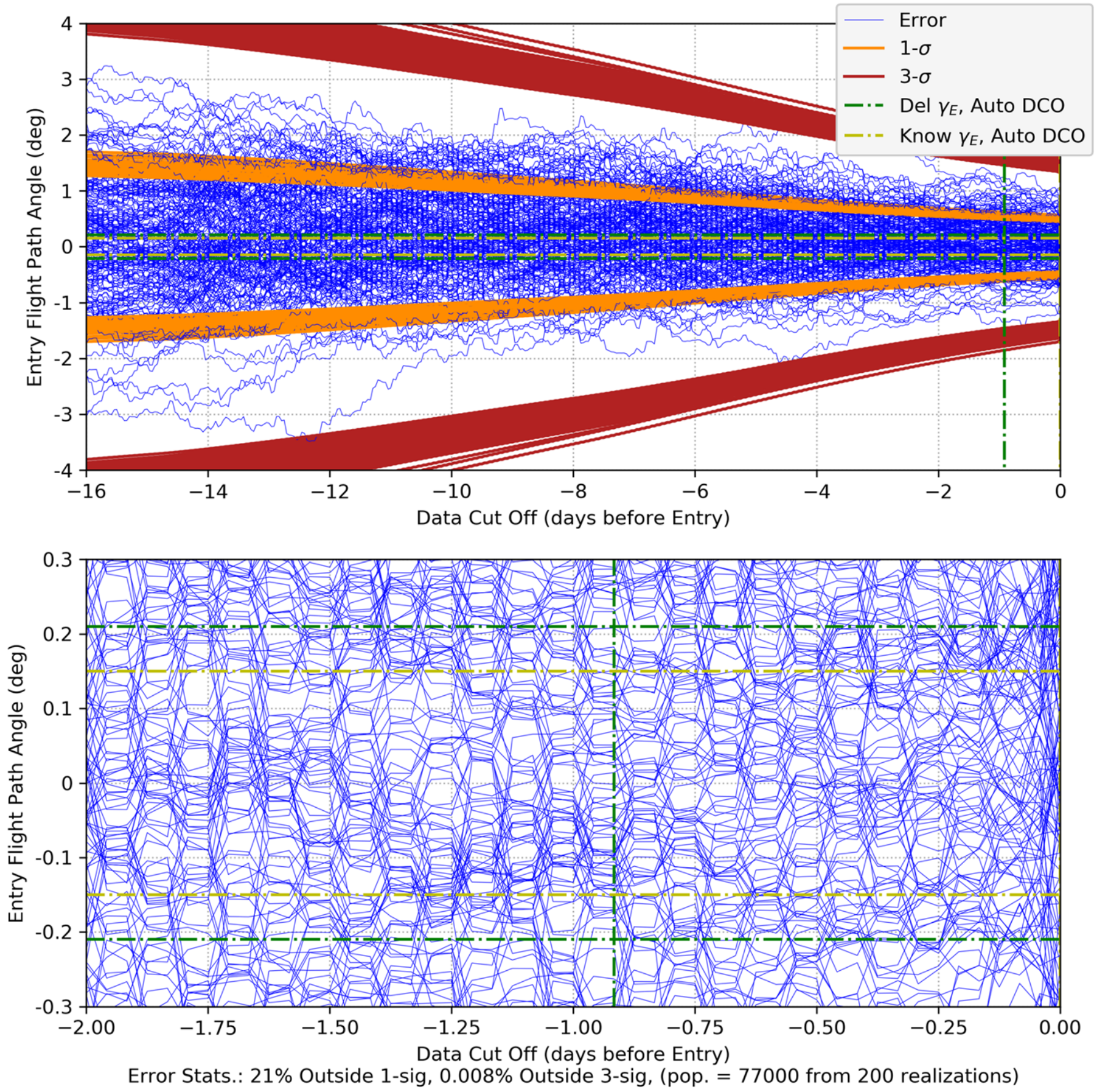

**Fig. 29** Mars entry flight path angle as a function of data cut off prior to entry with pulsar TOA tracking using the 'best 4' SEXTANT pulsars. Top plot spans from entry-16 days to entry. The bottom plot is for the final two days. Note that the vertical green dash-dot line represents the onboard DCO (just before TCM-6) for the delivery EFPA requirement and the vertical yellow dash-dot line represents the DCO (at entry) for the knowledge EFPA

**Comparing Pulsar TOA and CPF-Phase Measurement Information Content on Approach** The striking difference between the position uncertainty improvement during the final hours prior to entry exhibited by CPF-phase in Fig. 20 versus the position uncertainty growth exhibited by pulsar TOAs in Fig. 26 deserves further examination. We will, once again, turn to a semi-analytic analysis in the 2-d orbit plane to gain insight. Using the range measurement in Eq. (28) as a proxy for CPF-phase, the simplified pulsar TOA measurement in Eq. (26), and shifting the vectors from heliocentric to Mars-centric; we now compare variations in these measurements and formulate respective information matrices associated with the measurements in the final hours prior to entry, in particular the period $[t_E - 2\text{hr}, t_E]$. As illustrated in Fig. 30, we shift the origin from the Sun to Mars and align the x-axis with the eccentricity vector of the hyperbolic approach trajectory (which induces a rotation of the x-y axes). Doing so results in the following expressions for the vectors of interest:

1. The spacecraft position vector becomes $\mathbf{r} = \mathrm{r}(\theta)\cos\theta\hat{\mathbf{x}} + r(\theta)\sin\theta\hat{\mathbf{y}}$ where the Mars-relative trajectory is, to first-order, hyperbolic with $\mathrm{r}(\theta)$ defined as the radial distance from Mars, and $\theta$ is the true anomaly (for simplicity, the approach hyperbola and the x-y axes have been aligned to make this association possible).
2. The position vector to the Sun is $\mathbf{r}_\odot = r_\odot\xi\hat{\mathbf{x}} + r_\odot\sin\xi\hat{\mathbf{y}}$ with $r_\odot$ defined as the Mars-Sun distance and $\varsigma$ the angle between the x-axis and $\mathbf{r}_\odot$.
3. The position vector to the Earth tracking station becomes $\boldsymbol{r}_S = R_\oplus\cos\phi\hat{\mathbf{x}} + \boldsymbol{R}_\oplus\sin\phi\hat{\mathbf{y}} + \mathbf{r}_\oplus$ with the vector from Mars to Earth defined as $\mathbf{r}_\oplus = r_\oplus\cos\xi\hat{\mathbf{x}} + r_\oplus\sin\xi\hat{\mathbf{y}}$ where $r_\oplus$ is the Mars-Earth distance and $\xi$ the angle between the x-axis and $\mathbf{r}_\oplus$.

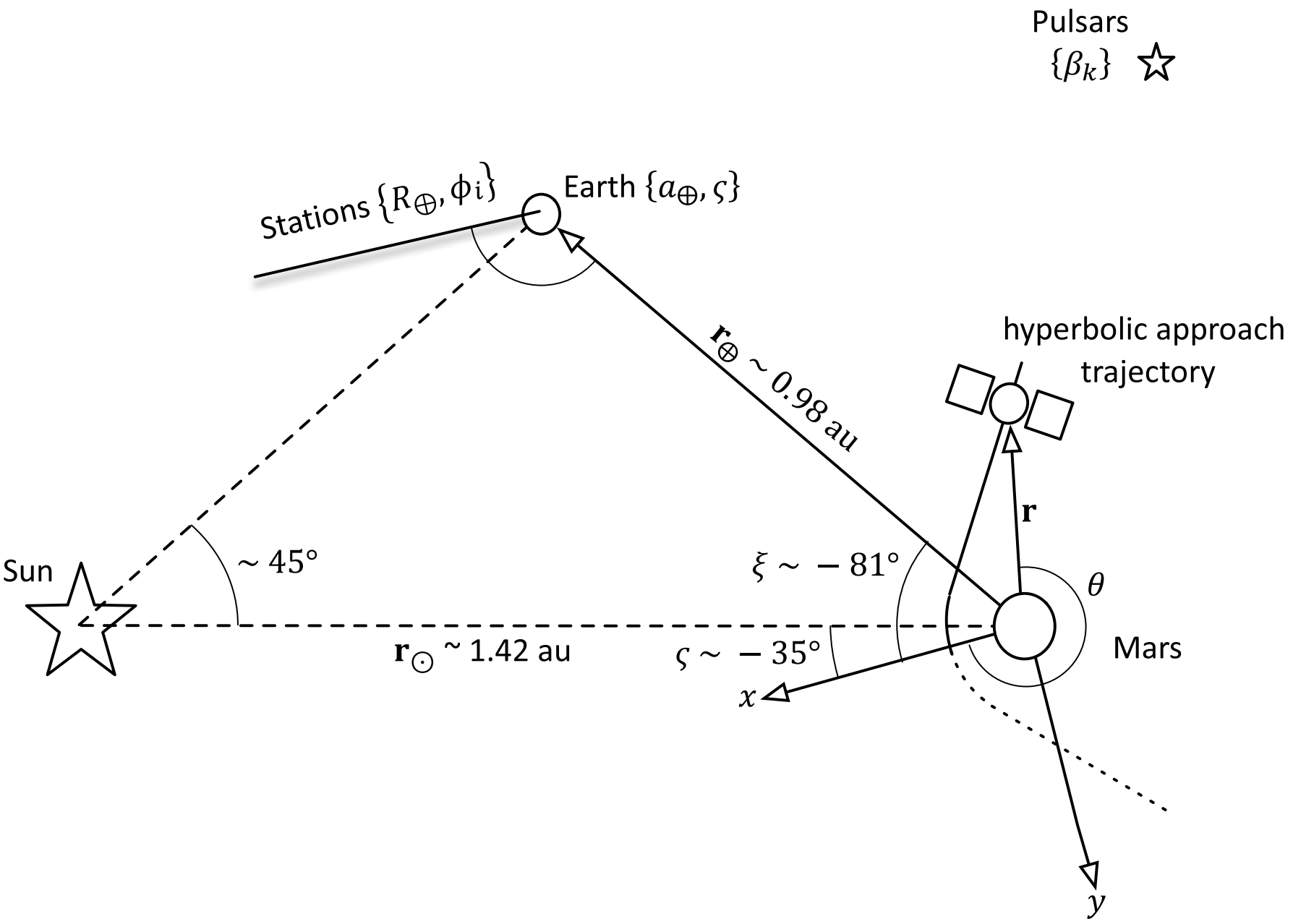

**Fig. 30** Geometry of a hyperbolic approach trajectory to Mars using InSight's approximate arrival conditions shown in Fig. 18 with the reference frame rotated so that the x-axis lies on Sun-Mars line and the origin shifted to Mars

4. Note that the unit vectors to the pulsars $\hat{\mathbf{n}}_{P_k}$ are inertial so remain they fixed by the change in origin of the axes; however, the value for $\beta_k$ does change because of the rotation of the x-y axes.

Recall that the measurement sensitivity equation for a generic scalar measurement $y$ at some time $t$ can be written as

$$\delta y = \mathbf{h}_y(t)\delta\mathbf{x}\left(t_E\right) = \frac{\partial y(t)}{\partial \mathrm{x}(t)}\mathbf{\Phi}\left(t, t_E\right)\delta\mathbf{x}\left(\mathrm{t}_E\right) \tag{129}$$

where, for the current problem, the estimate of the state deviation from nominal $\delta\mathbf{x}(t_E)$ at the time of entry $\mathrm{t}_E$ is of interest. We will formulate the aggregate information matrix that results from a set of measurements in the interval $(t_0,\ t_E)$ with the initial epoch $t_0$ set to 2 h prior to $\mathrm{t}_E$, and solve for $\delta\mathbf{x}(t_E)$. We examine the range measurement in Eq. (28) and recast it in Mars-centric vectors and squaring

$$(\rho)^2 = \left(\mathbf{r} - \mathbf{r}_\oplus - r_s\right) \bullet \left(\mathbf{r} - \mathbf{r}_\oplus - \mathbf{r}_s\right) = r^2 - 2\mathbf{r} \bullet \left(\mathbf{r}_\oplus - \mathbf{r_s}\right) + \left(\mathbf{r}_\oplus - \mathbf{r}_s\right)\left(\mathbf{r}_\oplus - \mathbf{r}_s\right) \tag{130}$$

Taking its variation yields

$$\rho\delta\rho = r\delta r - \delta\mathbf{r} \bullet \left(\mathbf{r}_\oplus - \mathbf{r}_s\right) \tag{131}$$

where variations in Earth location and station vector are assumed negligible (i.e., Mars relative locations are assumed to be known very well relative to other errors). Note the following first order approximations can be made

$$\frac{\left(\mathbf{r}_\oplus + \mathbf{r}_s\right)}{\rho} = \frac{\mathbf{r}_\oplus}{r_\oplus} + \mathbf{O}(\varepsilon) = \hat{\mathbf{r}}_\oplus + \mathbf{O}(\varepsilon)\ \&\ \frac{r}{\rho} = \mathbf{O}(\varepsilon) \tag{132}$$

and are used to obtain the following expression for measurement sensitivity with range

$$\delta\rho \cong \delta\mathbf{r} \bullet \hat{\mathbf{r}}_\oplus = \left(\hat{\mathbf{r}}_\oplus\right)^{\mathrm{T}}\mathbf{\Phi}\left(t, t_E\right)\boldsymbol{\delta}\mathbf{x}\left(\mathrm{t}_E\right) = \left(\hat{\mathbf{r}}_\oplus\right)^{\mathrm{T}}\left[\mathbf{\Phi}_{\mathbf{rr}}\left(t, t_E\right)\ \vdots\ \mathbf{\Phi}_{\mathbf{rv}}\left(t, t_E\right)\right]\begin{bmatrix}\delta\mathbf{r}\left(t_E\right)\\ \delta\mathbf{v}\left(t_E\right)\end{bmatrix} \tag{133}$$

For the pulsar TOA, we obtain its measurement sensitivity as

$$\delta(c\tau) = \boldsymbol{\delta}\mathbf{r} \bullet \hat{\mathbf{n}}_\beta = \left(\hat{\mathbf{n}}_\beta\right)^{\mathrm{T}}\mathbf{\Phi}\left(t, t_E\right)\boldsymbol{\delta}\mathbf{x}\left(\mathrm{t}_E\right) = \left(\hat{\mathbf{n}}_\beta\right)^{\mathrm{T}}\left[\mathbf{\Phi}_{\mathbf{rr}}\left(t, t_E\right) \vdots \mathbf{\Phi}_{\mathbf{rv}}\left(t, t_E\right)\right]\begin{bmatrix}\delta\mathbf{r}\left(t_E\right)\\ \delta\mathbf{v}\left(t_E\right)\end{bmatrix} \tag{134}$$

Comparing Eq. (133) to Eq. (134), the respective Mars-centric measurement sensitivities are functionally the same. Furthermore, both the unit vector to the pulsar $\hat{\mathbf{n}}_\beta$ and to, first-order, the unit vector to the Earth $\hat{\mathbf{r}}_\oplus$ are fixed over the short time scale $[t_E - 2\ \mathrm{hr}, t_E]$. We will develop a qualtitive understanding of Eqs. (133) and (134) using the methods utilized in Part 1. Once again, we formulate the aggregate information content for these measurement scenarios when only considering measurement noise and the simplified two-body dynamics in the 2-d plane of the orbit. The information matrix takes the following generic form for either range (equivalently CPF-phase) or pulsar TOA

$$\mathbf{I}^{\Sigma}(t_E) = \sum_{i=1}^{n} \frac{1}{\sigma_i^2} \begin{bmatrix} \mathbf{\Phi}_{\mathbf{rr}}^{\mathrm{T}}(t_i, t_E)\hat{\mathbf{u}}_i\hat{\mathbf{u}}_i^{\mathrm{T}}(t_i, t_E) & \mathbf{\Phi}_{\mathbf{rr}}^{\mathrm{T}}(t_i, t_E)\hat{\mathbf{u}}_i\hat{\mathbf{u}}_i^{\mathrm{T}}\mathbf{\Phi}_{\mathbf{rv}}(t_i, t_E) \\ \mathbf{\Phi}_{\mathbf{rr}}^{\mathrm{T}}(t_i, t_E)\hat{\mathbf{u}}_i\hat{\mathbf{u}}_i^{\mathrm{T}}(t_i, t_E) & \mathbf{\Phi}_{\mathbf{rr}}^{\mathrm{T}}(t_i, t_E)\hat{\mathbf{u}}_i\hat{\mathbf{u}}_i^{\mathrm{T}}\mathbf{\Phi}_{\mathbf{rv}}(t_i, t_E) \end{bmatrix} \tag{135}$$

where $\sigma_i$ is the measurement uncertainty and $\hat{\mathbf{u}}_i$ is the unit pointing vector associated with the given measurement type (CPF-phase or pulsar TOA) taken at time $t_i \in [t_E - 2\text{ hr}, t_E]$. For the CPF-phase the unit pointing vector is fixed; therefore, the numerical differences in the individual information matrices that make up the sum come from changes in the state transition submatrices from one measurement to the next. In the final two hours before entry, these changes are non-linear due to the orbit being hyperbolic and nearing periapsis (at entry the true anomaly is -19°), and are sufficient for the information matrix to attain full rank and become invertible. For the pulsar TOA measurements, a similar variation is present because the same state transition submatrices apply. Additionally, there are discrete pointing vector changes as the observed pulsar is changed (recall that this scenario is cycling through the set of the 'best 4' pulsars). So, with the structural similarities between CPF-phase and pulsar TOA why are their exhibited entry navigation knowledge results so qualitatively different? To answer this first note the measurement scenario details in the last two hours prior to entry as listed in Table 8. The distinguishable differences are the density of data collection (120 CPF-phase measurements versus 4 pulsar TOA measurements) and measurement uncertainty (0.5 m for CPF-phase versus 5 – 97 km for pulsar TOA). These are the fundamental drivers contributing to the observed differences.

To see this, we numerically compute for each scenario and then combine with the a priori state knowledge at $t_0 \equiv t_E - 2$ hr propagated to $t_E$, $\mathbf{I}^0(t_E)$, to obtain the total information content at entry. The a priori information at entry is found using

$$\mathbf{I}^0(t_E) = \left[\mathbf{\Phi}(t_E, t_0)\mathbf{P}(t_0)\mathbf{\Phi}^{\mathrm{T}}(t_E, t_0)\right]^{-1} \tag{136}$$

which, when added to $\mathbf{I}^{\Sigma}(t_E)$, can be inverted to obtain entry uncertainties as follows

$$\mathbf{P}(t_E) = \left[\mathbf{I}^0(t_E) + \mathbf{I}^{\Sigma}(t_E)\right]^{-1} \tag{137}$$

It is informative to examine the realized values of the relevant information matrices to determine how they combine to produce entry uncertainties. To begin, consider simplified, diagonal a priori uncertainties at two hours prior to entry that are representative and the same for each measurement scenario. Examination of Fig. 20 and Fig. 26 indicate 20 km (1-σ) is a representative position component uncertainty, and from Fig. 21 and Fig. 27 that 0.0001 km/s (1-σ) is a representative velocity component uncertainty. The growth in a priori position component uncertainties, when propagating from $t_0 \equiv t_E - 2$ hr to $t_E$, is shown in Fig. 31, where the a priori entry uncertainties become $\sigma_{xx} = 15.3$ km and $\sigma_{yy} = 36.6$ km. How well the addition of either CPF phase or pulsar TOA measurements impact these a priori uncertainties is now determined. Inverting, and mapping to entry using Eq. (136) yields the following a priori information matrix and associated Frobenius norm

$$\text{A priori}: \mathbf{I}^0(t_E) = \begin{bmatrix} 9.16\times10^1 & -6.59\times10^1 & -1.89\times10^4 & 1.75\times10^5 \\ -6.59\times10^1 & 6.46\times10^1 & 4.43\times10^4 & -1.29\times10^5 \\ -1.89\times10^4 & 4.43\times10^4 & 5.87\times10^7 & -4.15\times10^7 \\ 1.75\times10^5 & -1.29\times10^5 & -4.15\times10^7 & 3.35\times10^8 \end{bmatrix} \rightarrow \left\|\mathbf{I}^0(t_E)\right\|_F = 3.45\times10^8 \tag{138}$$

**Table 8** CPF-Phase and pulsar TOA measurement specifics in the final two hours prior to entry

| | CPF-Phase | Pulsar TOA |
|---|---|---|
| $t_i \in [t_E - 2\text{ hr}, t_E]$ | $t_i = t_{i-1} + 60$ sec | $t_i \in \{t_E - 90\text{min}, t_E - 60\text{min}, t_E - 30\text{min}, t_E\}$ |
| Source Signal | DSS-65 | {B0833–45, B1821–24, B1937+21, J0437–4715} |
| Angle defining $\hat{\mathbf{u}}_i$ | $\xi - 81^{\circ}$ | $\beta \in \{162^{\circ}, 283^{\circ}, 42^{\circ}, 46^{\circ}\}$ |
| $\sigma$ | 0.5 m | $\sigma_{(c\tau)} \in \{97\text{ km}, 5\text{ km}, 8\text{ km}, 87\text{ km}\}$ |

Calculating the information content from the pulsar TOA and CPF-phase measurement scenarios results in

$$\text{Pulsar TOA}: \mathbf{I}_\tau^\Sigma(t_E) = \begin{bmatrix} 5.21\times10^{-4} & 6.78\times10^{-4} & 8.56\times10^{-1} & 8.68\times10^{-1} \\ 6.78\times10^{-4} & 8.95\times10^{-4} & 1.19\times10^0 & 1.21\times10^0 \\ 8.56\times10^{-1} & 1.19\times10^0 & 1.88\times10^3 & 1.91\times10^3 \\ 8.68\times10^{-1} & 1.21\times10^0 & 1.91\times10^3 & 1.94\times10^3 \end{bmatrix} \rightarrow \left\|\mathbf{I}_\tau^\Sigma(t_E)\right\|_F = 3.82\times10^3 \tag{139}$$

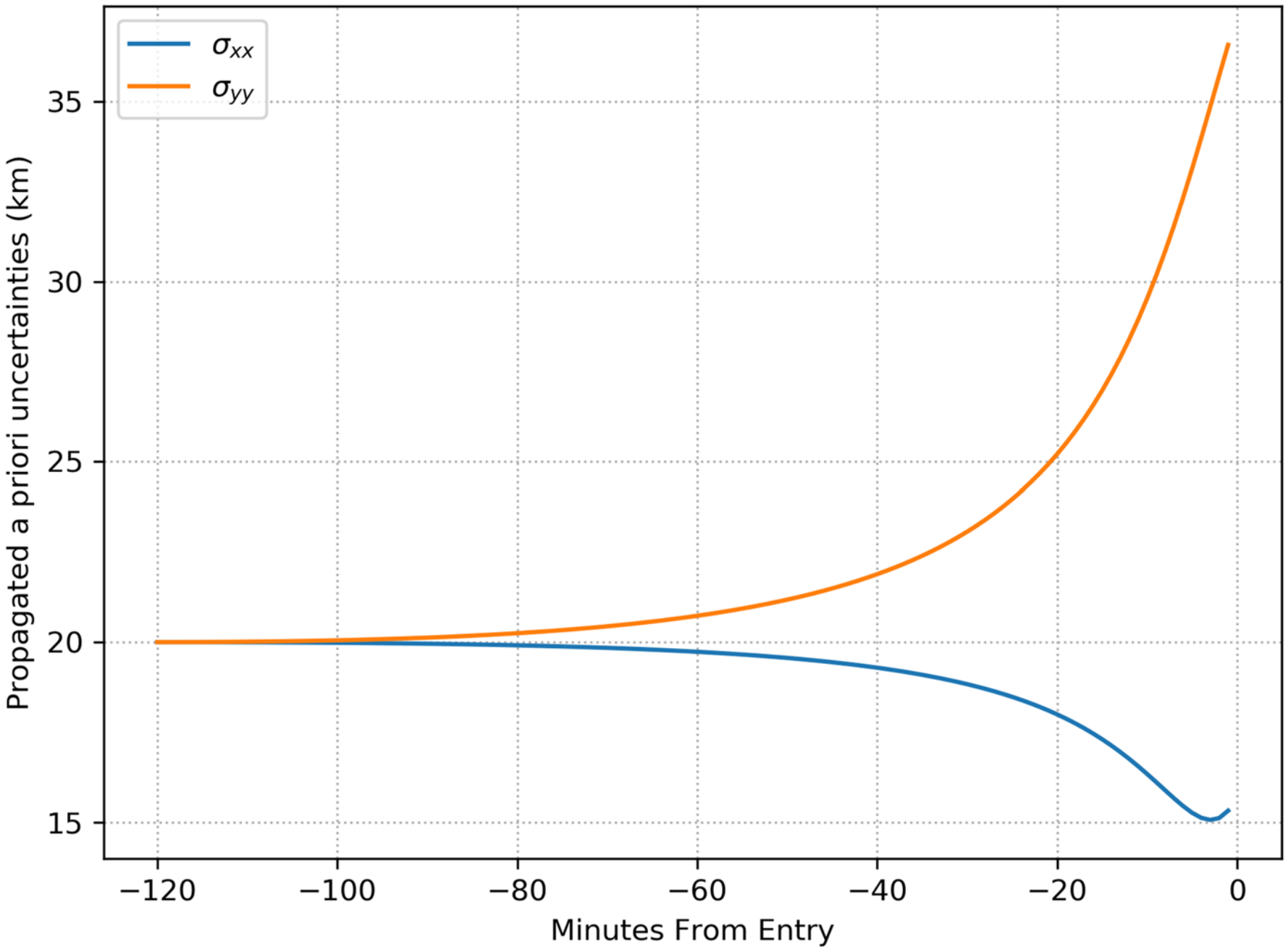

**Fig. 31** 2-d a priori position uncertainties propagated from $t_0 \equiv t_E - 2$ hr to $t_E$

$$\text{CPF} - \text{phase}: \mathbf{I}^{\Sigma}_{CPF}(t_E) = \begin{bmatrix} 1.01\times10^{13} & -5.27\times10^{13} & 3.40\times10^{16} & -1.54\times10^{17} \\ -5.27\times10^{13} & 3.18\times10^{14} & -2.35\times10^{17} & 1.12\times10^{18} \\ 3.40\times10^{16} & -2.35\times10^{17} & 1.96\times10^{20} & -9.59\times10^{20} \\ -1.54\times10^{17} & 1.12\times10^{18} & -9.59\times10^{20} & 4.76\times10^{21} \end{bmatrix} \rightarrow \left\| \mathbf{I}^{\Sigma}_{CPF}(t_E) \right\|_F = 4.95\times10^{21} \tag{140}$$

Comparing the values of Frobenius norms, it is clear that the pulsar TOA measurements can only marginally improve on the a priori information, while CPF-phase significantly increases the overall information. Applying Eq. (137) to determine the associated entry covariances yields the following component uncertainties for the two measurement schemes

$$\begin{aligned} &\text{pulsar TOA}: \sigma_{xx} = 14.8\ \text{km}, \quad \sigma_{yy} = 28.5\ \text{km} \\ &\text{CPF} - \text{phase}: \sigma_{xx} = 2.7\times10^{-6}\ \text{km}, \sigma_{yy} = 8.9\times10^{-7}\ \text{km} \end{aligned} \tag{141}$$

These results are qualitatively consistent with the high-fidelity results and confirm that the pulsar TOA measurements are only marginally effective in reducing trajectory uncertainties on entry, while CPF-phase have a significant benefit. Note that the CPF-phase values are overly optimistic because the other error sources have been neglected in this simple analysis. For the pulsar TOA measurements to be commensurate with the CPF-phase on entry, the pulsar TOA measurement uncertainty would have to significantly improve. For instance, if the pulsar TOA measurement uncertainty were as precise as the CPF-phase (i.e., $\sigma=0.5$ m) the resulting position uncertainties would be $\sigma_{xx}=4.4\times10^{-6}$ km and $\sigma_{yy}=4.6\times10^{-6}$ km, which, expectedly, are commensurate with the CPF-phase. Of course, pulsar's that admit this level of measurement uncertainty have not been discovered to date. Importantly, there was nothing unique to Mars in the preceding analysis of this entry problem that would qualitatively affect these conclusions. Therefore, we surmise that pulsar TOA measurements would yield similar insensitivities for approach and entry at other solar system bodies – including in the outer solar system.

## CPF-Phase Tracking and Optical Imaging

As our first measurement combination, we consider CPF-phase measurements combined with optical imaging using the high-end gimballed camera. The Monte Carlo position errors and associated uncertainties are shown in Fig. 32 and the velocity errors/uncertainties in Fig. 33. Comparing with the CPF-phase-only position results in Fig. 20 and the optical-only position results in Fig. 26, the combined data results in Fig. 32 are far more accurate. Indeed, the combined results quickly achieve <20 km uncertainties (3-σ) in each component, and by the last week <10 km (3-σ). The entry knowledge is <200 m (3-σ) – a slight degradation from the CPF-phase-only but not significant. The EFPA as a function of the data cut off (DCO) time prior to entry is shown in Fig. 34. The results in the final day show that all the realization errors (except for one) and associated 3-σ uncertainties lie within the 0.21° delivery limit by the time of TCM-6 (E-22 h). EFPA knowledge requirements are easily met, and, at <0.01° (3-σ), are an order of magnitude better than required. For this Mars

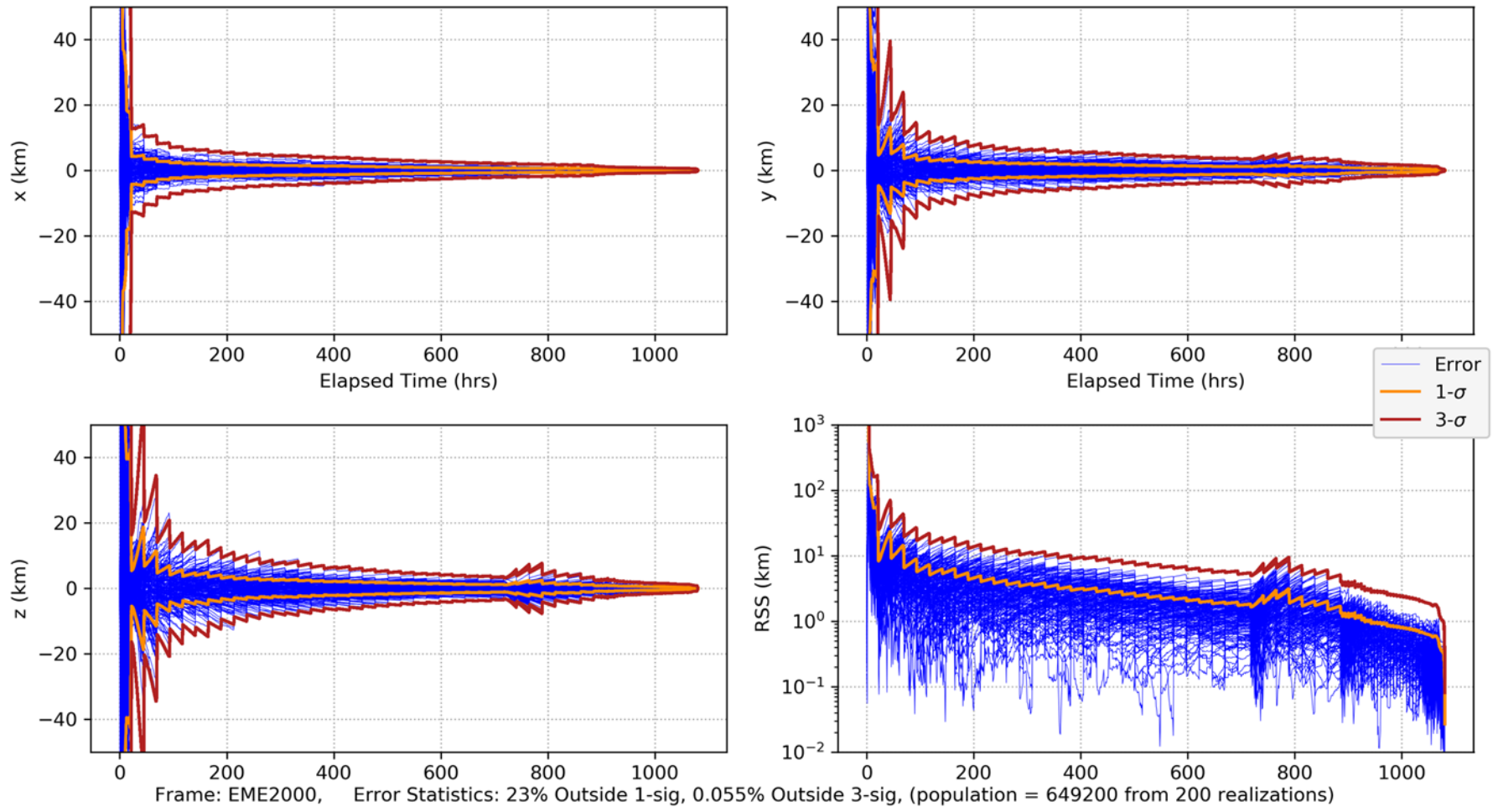

**Fig. 32** Mars approach and entry position errors for 200 realizations and uncertainties (1-sigma dark orange, 3-sigma brickred) with uplink one-way CPF-phase data from the DSN and optical imaging of main belt asteroids and Phobos and Deimos

approach and entry scenario, these results provide accurate solutions that readily meet all the requirements. Furthermore, use of two dissimilar data types provides a natural fault tolerant robustness in the event that one of the data types begins providing anomalous results. Indeed, a key strategy employed for ground-based navigation is to use two dissimilar data types (2-way range and Doppler coupled with double differenced one-way ranging (DDOR)) to ensure robust and accurate results. It is only natural to extend this strategy for onboard navigation. Next, we consider the combination of CPF-phase with pulsar TOA measurements, and compare with the CPF-phase/optical combination.

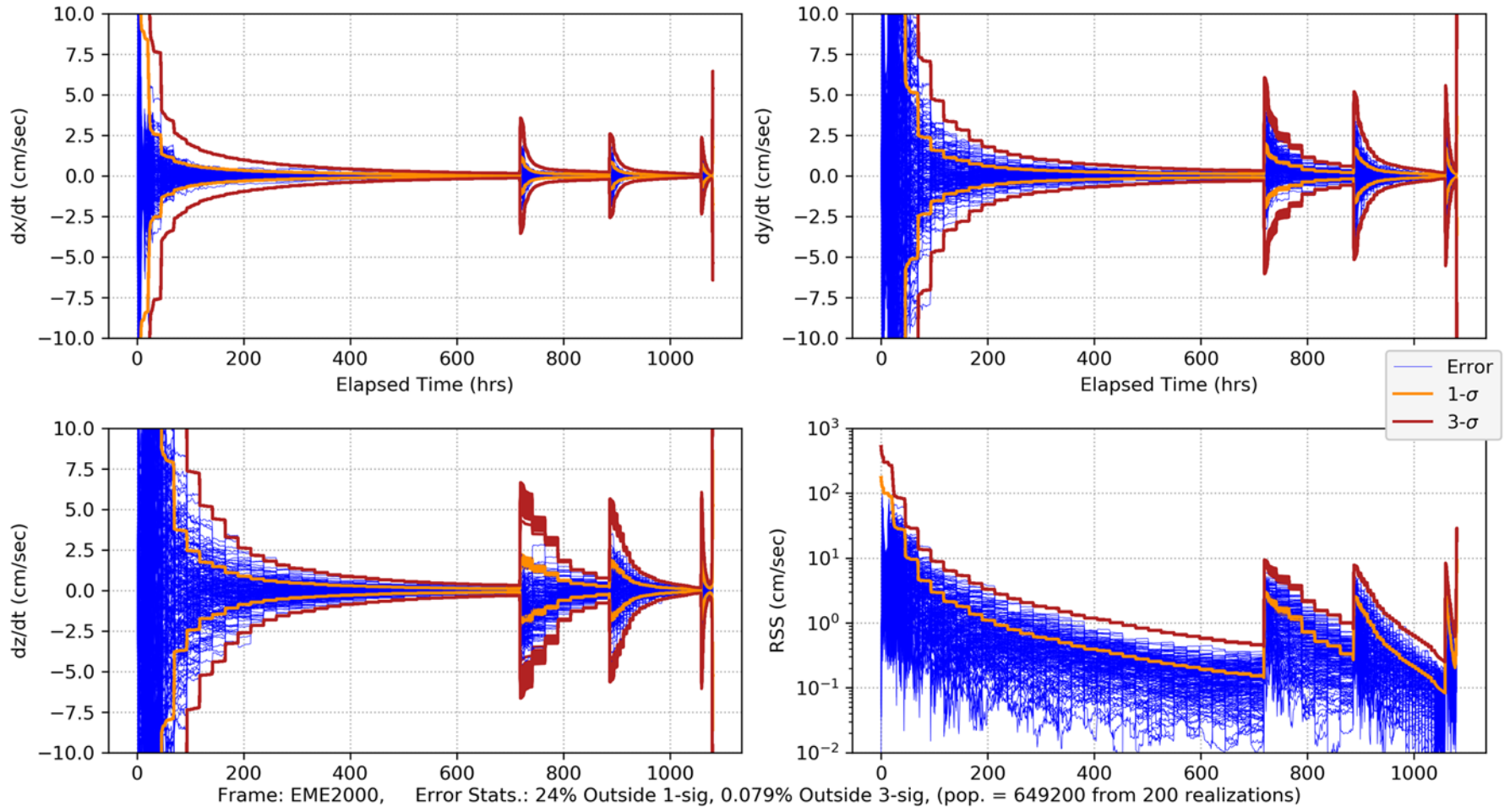

**Fig. 33** Mars approach and entry velocity errors for 200 realizations and uncertainties (1-sigma dark orange, 3-sigma brickred) with uplink one-way CPF-phase data from the DSN and optical imaging of main belt asteroids and Phobos and Deimos

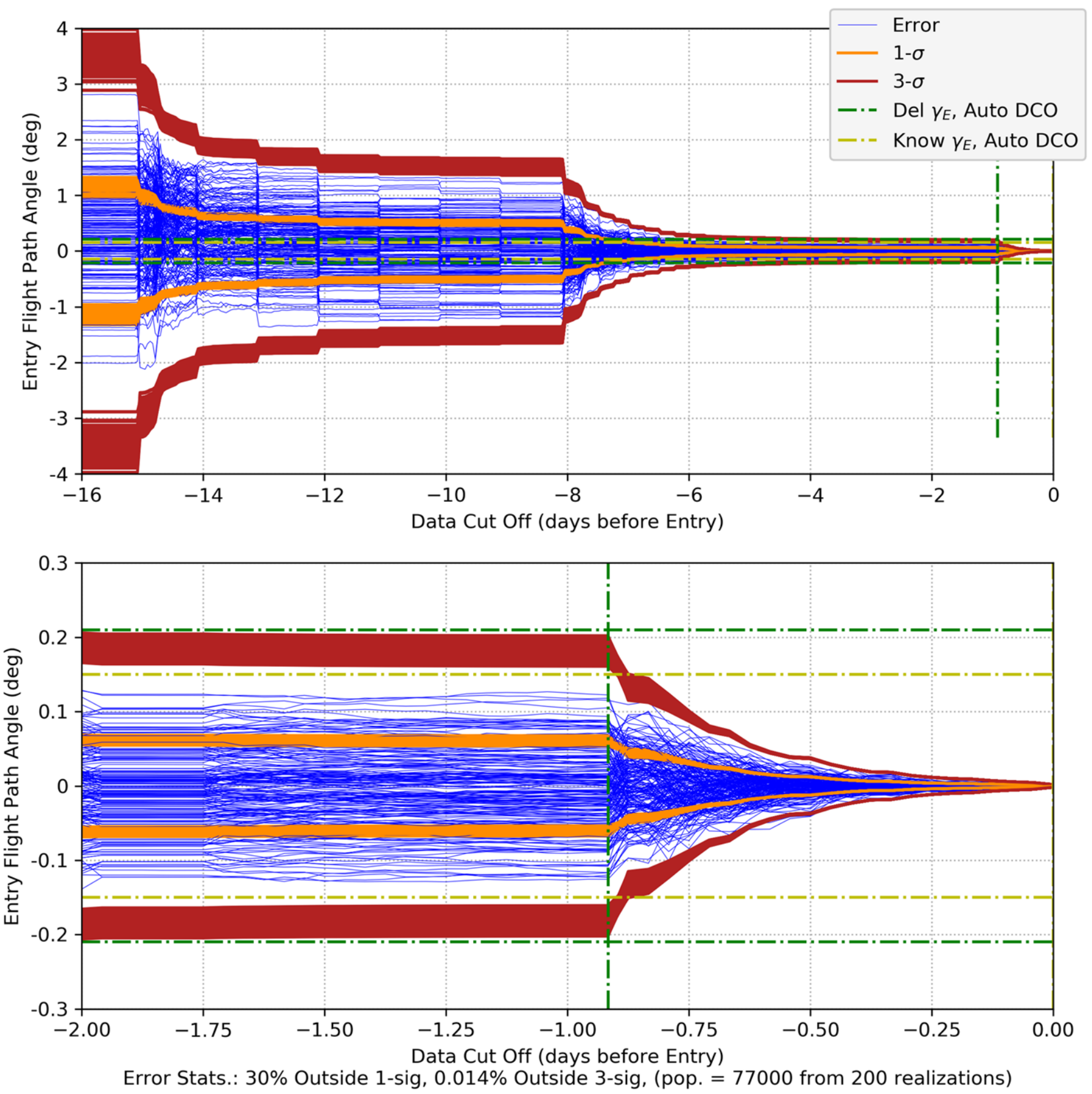

**Fig. 34** Mars entry flight path angle as a function of data cut off prior to entry with uplink one-way CPF-phase data from the DSN and optical imaging of main belt asteroids and Phobos and Deimos. Top plot spans from entry-16 days to entry. The bottom plot is for the final two days. Note that the vertical green dash-dot line represents the onboard DCO (just before TCM-6) for the delivery EFPA requirement and the vertical yellow dash-dot line represents the DCO (at entry) for the knowledge EFPA

## CPF-Phase and Pulsar TOA Tracking

We now examine combing CPF-phase with pulsar TOA tracking, this is a practical combination as an extremely precise, stable clock is needed for both measurement types. Practically, it would also be complex to deploy three external hardware systems (as would be the case with a separately gimballed camera and gimballed pulsar detector

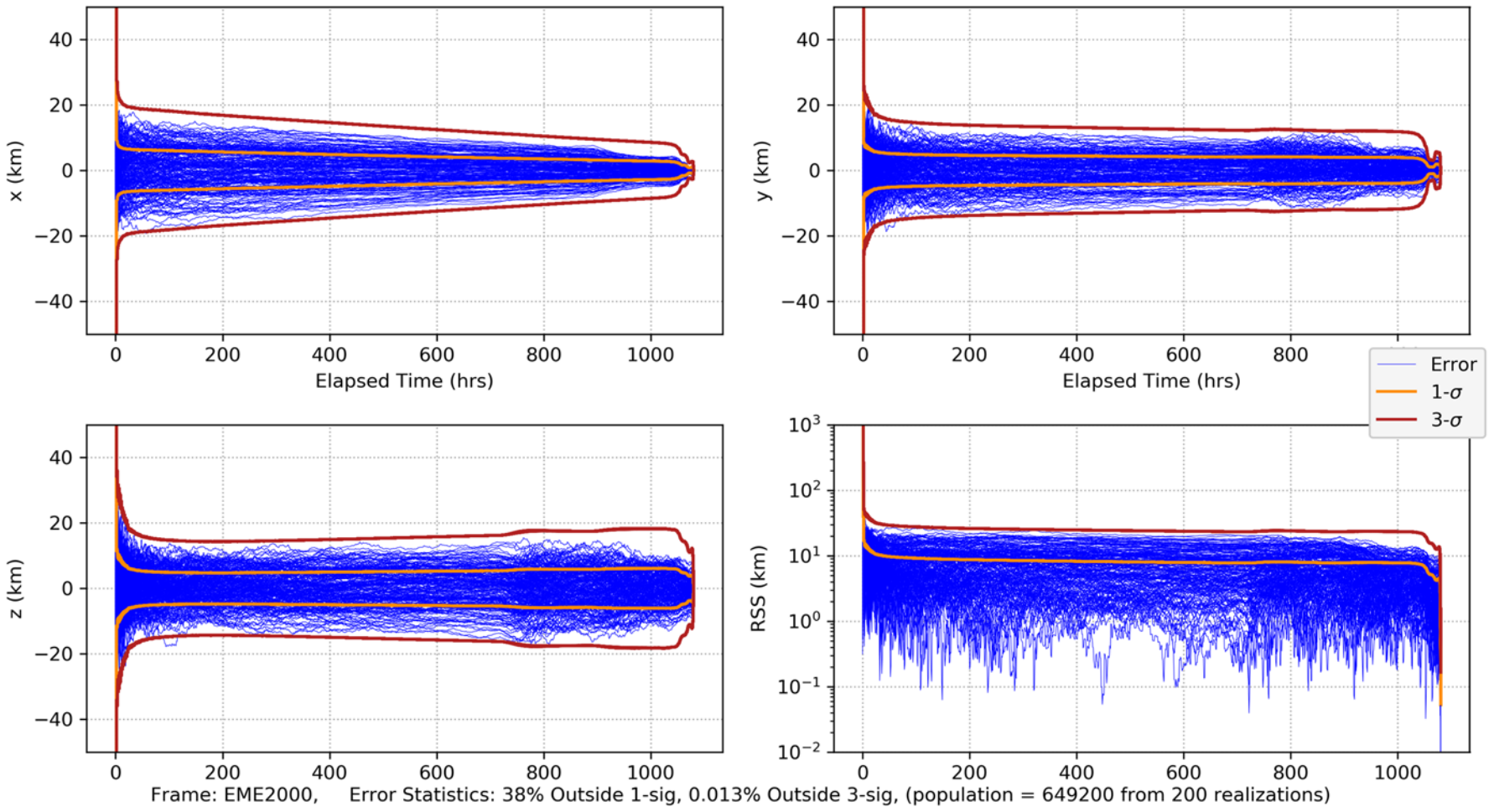

**Fig. 35** Mars approach and entry position errors for 200 realizations and uncertainties (1-sigma dark orange, 3-sigma brickred) with uplink one-way CPF-phase data from the DSN and pulsar TOA tracking using the 'best 4' SEXTANT pulsars

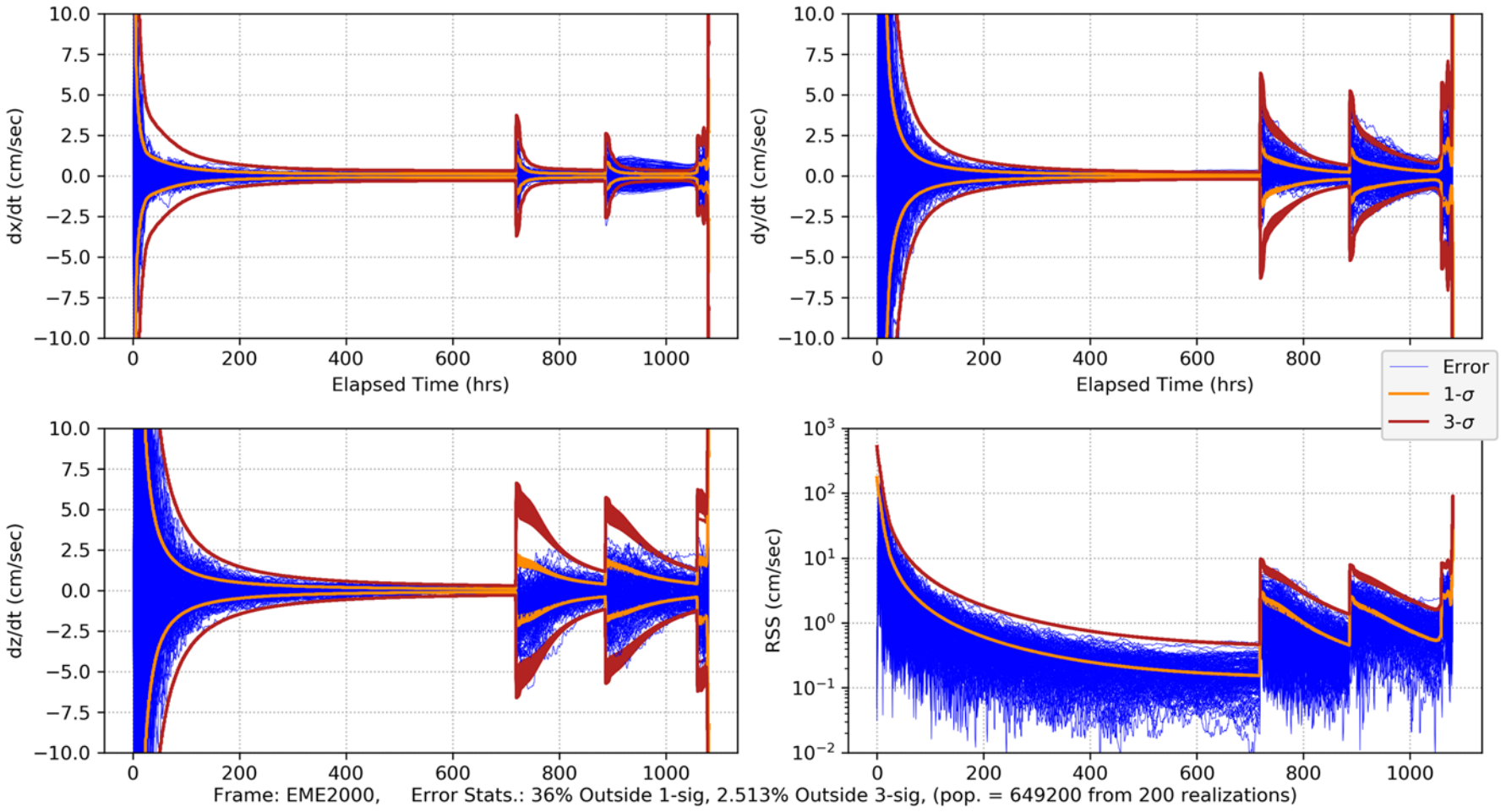

**Fig. 36** Mars approach and entry velocity errors for 200 realizations and uncertainties (1-sigma dark orange, 3-sigma brickred) with uplink one-way CPF-phase data from the DSN and pulsar TOA tracking using the 'best 4' SEXTANT pulsars

competing with already manifested communications antennas).[7] As with the CPF-phase and optical combination, the CPF-phase and pulsar TOA is a combination of dissimilar data types so it is anticipated that accuracy will improve via the combination of the two. This is evident in the position and velocity results shown in Fig. 35 and Fig. 36 where, as before, the pulsar TOA measurements are from the 'best 4' SEXTANT pulsars. Comparing the RSS position results in Fig. 35 after the initial transient dies off to the CPF-phase only results (Fig. 20 with a range of 40–60 km 3-σ) and the pulsar TOA only results (Fig. 27 with a range of 40–50 km 3-σ and a 'peak' at entry), there is an overall improvement in the magnitude of the position uncertainties with a steady state near 25 km (3-σ) and entry knowledge of 150 m (3-σ) (similar to the CPF-phase only results). There is also an improvement with the combined velocity results (Fig. 36 compared with Fig. 21 and Fig. 28); however, in the final 15 days during the TCM sequence the filter is not producing statistics that are consistent with a converged Kalman filter producing Gaussian-like statistics. There are numerous exceedances beyond 3-σ that can be attributed to considering pulsar location errors. The filter is unable to optimally estimate for the velocities in the presence of significant pulsar location errors that are not corrected for via an estimation process. A potential solution to this is to select pulsars with the smallest location errors. Doing so could be necessary for use of the pulsar TOA's further out into the solar system as location errors grow with the distance from the solar system barycenter.

[7] If it became possible to deploy both the pulsar detector and the optical camera on the same gimballed platform then consideration of the combined optical and pulsar TOA measurements becomes a distinct possibility, but this would require advances in X-ray detector technology to reduce SWaP and subsequent integration of the two disparate systems. This approach would also have the distinct advantage of not requiring any Earth-based signals; hence, improving overall autonomy.

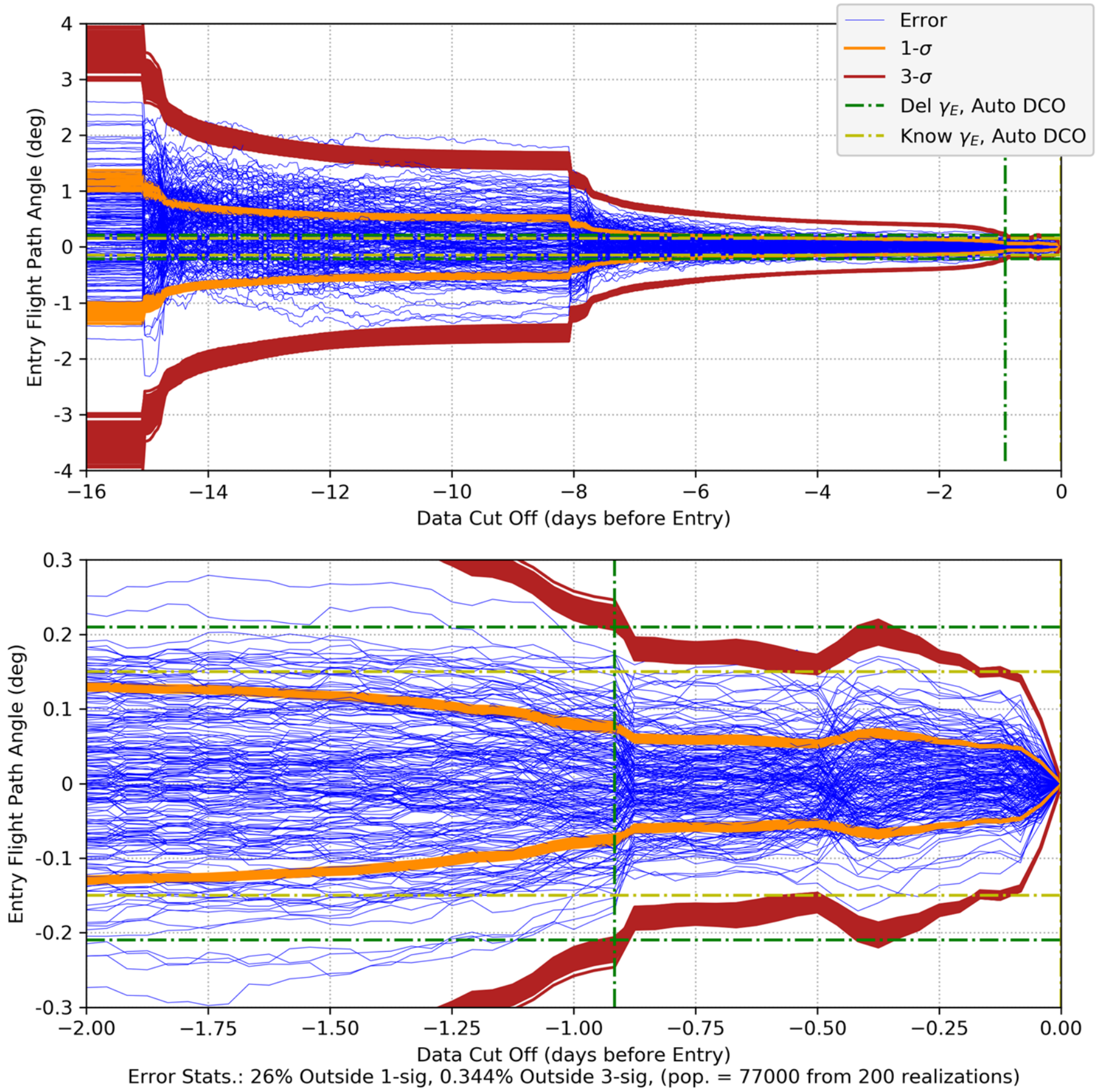

**Fig. 37** Mars entry flight path angle as a function of data cut off prior to entry with uplink one-way CPF-phase data from the DSN and pulsar TOA tracking using the 'best 4' SEXTANT pulsars. Top plot spans from entry-16 days to entry. The bottom plot is for the final two days. Note that the vertical green dash-dot line represents the onboard DCO (just before TCM-6) for the delivery EFPA requirement and the vertical yellow dash-dot line represents the DCO (at entry) for the knowledge EFPA

Now, comparing the current results with those in the prior case that combined CPF-phase and optical measurements in Fig. 33, it is seen that the CPF-phase/optical combination yields consistently more accurate results than with the CPF-phase/pulsar TOA combination. Considering the EFPA results, the CPF-phase/pulsar TOA solutions shown in Fig. 37 easily meet the knowledge requirement at entry (because of the CPF-phase) and the delivery uncertainties at a E-0.9 day DCO (i.e., TCM6) improve relative to CPF-phase or pulsar TOA by themselves. However, they are not as accurate as those exhibited by the CPF-phase/optical combination. Additionally, the uncertainties for the delivery results fall just outside of the requirement. While it is possible to consider different TCM strategies to pull the performance into compliance or to select pulsars with smaller location uncertainties, these results show that the CPF-phase/optical combination has superior observability as compared to the CPF-phase/pulsar TOA combination.

## Comment on Outer Solar System Navigation

Even though the preceding analysis was for navigating to Mars, an inner solar system planet, the results are still relevant for navigating in the outer solar system. A primary difference that should be considered for navigation to outer solar system bodies versus to inner solar system bodies is the error in their ephemeris knowledge. Vallisneri [53] surveyed the accuracy of recent solar system ephemerides and finds that the inner planet orbits relative to the SSB are known to better than a kilometer, Jupiter and Saturn to tens of kilometers, and Uranus and Neptune to thousands of kilometers. Similarly, as noted in Part 1 for the optical navigation analysis, small body ephemeris knowledge errors range from kilometers to thousands of kilometers. Hence, any mission to a solar system object must contend with these location errors, and is a significant reason that they carry cameras so that they can obtain target relative OpNav measurements to reduce the impact of inertial location errors of these bodies as they approach and orbit them or fly by them. As an example, optical imaging by the NASA Cassini mission was critical to determining the orbits of Saturn's moons as it toured the Saturn system [14]. Furthermore, the mission will always have some form of telecom system for communications, so utilizing this same system for radiometric tracking has been the traditional, resource efficient method for obtaining absolute navigation solutions. Having a stable clock enables this navigation methodology to be resident onboard the spacecraft. As noted previously, pulsar TOA measurements are not well suited for precision approach or flyby navigation because of each measurement's large magnitude noise level. This insensitivity is exacerbated by the target body's ephemeris uncertainty. Hence, the combination of radio data and OpNav provides for both absolute and target relative navigation information that would be needed for encounters with outer solar system bodies, especially those with uncertain ephemerides.

Examples of the strength of OpNav image data is illustrated with the recent New Horizons' flybys of Pluto and Arrokoth; detailed results are documented in Ref. [54–57]. With the Pluto encounter, the goal was to deliver the spacecraft to a flyby altitude of 12,540 km. To satisfy science instrument pointing constraints, the spacecraft's final TCM on approach had to deliver the spacecraft to within a rectangular box 150 km long by 100 km wide. Furthermore, the spacecraft needed to know the time of the closest approach to within 150 s. Given that the a priori ephemeris of Pluto was not known to better than ~1000 km, these goals could not have been accomplished without the use of OpNav images on approach. Therefore, a series of OpNav campaigns imaging the Pluto system was undertaken starting a year before and continuing up to 2 days prior to the encounter. The images were used to progressively improve the spacecraft's location relative to the Pluto system and design TCMs to target the flyby location, with the final TCM being performed a little over 2 weeks prior to encounter. However, the timing update was only satisfied after the last of the OpNav images were taken, where imaging the satellites relative to Pluto provided enough information along the spacecraft's velocity vector to satisfy the timing requirements. It was a very similar situation for Arrokoth, a Kuiper Belt Object. Having only been discovered in 2014, there has been little data collected to determine Arrokoth's orbit, so its ground-based ephemeris error was over 3000 km prior to the flyby. As with Pluto, OpNav images were used to continually improve the planetoid relative ephemeris knowledge of the spacecraft to safely attain the targeted flyby altitude of roughly 3500 km.

In scenarios where a mission does not have any encounters or the encounter navigation requirements are sufficiently relaxed, then pulsar TOA measurements could provide a sufficient navigation capability. Pulsar TOA measurements have superior geometric diversity for absolute cruise phase navigation as compared to radio data, and are not limited by the availability of visible optical 'beacons' as with OpNav, so could offer a flexible navigation capability in such cases. Enabling this requires selection of the most stable pulsars with well-known locations (to reduce the effect of this error from dominating the trajectory solution error).

## Conclusions

The relative navigation performance of radiometric tracking (via range and a derived data type called CPF-phase), optical imaging (aka OpNav) of solar system bodies, and pulsar time-of-arrival tracking has been explored using both analytic methods and high-fidelity, realistic Monte Carlo simulations of approach and entry to Mars. A key goal of this analysis, especially with the analytic explorations, was to compare each data type using common assumptions on a tractable problem so that the relative merits of each data type could be easily examined. The analytical analysis assumed aspirational best-case measurement capabilities (i.e., gimbaled, highly sensitive cameras, reduced SWaP pulsar detectors, or the ready availability of atomic clocks) with only measurement noise present (or, when necessary, source location errors) and no other systematic errors, so that the geometric sensitivities of each data type could be readily examined. In turn, the high-fidelity simulations introduced realistic errors that affect spacecraft navigation using typical operational models and assumptions often used for navigating to Mars. This put each measurement type into context for potential onboard, autonomous use in navigating a Mars lander to entry at the top of the Martian atmosphere using realistic model effects and error processes, so that their impact on the capability of the selected measurement type for navigation could be assessed in a more quantitative way (i.e., could a representative delivery entry flight path angle be met).

The key analytical findings include:

1. Ranging is the most precise of the three measurement types; however, geometrically it has weaker observability (as determined by the averaged GDOP) than the other two. However, this can be compensated by tracking for longer periods of time and, when combined with its superior measurement precision, yields the least uncertainty of the three types in the analytic cruise phase initial position determination problem. Onboard use does require an extremely stable clock, such as DSAC, so continued improvements in clock and radio technology are needed to best utilize this data type.
2. Depending on the selected pulsars, the pulsar TOA is the least precise measurement of the three and is strongly affected by errors in pulsar location knowledge that grows with increasing distance from the Sun. Compared to the other three data types, it has the best average GDOP, but the large measurement error degrades the overall position knowledge. This could be improved by expanding the set of known pulsars and reducing pulsar location errors. Also, the pulsar TOA detector requires the most size, weight, and power (SWaP) to be practical for future deep space use. If the detector SWaP could be reduced, pulsar TOA measurements would yield consistent orbit determination errors at the several 10s of kilometer level during cruise phase whereas the ranging and optical measurements would result in greater variability.
3. Navigation using optical imaging of asteroids benefits from an extensive catalog of known asteroids that makes it ideal for onboard use in the inner solar system. In the outer solar system, onboard optical navigation is limited by the sensitivity of existing camera systems and/or knowledge of other suitable solar system reference bodies. Improving camera technology and extending the catalog of known bodies (such as including planets, other asteroid families and/or KBOs) would facilitate optical navigation use in the outer solar system.

Key findings from the high-fidelity analysis include:

1. Optical navigation provides the best relative information for navigating to a target destination. Of the three data types, it is the only one that provides a direct relative measure of the target body. As demonstrated in this paper for planetary entry (or for close flybys or a very low altitude orbit insertion), having accurate target relative knowledge is important for meeting target conditions as well as early target relative knowledge for minimizing flight path control costs. In the case studied here, CPF-phase was target sensitive in the final hours prior to entry, but optical navigation achieved this a day prior. Pulsar TOA was insensitive to the target.

2. It was shown in an encounter (where the target has a well-known ephemeris) that pulsar TOA measurements and CPF-phase have similar geometric contributions to the information matrix, but significant pulsar TOA measurement noise limits its utility for approach and entry navigation.
3. For the Mars case, CPF-phase provides the best entry knowledge, but this is only in the final few hours prior to entry. Optical navigation has superior sustained improvement on approach, and was the only data type to meet both the delivery and entry knowledge requirements.
4. For less stringent orbit insertions, flybys, or missions that are primarily in cruise, use of only pulsar TOA data may be sufficient assuming the detector SWaP can be reduced to practical levels. For example, the orbit insertion requirements of MRO could have been achieved with each data type by itself: pulsar TOA, CPF-phase, or optical.
5. CPF-phase and pulsar TOA provide absolute navigation knowledge enroute to the target body at similar performance levels (many 10's of kilometers for the Mars case), while optical is degraded by several factors until it nears the target.
6. *The performance of the combined data types far exceeds that of any one data type by itself.* For onboard, autonomous navigation it is recommended that this strategy be used to not only improve accuracy, but increase solution robustness that is tolerant to failures of any one data type. The combination of CPF-phase and optical performed the best across all mission phases all the way to delivery to the top of Mars' atmosphere with results that were nearly an order of magnitude more accurate and stable than the data types by themselves, and several factors better than the CPF-phase/pulsar TOA combination.

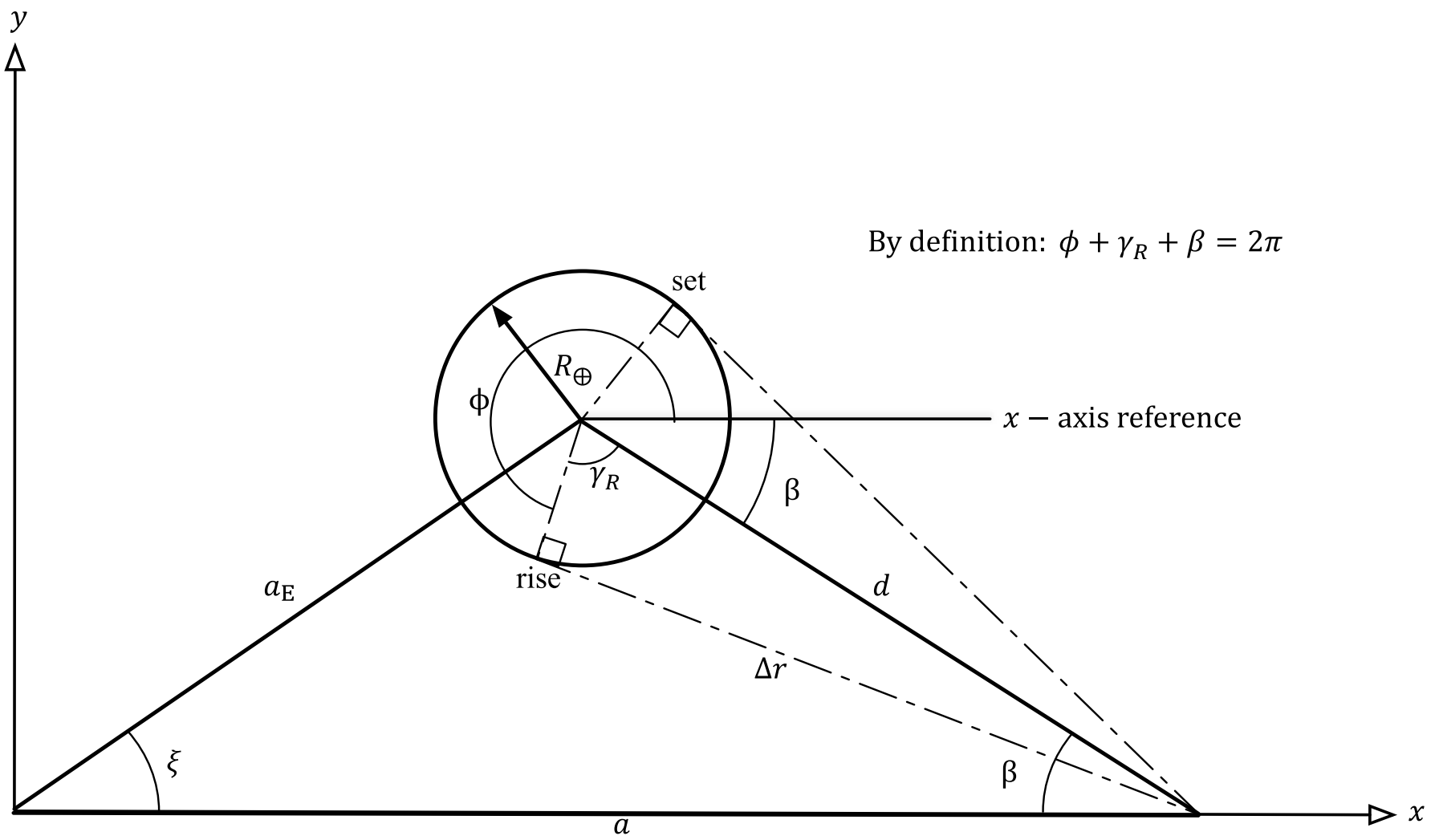

**Fig. 38** 2-d Earth station viewing geometry

## Appendix 1: Two-dimensional Earth Station Viewing Geometry

To develop a simple expression that constrains when an Earth station is in view of the spacecraft, a simplified 2-d Earth station viewing geometry is shown in Fig. 38. Note that the point at which an Earth station comes in view of the spacecraft is denoted 'rise' and when the station has rotated out of view is 'set.' Also note that, to simplify the drawing for the station setting, the spacecraft is assumed fixed in space. As will be shown, this assumption is a natural outcome of a the first-order analysis that we will conduct. From the drawing, the following planar geometry facts can be ascertained (where note that angles are measured as positive counterclockwise)

$$\begin{aligned} d^2 &= a_\oplus^2 + a^2 - 2a_\oplus a\cos\xi \text{ (cosine law)} \\ d^2 &= R_\oplus^2 + \Delta r^2 \text{ (Pythagorean theorem)} \end{aligned} \tag{142}$$

By combining these two relations we arrive at the following relationship for the distance between the Earth station at rise and the spacecraft

$$\Delta r^2 = a_\oplus^2 + R_\oplus^2 - 2a_\oplus a\cos\xi \tag{143}$$

These lead to the following approximate trigonometric relations (where the ordering parameter $\varepsilon$ has been inserted for tracking order position) for the central angle at Earth station rise

$$\sin\gamma_R \equiv \frac{\Delta r}{d} = \sqrt{\frac{a_\oplus^2 + a^2 - R_\oplus^2 - 2a_\oplus a\cos\xi}{a_\oplus^2 + a^2 - 2a_\oplus a\cos\xi}} \cong 1 - \varepsilon^2\frac{1}{2}\frac{R_\oplus^2}{a_\oplus^2 + a^2 - 2a_\oplus a\cos\xi} + O(\varepsilon^2) \tag{144}$$

and

$$\cos\gamma_R \equiv \frac{R_\oplus}{d} = \varepsilon\frac{R_\oplus}{\sqrt{a_\oplus^2 + a^2 - 2a_\oplus a\cos\xi}} \text{ (exact)} \tag{145}$$

Thus, to third order we have, we have the following approximate value for $\gamma_R$

$$\gamma_R \equiv \tan^{-1}\left(\frac{\sin\gamma_R}{\cos\gamma_R}\right) = \frac{\pi}{2} - \varepsilon\frac{R_\oplus}{\sqrt{a_\oplus^2 + a^2 - 2a_\oplus a\cos\xi}} + O(\varepsilon^3) \tag{146}$$

where note that the zeroth order approximation is $\pi/2$. Similarly, assuming the spacecraft is fixed for the observation period, we arrive at the central angle at Earth station set is $\gamma_S = \pi/2 + O(\varepsilon)$. For this first order analysis, we will simplify the central angle of the tracking view period $\gamma_S + \gamma_R = \pi + O(\varepsilon)$ and the time interval for this track is $\pi/\omega_\oplus (= T/2)$.

We now seek to develop an expression that relates the position of the Earth as defined by $\xi$ to the location of the Earth tracking station $i$ at rise, that is $\phi_{0,i}(\xi)$. We start with the following geometric fact

$$a_{\oplus}^2 = d^2 + a^2 - 2ad\cos\beta \text{ (cosine law)} \tag{147}$$

Substituting Eq. (147) into Eq. (142) and reducing leads to the following expression

$$\cos\ \beta = \frac{a - a_{\oplus} cos\xi}{\sqrt{a_{\oplus}^2 + a^2 - 2a_{\oplus} a\cos\xi}}, \tag{148}$$

and, using the sine law expression $\sin\beta/a_{\oplus} = \sin\xi/d$, yields

$$\sin\ \beta = \frac{a_{\oplus}\sin\xi}{a_{\oplus}^2 + a^2 - 2a_{\oplus} a\cos\xi} \tag{149}$$

Now, using $\gamma_R \cong \pi/2$, $\phi + \gamma_R + \beta$, Eqs. (148) and (149) leads to

$$\sin\phi = -\frac{a - a_{\oplus} cos\xi}{a_{\oplus}^2 + a^2 - 2a_{\oplus} a\cos\xi}, \cos\phi = -\frac{a - a_{\oplus} sin\xi}{a_{\oplus}^2 + a^2 - 2a_{\oplus} a\cos\xi} \tag{150}$$

## Appendix 2: Asymptotic Analysis of the Radio Information Matrix

To determine the invertibility of the aggregate information matrix when using ranging data $\mathbf{I}_{\rho}^{\Sigma}$ in Eq. (119) we examine the eigenvalues of the matrix to determine rank and numerical stability with different asymptotic expansion orders. To begin, the eigenvalues from the full nonlinear information matrix using one day

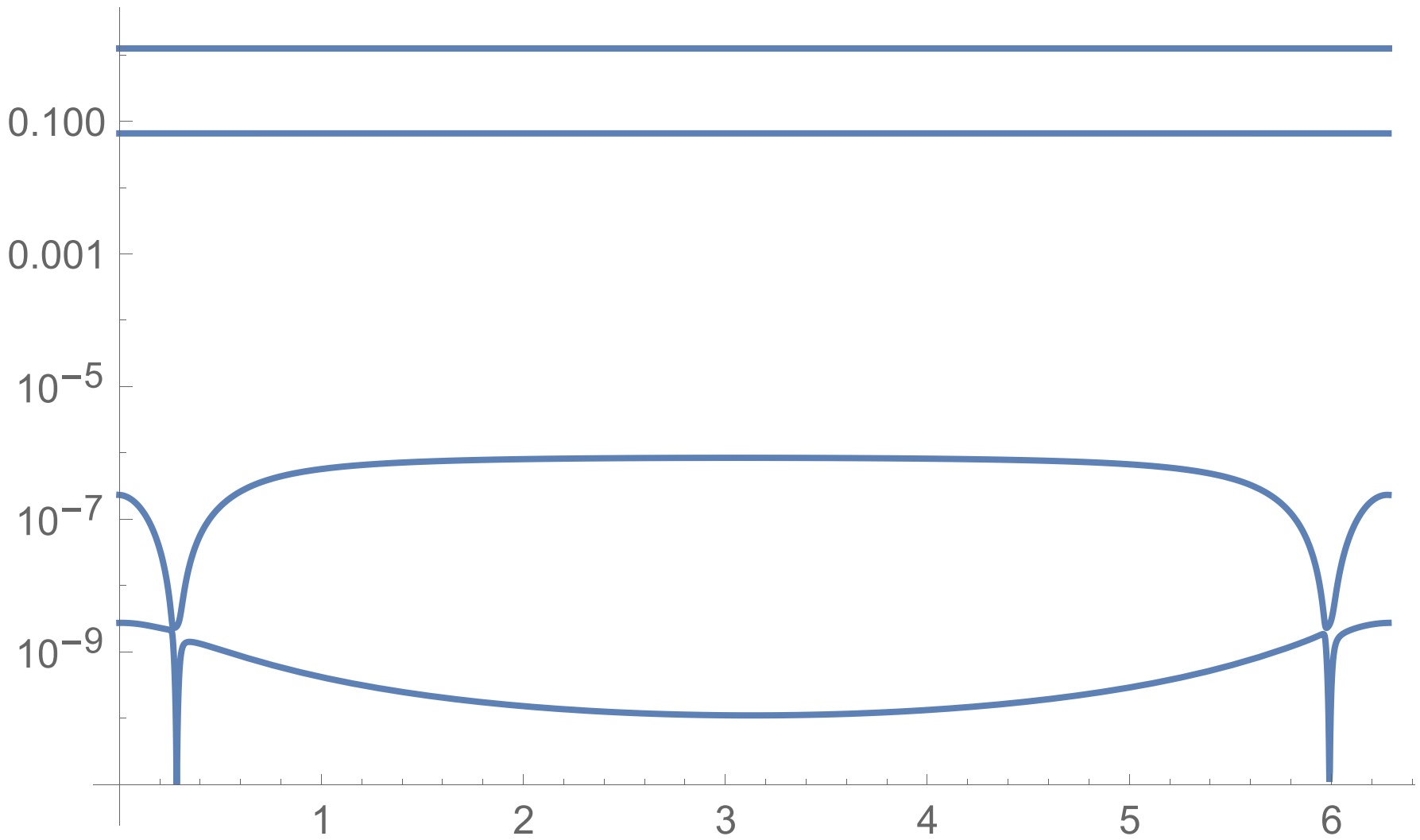

**Fig. 39** Eigenvalues of $\mathbf{I}_{\rho}^{\Sigma}$ for the full nonlinear case as a function of $\xi_0$

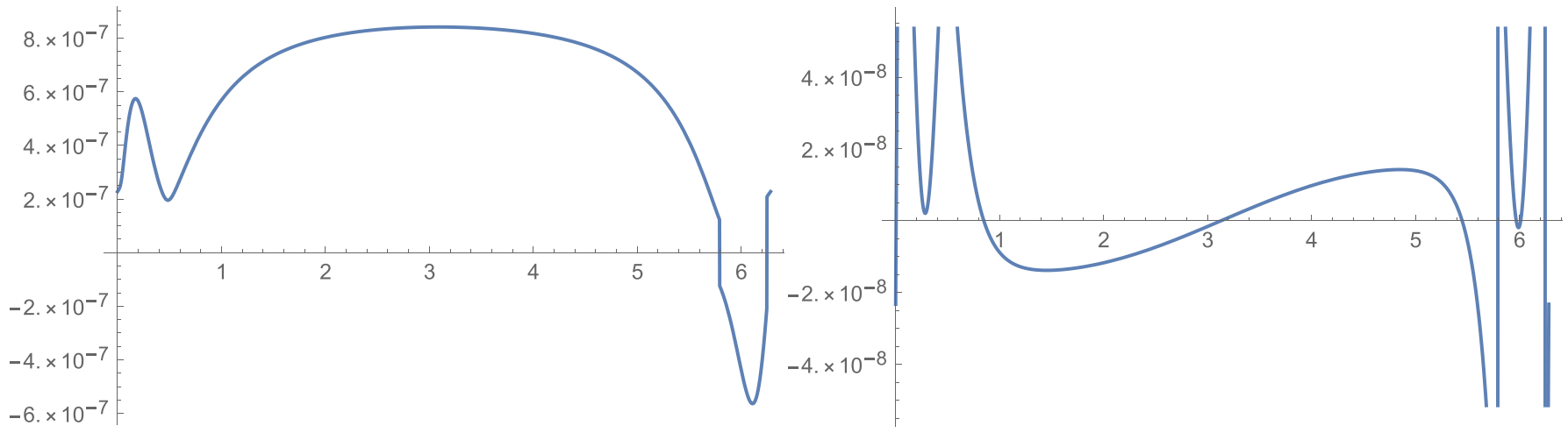

**Fig. 40** The third and fourth eigenvalues from the expansion $\mathbf{I}_{\rho}^{\Sigma(2)}$as a function of $\xi_0$

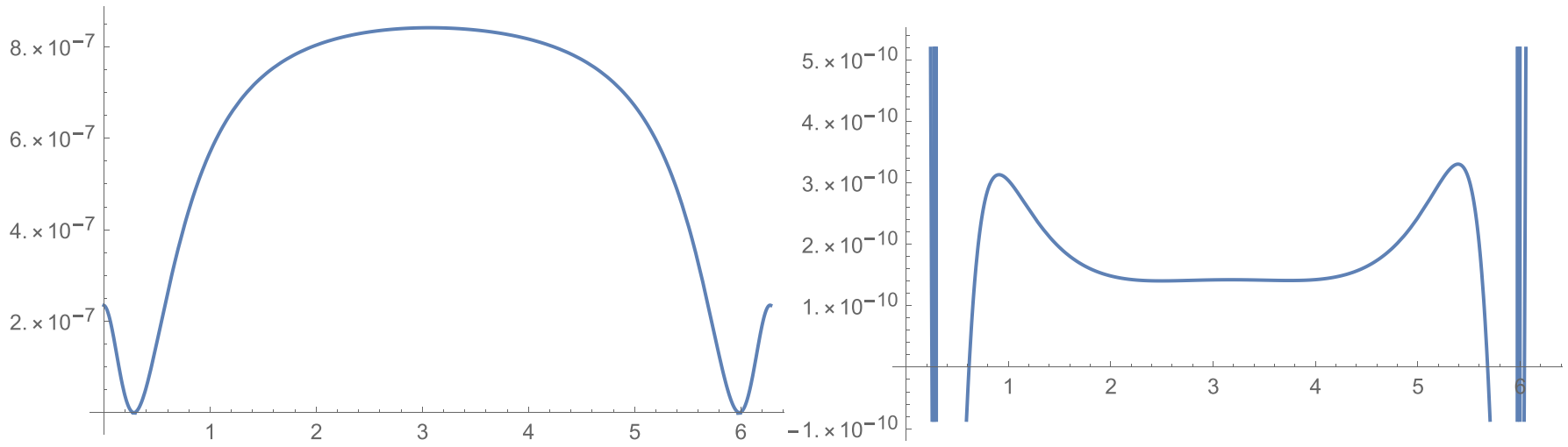

**Fig. 41** The third and fourth eigenvalues from the expansion $\mathbf{I}_{\rho}^{\Sigma(3)}$as a function of $\xi_0$

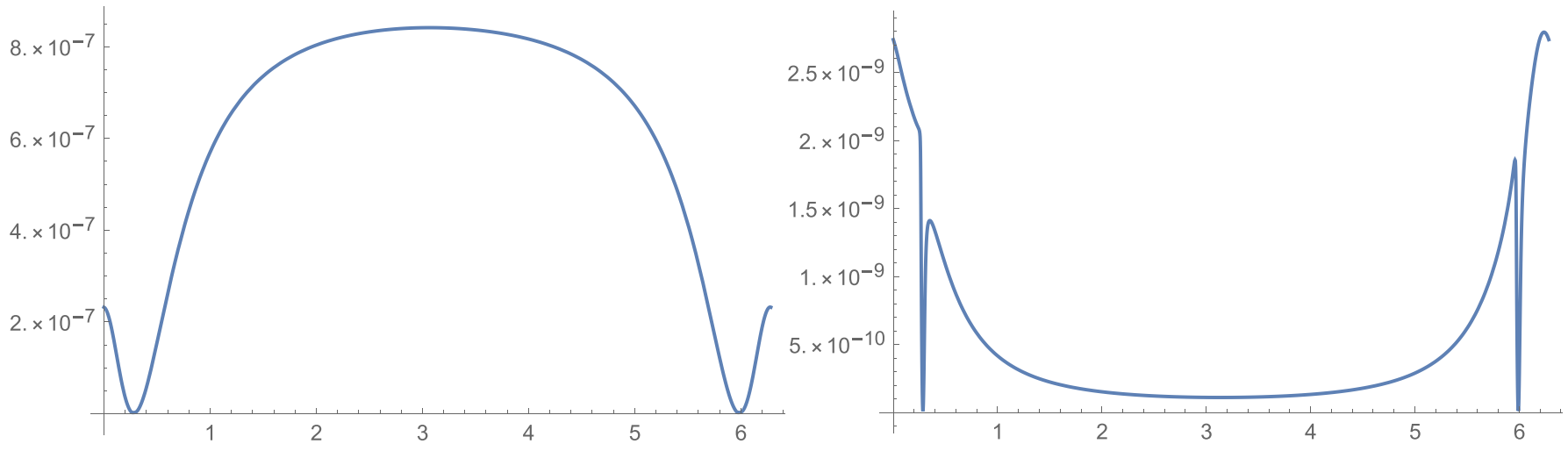

**Fig. 42** The third and fourth eigenvalues from the expansion $\mathbf{I}_{\rho}^{\Sigma(4)}$as a function of $\xi_0$

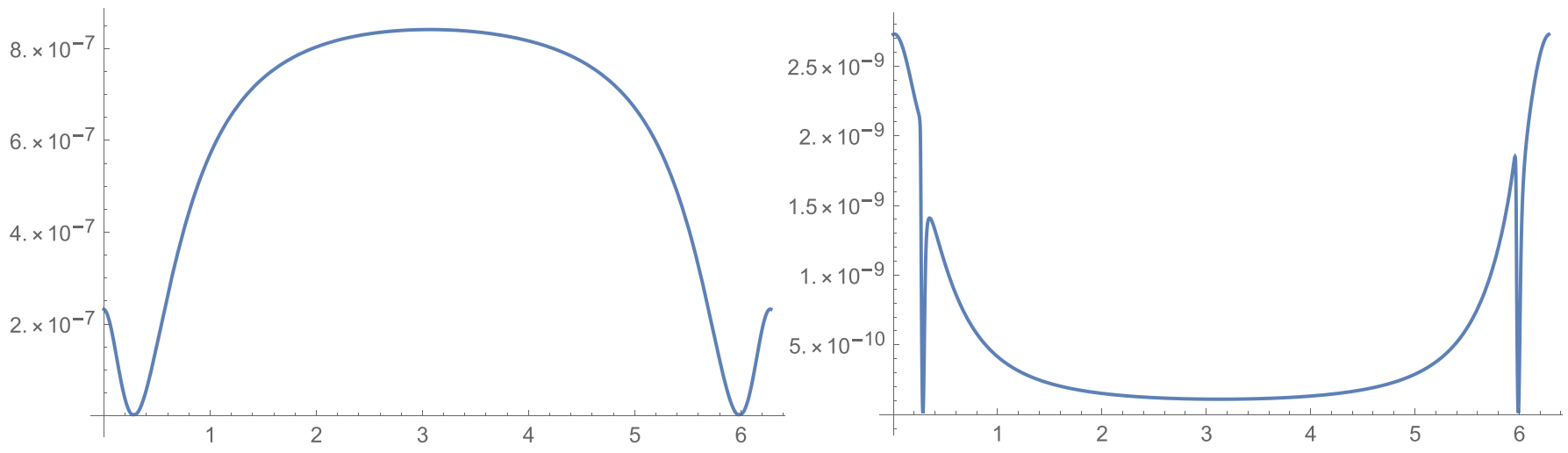

**Fig. 43** The third and fourth eigenvalues of $\mathbf{I}_{\rho}^{\Sigma}$ for the full nonlinear case as a function of $\xi_0$. Note the good agreement with the third and fourth eigenvalues of $\mathbf{I}_{\rho}^{\Sigma(4)}$ shown in Fig. 42

of tracking (i.e., $p=1$) are shown in log scale on Fig. 39. All the eigenvalues are positive and non-zero, thus conforming to the constraints of a covariance matrix that is symmetric and positive definite. The largest eigenvalues yield consistent results across all values of $\xi_0$; however, the two smallest approach zero and negatively affect the conditioning of the matrix (with condition number values ranging between $5\times10^9$ to $5\times10^{10}$). It is these two smallest eigenvalues that any asymptotic expansion will need to capture, at least qualitatively (positive and sufficiently far from zero), for the approximation to be valid. Indeed, the minimal asymptotic expansion necessary for full rank would be good to 3rd- order, i.e., $\mathbf{I}_\rho^\Sigma = \mathbf{I}_\rho^{\Sigma(2)} + \mathbf{O}\left(\varepsilon^3\right)$. Analysis has shown that the largest two eigenvalues are well recovered by $\mathbf{I}_\rho^{\Sigma(2)}$, but that the third and fourth eigenvalues are not, as exhibited in Fig. 40 where for some $\xi_0$ they are negative valued. For $\mathbf{I}_\rho^{\Sigma(3)}$the third smallest eigenvalue is now well behaved (always positive) as seen in Fig. 41. Finally, for $\mathbf{I}_\rho^{\Sigma(4)}$ the fourth smallest eigenvalue now conforms to the positive constraint, shown in Fig. 42, and compare well to the associated eigenvalues form the full nonlinear matrix $\mathbf{I}_\rho^\Sigma$ as seen in Fig. 43.

**Acknowledgments** This research was carried out at the Jet Propulsion Laboratory, California Institute of Technology, under a contract with the National Aeronautics and Space Administration. The authors would like to thank the support and feedback from Joseph Guinn and Jill Seubert while conducting this research.

On behalf of all authors, the corresponding author states that there is no conflict of interest.